\documentclass[preprint,preprintnumbers,amsmath,amssymb,12pt]{revtex4}

\usepackage{graphicx}
\usepackage{epsfig}
\usepackage{bm}
\usepackage{amsfonts}
\usepackage{subfigure}
\usepackage{multirow}
\usepackage{color}
\usepackage{float}
\usepackage{xcolor, soul}
\usepackage{hyperref}
\hypersetup{colorlinks=True,citecolor=blue}

\graphicspath{{./figs/}}

\sethlcolor{green}
\providecommand{\tabularnewline}{\\}

\newcommand*{\Mpl}{M_{\rm Pl}}

\newcommand{\e}{\epsilon}

\newcommand{\rd}{{\rm d}}

\def\thickhline{\noalign{\hrule height1.9pt}}

\begin{document}

\title{\textbf{Primordial black holes in non-minimal Gauss-Bonnet inflation in light of the PTA data}}

\author{Milad Solbi\footnote{miladsolbi@gmail.com} and Kayoomars Karami\footnote{kkarami@uok.ac.ir}}
\affiliation{\small{Department of Physics, University of Kurdistan, Pasdaran Street, P.O. Box 66177-15175, Sanandaj, Iran}}

\date{\today}

\begin{abstract}
Here, we investigate the formation of primordial black holes (PBHs) in non-minimal coupling Gauss-Bonnet inflationary model in the presence of power-law potentials.
We employ a two part coupling function to enhance primordial curvatures at small scales as well as satisfy Planck measurements at the CMB scale.
Moreover, our model satisfies the swampland criteria.
We find PBHs with different mass scales and demonstrate that PBHs with masses around $\mathcal{O}(10^{-14})M_{\odot}$ can account for almost all of the dark matter in the universe. In addition, we investigate the implications of the reheating stage and show that the PBHs in our model are generated during the radiation-dominated era.
Furthermore, we investigate the production of scalar-induced gravitational waves (GWs).
More interestingly enough is that, for the specific cases $D_{\rm n}$ in our model, the GWs can be considered as a source of PTA signal.
Also, we conclude that the GWs energy density parameter at the nano-Hz regime can be parameterized as $\Omega_{\rm GW_0} (f) \sim f^{5-\gamma}$, where the obtained $\gamma$ is consistent with the PTA Observations.
 \end{abstract}



\maketitle

\newpage
\section{Introduction}
Primordial black holes (PBHs), an intriguing topic in cosmology, have achieved the most attention in recent years.
For the first time, Zel'dovich and Novikov brought up the PBH existence possibility \cite{zeldovich:1967}.
After that, based on Hawking's studies, it was clarified that PBHs with mass scale less than $ 10^{14}g$ are evaporated, but the heavier ones did not evaporate entirely and still exist \cite{Hawking:1971,Carr:1974,Carr:2008}.
PBHs are formed in the early universe and can be considered as one of the dark matter candidates.
Recent studies show that the PBHs can comprise all or a part of the total dark matter in the universe \cite{Ivanov:1994,Khlopov1:2005,Frampton:2010,Belotsky:2014,Clesse:2015,Carr:2016,Inomata:2017,Nojiri:2011,Faraoni:2010,Capozziello:2011,Nojiri:2017}.

The detection of gravitational waves (GWs) by the LIGO-Virgo collaboration in 2015 renewed interest in PBHs \cite{Abbott:2016-a,Abbott:2016-b,Abbott:2017-a,Abbott:2017-b,Abbott:2017-c,Abbott:2019}.
The detected GWs were emitted after the merging of two heavy black holes (BHs).
The mass of the BHs involved in the LIGO/Virgo event was outside the astrophysical BHs mass spectrum.
Due to the PBHs can be formed in a wide mass range, the idea was raised that the source of detected GWs might be the PBHs. \cite{Bird:2016,Clesse:2017,Sasaki:2016}.

Investigating PBHs in the recent years has led to constrain their abundance.
As mentioned before, PBHs can form in a wide mass range.
Almost, all of this range is constrained by the observational upper limits \cite{Clark,Laha:2019,EGG,subaro,Icarus,Kepler,EORS,Boehm:2021,CMB,Kavanagh:2018,Chen:2022,Alcock:2001}.
These constraints represent the maximum fraction of total dark matter that a PBH can contain.
In addition to the upper limits, there is an allowed region in OGLE data \cite{OGLE}.
Six microlensing events with ultra-short timescales have been detected in OGLE data, which provide the allowed region for PBH formation. Therefore, PBHs with a mass scale of ${\mathcal{O}(10^{-5})}M_\odot $ can make up $\mathcal{O}(10^{-2})$ of total dark matter \cite{OGLE}.
The most impressive case in PBHs studies pertains to the mass range between ${\mathcal{O}(10^{-17})}M_\odot$ and ${\mathcal{O}(10^{-13})}M_\odot$.
The observations have not yet been able to set a limit on this mass interval.
Thus, PBHs with these mass scales could encompass all of the dark matter in the universe.

Early studies revealed that primordial curvature perturbations could generate overdense regions in the early universe.
These areas might collapse upon re-entry of the horizon and produce PBHs \cite{Hawking:1971,Carr:1974}.
The PBH formation with this procedure needs significant growth in the amplitude of the primordial curvature ${\cal P}_{s}(k)$ at small scales to the order of $\mathcal{O}(10^{-2})$ \cite{sato:2019}.
Furthermore, based on CMB measurements, the exact value of ${\cal P}_{s}(k)$ at the pivot scale ($k_{*}=0.05~{\rm Mpc}^{-1}$) has been calculated to be $2.1 \times 10^{-9}$ \cite{akrami:2018}.
Thus, the amplitude of primordial curvature must increase by seven orders of magnitude at small scales to be suitable for PBH formation.
It is noteworthy that increasing primordial curvature at small scales must not lead to inconsistency between the inflationary model and Planck observations at the CMB scale \cite{akrami:2018,bk18}.

On the other hand, one of the theoretical criteria that can assess inflationary models is recognized as the swampland criteria.
The swampland criteria are derived from string theory and consist of the distance conjecture and the de Sitter conjecture \cite{Garg:2019,Ooguri:2019,Kehagias:2018,Das:2019,Vagnozzi:2019add}.
The validity of the de Sitter conjecture contradicts the slow-roll condition in the standard inflationary model.
Therefore, examining the swampland conjectures could be an appropriate method for evaluating modified inflationary models.

Additionally, the formation of PBHs is accompanied by the production of scalar-induced gravitational waves \cite{Matarrese:1998,Mollerach:2004,Saito:2009,Garcia:2017,Cai:2019-a,Cai:2019-b,Cai:2019-c,Bartolo:2019-a,Bartolo:2019-b,Wang:2019,Fumagalli:2020b,Domenech:2020a,Domenech:2020b,Hajkarim:2019,Kohri:2018,Xu:2020,Fu:2020,GDomenech:2021a,GDomenech:2021b,MRGangopadhyay:2023b,SChoudhury:2023e,GBhattacharya:2023,rezazadeh:2022,Braglia:2021add,Papanikolaou:2021add,Papanikolaou:2022add,Papanikolaou:2024add}.
In other words, the collapse of the enhanced curvature perturbations at horizon re-entry leads to the generation of considerable metric perturbations.
At the second order, the scalar and tensor perturbations can be coupled, unlike the linear level.
Consequently, the scalar metric perturbations at the second order can lead to stochastic GWs background production \cite{Cai:2019-a,Cai:2019-b,Cai:2019-c,Bartolo:2019-a,Bartolo:2019-b,Wang:2019,Fumagalli:2020b}.
Therefore, detecting induced GWs signals can provide a novel approach to discovering the existence of PBHs.
Recently, the pulsar timing array teams  have discovered  low-frequency stochastic gravitational wave background in nano-hertz regime \cite{NG15a,NG15b,NG15c,NG15d,epta1:add,epta2:add,epta3:add,epta4:add,epta5:add}. In the meantime, the most recent data reported by the PTA signal observation have attracted considerable attention \cite{Vagnozzi:2021add,Vagnozzi:2022add,Basilakos:a2024add,Basilakos:b2024add,Oikonomou:2023add}.
Although the nature of the PTA signals has not yet to be confirmed, given the wide range of implications that such a signal could have for early universe cosmology, it is worth seriously analyzing the signal and proposing suitable mechanisms that could generate.
One of the most interesting explanations for the source of the detected signal is the scalar-induced gravitational waves propagated simultaneously with the PBH generation.

As mentioned above, PBH formation requires significant growth in the scalar power spectrum at small scales.
In the literature, various mechanisms have been suggested for increasing the amplitude of the scalar power spectrum \cite{Cai:2018,Ballesteros:2019,Ballesteros:2020a,Ballesteros:2020b,Kamenshchik:2019,Inomata:2018,Ezquiaga:2018,Germani:2017,Di:2018,Ballesteros:2018,Dalianis:2019,chen:2019,Ozsoy:2018,Atal:2019,mishra:2020,fu:2019,lin:2020,Khlopov:2010,Belotsky1:2014,Belotsky:2019,Braglia:2020,Braglia2:2020,shiPi:2018,Fumagalli:2020a,Sypsas:2020,Dalianis:2020,Heydari:2023a,Heydari:2023b,Solbi-a:2021,Solbi-b:2021,Teimoori-b:2021,Teimoori:2021,Heydari:2022,Heydari-b:2022,Rezazadeh:2021,Kawaguchia:2023,Kawai:2021,Zhang:2022,Ashrafzadeh:2023,mahbub:2020,SChoudhury:2014,YCai:2021,AChakraborty:2022,TPapanikolaou:2022,TPapanikolaou:2023a,TPapanikolaou:2023b,GDomenech:2023,YCai:2023,Laha:2020,Ali-Haimoud:2017,Wang:2018,Braglia:b2021add,Choudhury:a2023add,Choudhury:b2023add,Choudhury:c2023add,Choudhury:a2024add,Choudhury:b2024add,Choudhury:c2024add,Sharma:2024add}.
It is worth noticing that the scalar power spectrum cannot sufficiently increase under the slow-roll conditions \cite{Ballesteros:2019,Kamenshchik:2019}.
Indeed, around the PBH formation region, the slow-roll conditions are violated, and inflaton goes under the ultra-slow-roll (USR) regime.
For instance, if a single-field inflationary potential has an inflection point, considerable growth in the power spectrum can be possible.
This happens because the inflaton experiences an USR phase in the vicinity of the inflection point, which causes a significant decrease in the inflaton velocity.
As a result, curvature perturbation increases significantly  \cite{Germani:2017,Di:2018,Ezquiaga:2018,Wang:2018,Dalianis:2019}.
Also, the oscillation of sound speed may lead to increase of the primordial curvature fluctuations \cite{Cai:2018,chen:2019}.
Additionally, in a single field model with a non-canonical kinetic term, the scalar power spectrum can be significantly increased if the sound speed tends to zero.
The PBHs can be produced using both of these mechanisms \cite{Ballesteros:2019,Kamenshchik:2019}.
It was pointed out that for sufficient growth of the scalar power spectrum, fine-tuning of the model parameters is required in all of the mentioned mechanisms.

Recently, PBH formation in the Gauss-Bonnet (GB) framework has been studied in \cite{Kawaguchia:2023,Kawai:2021,Zhang:2022,Ashrafzadeh:2023}.
The PBH production in the GB gravity with the natural potential has been investigated in \cite{Kawai:2021}.
%
Furthermore, the formation of PBHs from the E-model with a GB term has been investigated in \cite{,Zhang:2022}.
Moreover, the formation of PBHs from Higgs inflation with a GB coupling has been explored in \cite{Kawaguchia:2023}.
In the proposed model, the authors have considered two couplings for the scalar field, including a non-minimal coupling to gravity as well as a GB coupling.
They have utilized the inflaton non-minimal coupling to gravity function to enhance the consistency of the model with observations at the CMB scale.
In addition, by using the GB coupling function, they showed that the proposed mechanism can enhance the primordial curvatures at small scales and consequently lead to PBH generation.
Additionally, the possibility of PBHs generation in a scalar field inflationary model coupled to the GB term
has been investigated in \cite{Ashrafzadeh:2023}.
%

All mentioned in above motivate us to study the formation of PBHs and scalar-induced GWs in non-minimal coupling GB inflation driven by power-law potentials. To do this, we try to employ a suitable coupling function to the GB term to remedy the predictions of power-law potentials on CMB scale as well as producing PBHs on smaller scales. Additionally,
one of our main goals is to examine the viability of the model in light of the PTA observations. The structure of paper is as follows. In
Section \ref{sec2}, at first we review the GB inflationary model. In
Section \ref{sec3}, we introduce the mechanism of PBH formation in our model.
In Section \ref{sec4}, we investigate the reheating considerations.
Section \ref{sec5} focuses on estimating the abundance of PBHs, while Section \ref{sec6} examines the scalar-induced GWs.
Finally, in Section \ref{sec7}, we present our conclusions.

\section{Non-minimal coupling Gauss-Bonnet inflationary model}
\label{sec2}
The Einstein-Gauss-Bonnet action for the non-minimal coupling GB inflationary model is considered as follows \cite{Jiang:2013,Koh:2014,Guo:2010,Odintsov:2020,Gao:2020,RashidiNozari:2020,AziziNozari:2022,Oikonomou:2021kql}
%
%
\begin{align}
	{\cal S}=\frac{1}{2}\int {\rm d}^4 x \sqrt{-g} \bigg[M_{\rm Pl}^{2}R
	-g^{\mu \nu} \partial_{\mu}\phi
	\,\partial_{\nu}\phi- 2 V(\phi)-\xi(\phi)\,\mathcal{G}\bigg],
	\label{action1}
	\end{align}
where $R$ denotes the Ricci scalar, $\phi$ is the scalar field, $\mathcal{G}$ is the invariant scalar GB term
and $\xi(\phi)$  describes the coupling function between the scalar field and GB term.
Here $M_{\rm Pl}=(8\pi G)^{-1/2}$ stands for the reduced Planck mass.

From the action (\ref{action1}), for a spatially flat Friedmann-Robertson-Walker (FRW) universe, the Friedmann equations
take the forms 
\begin{eqnarray}
\label{bge1}
& 3{M_{\rm Pl}^{2}}H^2 = \frac{1}{2}\dot{\phi}^2+V(\phi)+12\dot{\xi}H^3, \\
\label{bge2}
& -2\Mpl^{2}\dot{H} =\dot{\phi}^2 - 4\ddot{\xi}H^2 - 4\dot{\xi}H\big(2\dot{H} - H^2\big),
\end{eqnarray}
where $H\equiv\dot{a}/a$ is the Hubble parameter, $a$ is the scale factor and dot shows the time derivative.
In addition, the scalar field equation of motion can be obtained by variation of the action (\ref{action1}) with respect to the scalar field $\phi$ as follows 
\begin{align}\label{bge3}
\ddot{\phi}+3H\dot{\phi}+ V_{,\phi}= -12 \xi_{,\phi}H^2\big(\dot{H}+H^2\big),
\end{align}
where the subscript $({,\phi})$ describes derivative with respect to $\phi$.
It is necessary to mention that by utilizing the first and second Friedmann equations (\ref{bge1}) and (\ref{bge2}), one can derive the following equation
\begin{equation}
\dot{H}=\frac{1}{2{\cal D}}
\Big[ -4 \xi_{,\phi}H^2 \big(V_{,\phi}+12 \xi_{,\phi}H^4 \big)-\dot{\phi}^2
\big(1-4 \xi_{,\phi \phi}H^2 \big)-16 \dot{\phi} \xi_{,\phi}H^3 \Big],
\label{Hdot}
\end{equation}
where ${\cal D} \equiv \Mpl^{2} - 4\xi_{,\phi} H \big(\dot{\phi}-6 \xi_{,\phi} H^3 \big)$.
This equation is employed for subsequent calculations.
In addition, we can accurately determine the value of the curvature power spectrum by numerical solving of the Mukhanov-Sasaki (MS) equation given by \cite{Kawaguchia:2023,Kawai:2021,Zhang:2022}
\begin{equation}
u_k''+\left( k^2-\frac{Z_s''}{Z_s}\right)u_k=0,
\label{M.S}
\end{equation}
where $u_k=Z_s \zeta_k$, $Z_s=a \sqrt{2Q_s c_s}$.
Here, the parameters $Q_s$ and $c_s$ are given by
\begin{eqnarray}
	Q_s=16\frac{\Sigma}{\Theta^2}Q_t^2+12Q_t,\qquad
	c_s^2=\frac{1}{Q_s}\left[ \frac{16}{a}\frac{\rd}{\rd t}
	\left(\frac{a}{\Theta}Q_t^2\right)-4c_t^2 Q_t\right],
\end{eqnarray}
where
\begin{eqnarray}
&\Sigma=\frac{1}{2}\dot{\phi}^2-3 \Mpl^2 H^2+24 H^3 \xi_{,\phi}\dot{\phi},\qquad\quad
\Theta=\Mpl^2 H-6 H^2 \xi_\phi\dot{\phi},\nonumber\\\label{cT}
&Q_t=\frac{1}{4}\big(-4H\xi_{,\phi}\dot{\phi}+\Mpl^2 \big),\quad c_t^2=\frac{1}{4Q_t}\big(\Mpl^2-4\xi_{,\phi\phi}\dot{\phi}^2-4\xi_{,\phi}\ddot{\phi}\big).
	 \label{cs}
\end{eqnarray}
%
%
%
Consequently, the scalar power spectrum can be obtained as
\begin{align}
	{\cal P}_{s}(k)\equiv \frac{k^3}{2\pi^2} \big| \zeta_k (\tau_{s},{ k})\big|^2.\label{powerspectraZeta}
\end{align}
It is noteworthy that the parameters $c_s^2$, $Q_s$, $c_t^2$, and $Q_t$ must possess positive values to prevent the occurrence of ghost and Laplacian instabilities.
Similarly, the tensor power spectrum can be calculated numerically using the following MS equation \cite{Kawaguchia:2023,Kawai:2021}
\begin{equation}
u_k''+\left( k^2-\frac{Z_t''}{Z_t}\right)u_k=0,
\label{M.St}
\end{equation}
where $u_{k}=Z_{t}\zeta_{t}$, and $Z_{t}=a \sqrt{2Q_t c_t}$.
Therefore, the tensor power spectrum can be estimated by
\begin{align}\label{powerspectraZetaT}
 {\cal P}_{t}(k)\equiv 4\frac{k^3}{2\pi^2} \left| \zeta_t (\tau_{t},{ k})\right|^2.
\end{align}
Note that prime notations in Eqs. (\ref{M.S}) and (\ref{M.St}) denote derivatives with respect to conformal time; $\tau_s=\int\left({c_s}/{a}\right){\rm d}t$ for the scalar curvature and $\tau_t=\int\left({c_t}/{a}\right){\rm d}t$ for the tensor perturbations.

In non-minimal coupling GB gravity, the slow-roll parameters are defined as follows 
    %
	%
\begin{align}\label{SRH}
\epsilon_1\equiv\frac{-\dot{H}}{H^2},~~\epsilon_2\equiv\frac{\dot{\epsilon_1}}{H \epsilon_1},~~ \delta_1\equiv4\dot{\xi}H,~~\delta_2\equiv\frac{\dot{\delta_1}}{H \delta_1}.
\end{align}
In the slow-roll approximation $|\epsilon_{i}|\ll 1$ and $|\delta_{i}|\ll 1$, the background Eqs. (\ref{bge1})-(\ref{bge3}) are simplified into
\begin{eqnarray}
	\label{sre1}
& H^2 \simeq \frac{V}{3\Mpl^{2}},\\
	\label{sre2}
& -2{\Mpl^{2}}\dot{H} \simeq \dot{\phi}^2 + 4\dot{\xi}H^3,\\
	\label{sre3}
& 3H\dot{\phi} +V_{,\phi}\simeq  - 12\xi_{,\phi}H^4.
\end{eqnarray}
Consequently, using the above equations, the slow-roll parameters (\ref{SRH}) can be written as 
\begin{eqnarray}\label{SRV}
  & \epsilon_1 \simeq \frac{Q}{2} \frac{V_{,\phi}}{V}, ~~\qquad \epsilon_2 \simeq -Q \bigg( \frac{V_{,\phi\phi}}{V_{,\phi}} -\frac{V_{,\phi}}{V} + \frac{Q_{,\phi}}{Q} \bigg), \nonumber \\
   & \delta_1 \simeq -\frac{4Q}{3} \frac{\xi_{,\phi}V}{\Mpl^4}, \quad \delta_2 \simeq -Q \bigg( \frac{\xi_{,\phi\phi}}{\xi_{,\phi}} +\frac{V_{,\phi}}{V} + \frac{Q_{,\phi}}{Q} \bigg),
\end{eqnarray}
where
$Q \equiv \Mpl^2\Big(\frac{V_{,\phi}}{V} + \frac{4 \xi_{,\phi}V}{3\Mpl^4}\Big) $.
Furthermore, under the slow-roll approximation, the scalar and tensor power spectrum can be expressed as \cite{Kawaguchia:2023,Felice:2011}
\begin{align}
{\cal P}_{s} = \frac{H^2}{8\pi^2 Q_s c_s^3}\biggr|_{c_s k=a H},
\quad
{\cal P}_{t}=\frac{H^2}{2\pi^2 Q_t c_t^3}\biggr|_{c_t k=a H}.
\label{ps-sr}
\end{align}
	%
	%
According to Planck 2018 measurements, the amplitude of curvature perturbations  at the pivot scale ($k_{*}=0.05~\rm Mpc^{-1}$) is determined to be ${\cal P}_s(k_{*})=2.1\times{10}^{-9}$ \cite{akrami:2018}.

The scalar spectral index $n_s$ and the tensor-to-scalar ratio $r$ can be represented as follows \cite{Jiang:2013}
%
\begin{equation}
n_s-1 \equiv \frac{\rd \ln {\cal P}_s}{\rd \ln k}\biggr|_{c_s k=aH} \simeq -2\epsilon_{1} - \left(\frac{2\epsilon_1 \epsilon_2 - \delta_1 \delta_2}{2\epsilon_1 - \delta_1}\right), \label{ns-r1}
\end{equation}
\begin{equation}
r \equiv \frac{{\cal P}_t}{{\cal P}_s} \simeq 8 \left( 2\epsilon_1 - \delta_1 \right).
\end{equation}
The value of the scalar spectral index and the upper limit on the tensor-to-scalar ratio, derived from Planck 2018 TT, TE, EE + lowE + lensing + BK18 + BAO \cite{akrami:2018, bk18}, are given by $n_s = 0.9649 \pm 0.0042$ (68\% CL) and $r < 0.036$ (95\% CL), respectively.

\section{PBH formation mechanism}\label{sec3}
To generate PBHs, a significant enhancement in the primordial curvature at small scales is necessary. In non-minimal coupling GB inflation, selecting an appropriate coupling function can lead to a suitable enhancement in the scalar power spectrum \cite{Kawaguchia:2023,Kawai:2021,Zhang:2022,Ashrafzadeh:2023}.
The proposed coupling function must satisfy two objectives containing compatibility of the model with the Planck observations at the CMB (large) scale and formation of PBHs at small scales.
To achieve both of these aims, we introduce the following non-minimal coupling function as
\begin{align}\label{xip}
\xi (\phi) = \xi_{1}(\phi) \big(  1 + \xi_{2}(\phi) \big),
\end{align}
where
\begin{eqnarray}
&\xi_{1} (\phi) =M \exp{\big(-\alpha \phi  / \Mpl\big)},\\
&\xi_{2} (\phi) = d\tanh\left[c \left(\phi-\phi_c \right)\right].
\end{eqnarray}
%
In the above,
$\xi_1(\phi)$ ensures CMB compatibility via its exponential form, similar to successful GB inflation models \cite{Jiang:2013}.
However, PBH formation requires a different behavior at small scales. Here, the critical term is $\xi_{2} (\phi)$, particularly its derivative is important.
The chosen function acts as a switch, enhancing the derivative of $\xi(\phi)$ (which appears in background Eqs. (\ref{bge3})-(\ref{Hdot})) near a specific field value ($\phi = \phi_c$).
This localized amplification is essential for PBH formation. The parameters appeared in $\xi_{2} (\phi)$ control this enhancement, allowing model calibration for desired PBH abundance.


Here, $\alpha$, $M$, and $d$ are dimensionless variables, while  $\phi_c$ and $c$ have dimension of mass and inverse of mass, respectively.
Furthermore, $d$ and $c$ represent height and width of the scalar power spectrum at the peak position, respectively.

\begin{figure}[H]
\centering
\begin{minipage}[b]{1\textwidth}
\subfigure[\label{fig-phi} ]{ \includegraphics[width=0.44\textwidth]%
{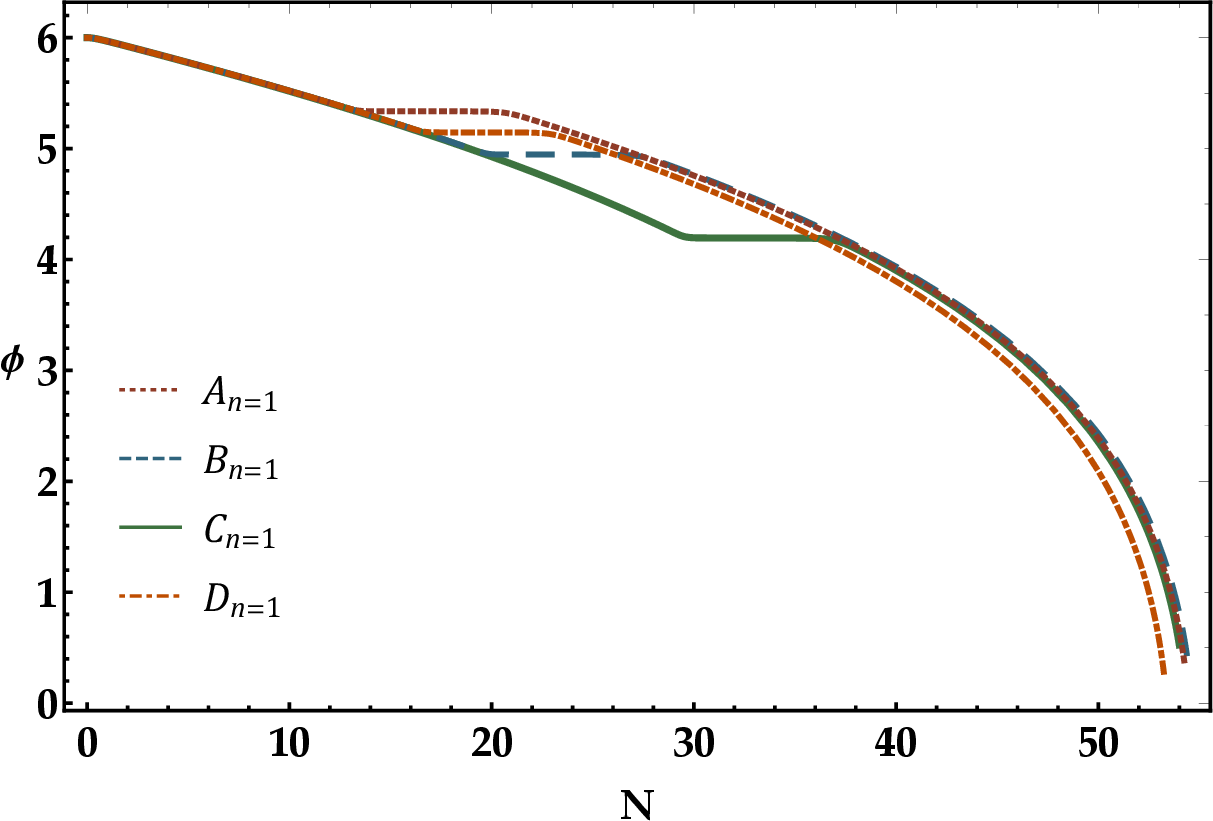}}\hspace*{0.3cm}
\subfigure[\label{phi-e1n} ]{ \includegraphics[width=0.48\textwidth]%
{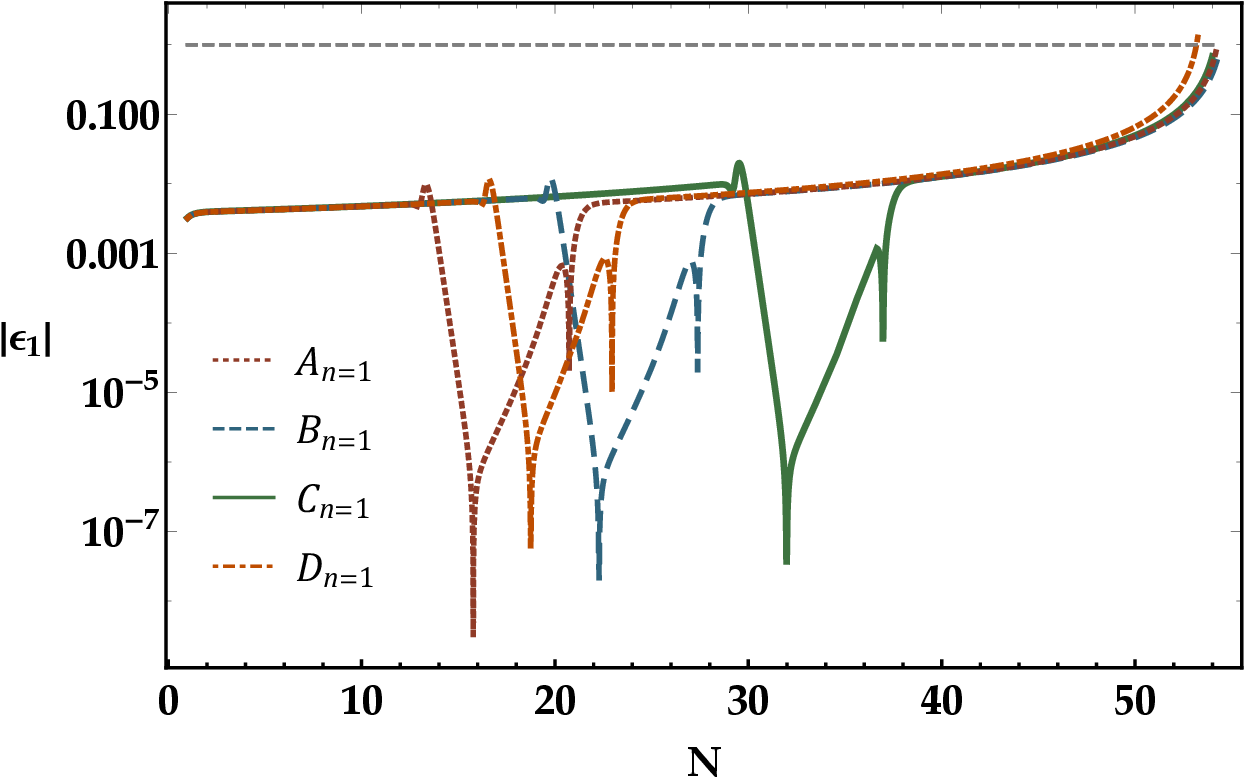}}\\
\hspace*{-0.5cm}
\subfigure[\label{phi-e2}]{
 \includegraphics[width=.46\textwidth]%
{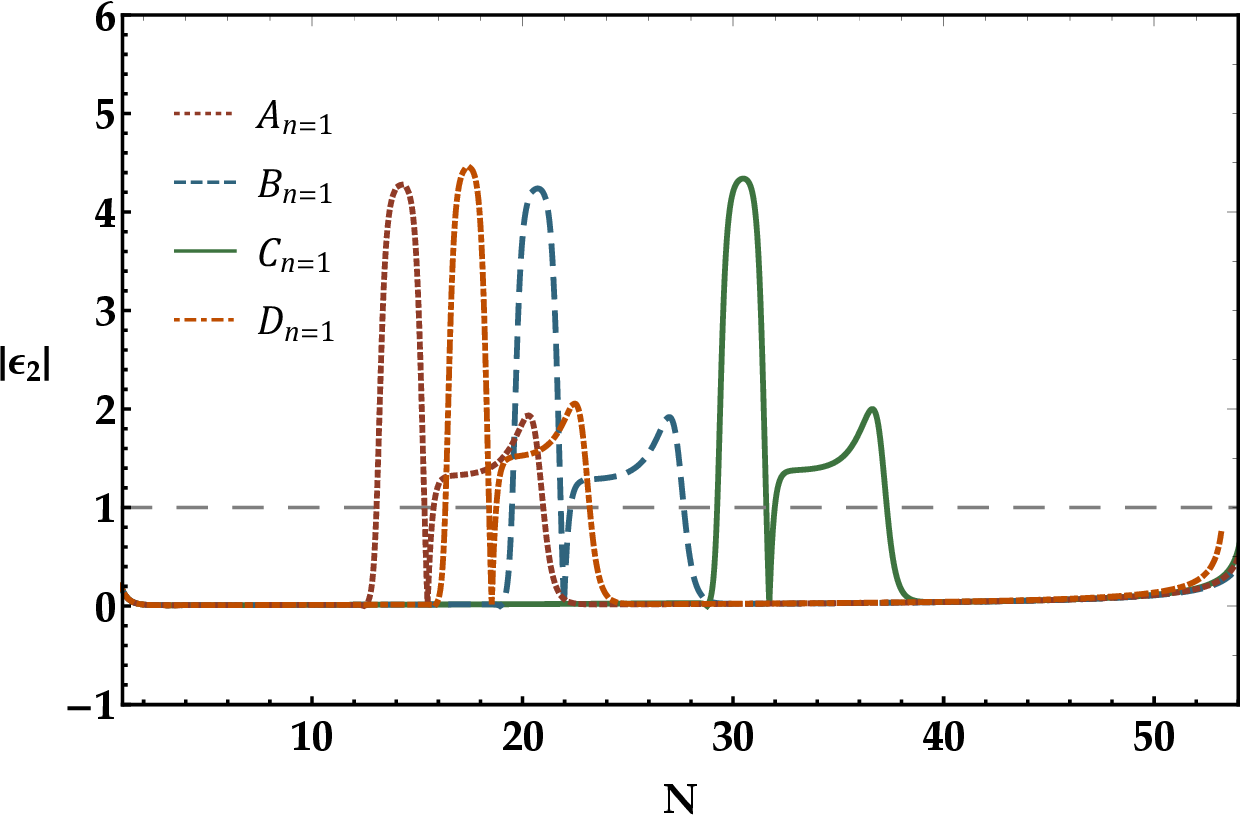}}\hspace*{0.3cm}
\subfigure[\label{phi-cs}]{
 \includegraphics[width=.46\textwidth]%
{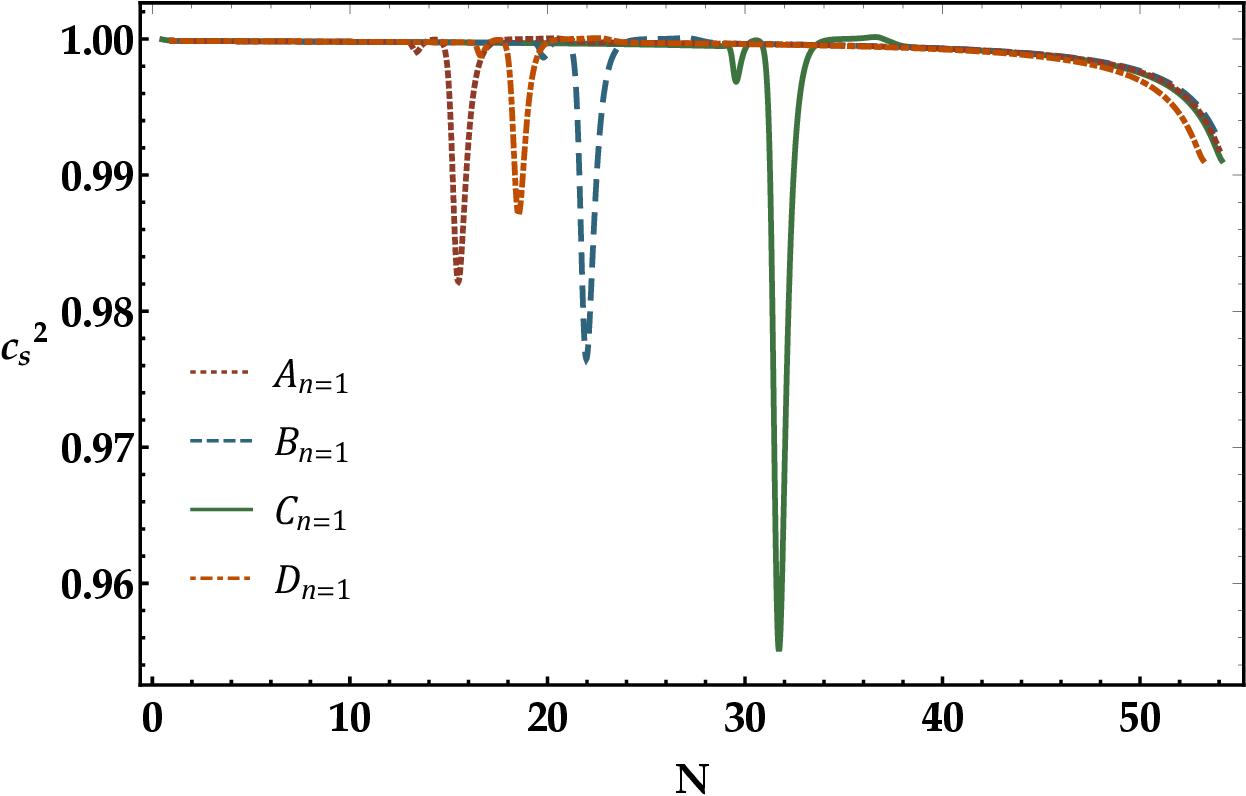}}
\end{minipage}
\caption{Evolutions of (a) the scalar filed $\phi$, (b) the first slow-roll parameter $\epsilon_1$, (c) the second slow-roll parameter $\epsilon_2$, and (d) the square sound speed $c_s^2$ versus the $e$-fold number $N$ for our model. In panel (c), the grey dashed line indicates the slow-roll conditions are violated. These evolutions are obtained using parameter sets of cases $A_{\rm n=1}$, $B_{\rm n=1}$, $C_{\rm n=1}$, and $D_{\rm n=1}$ in Table \ref{tab1-phi}, with $N_{*}=0$ at the horizon exit.}\label{linear}
\end{figure}

\begin{figure}[H]
\centering
\begin{minipage}[b]{1\textwidth}
\subfigure[\label{fig-phi} ]{ \includegraphics[width=0.44\textwidth]%
{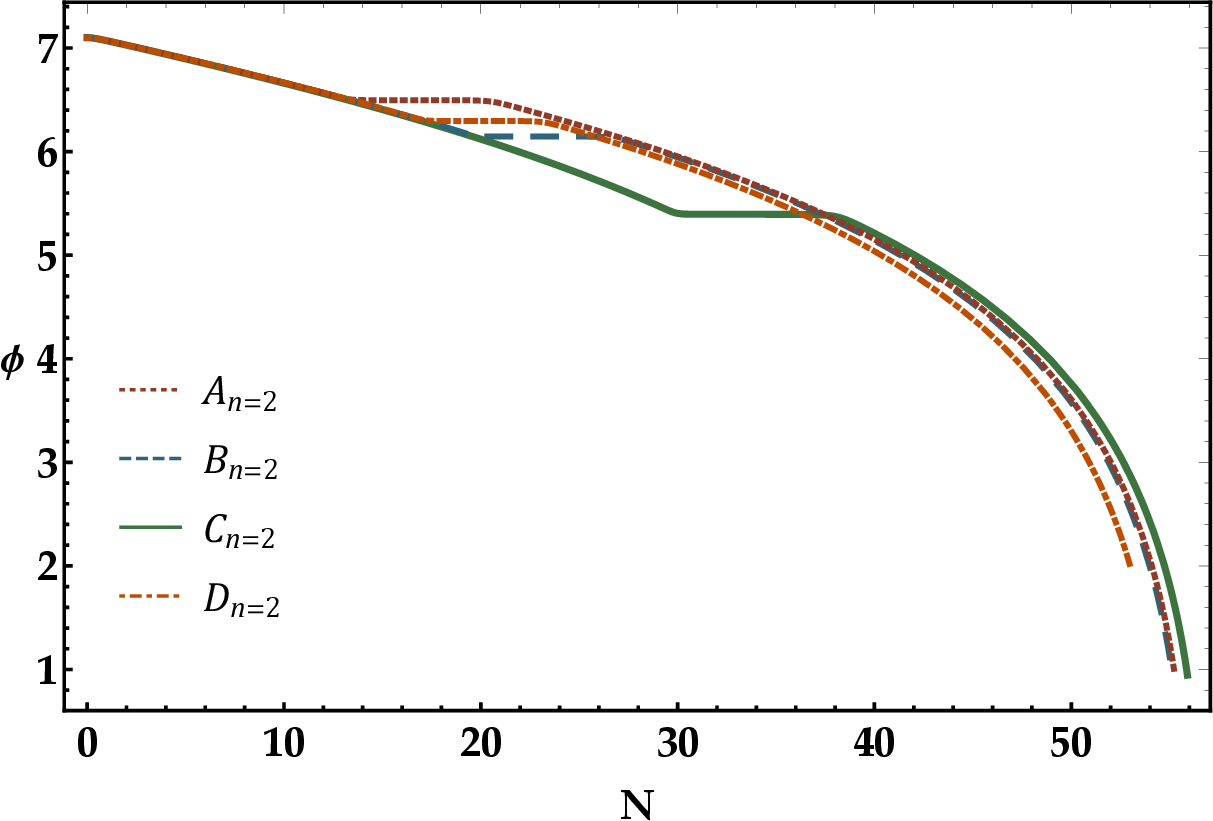}}\hspace*{0.3cm}
\subfigure[\label{phi-e1n} ]{ \includegraphics[width=0.48\textwidth]%
{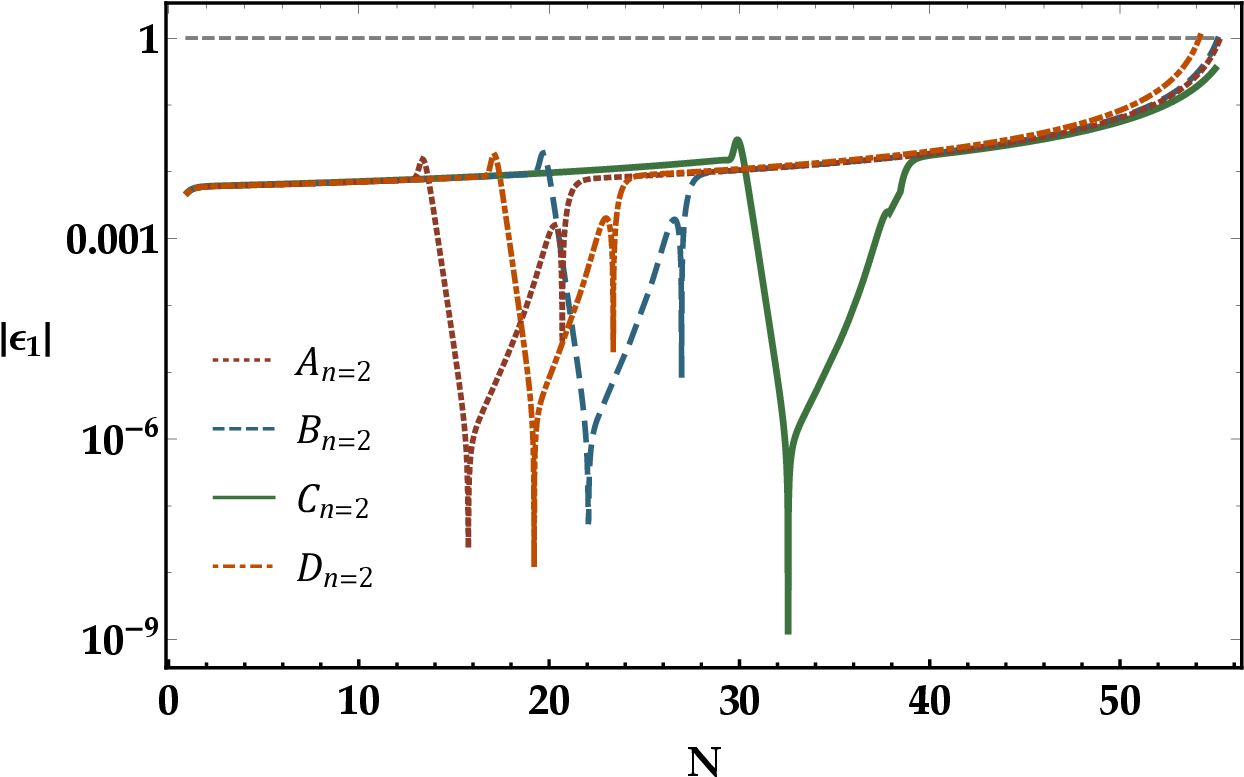}}\\
\hspace*{-0.5cm}
\subfigure[\label{phi-e2}]{
 \includegraphics[width=.46\textwidth]%
{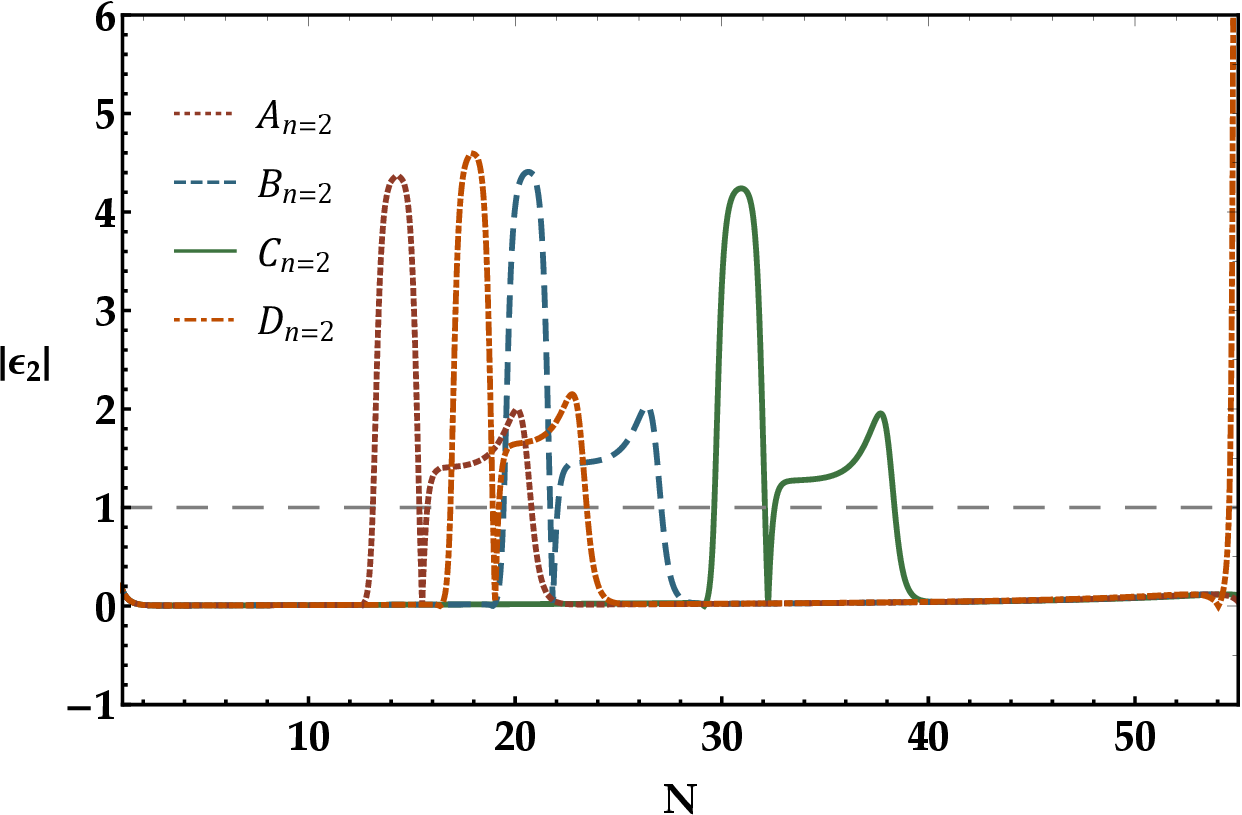}}\hspace*{0.3cm}
\subfigure[\label{phi-cs}]{
 \includegraphics[width=.46\textwidth]%
{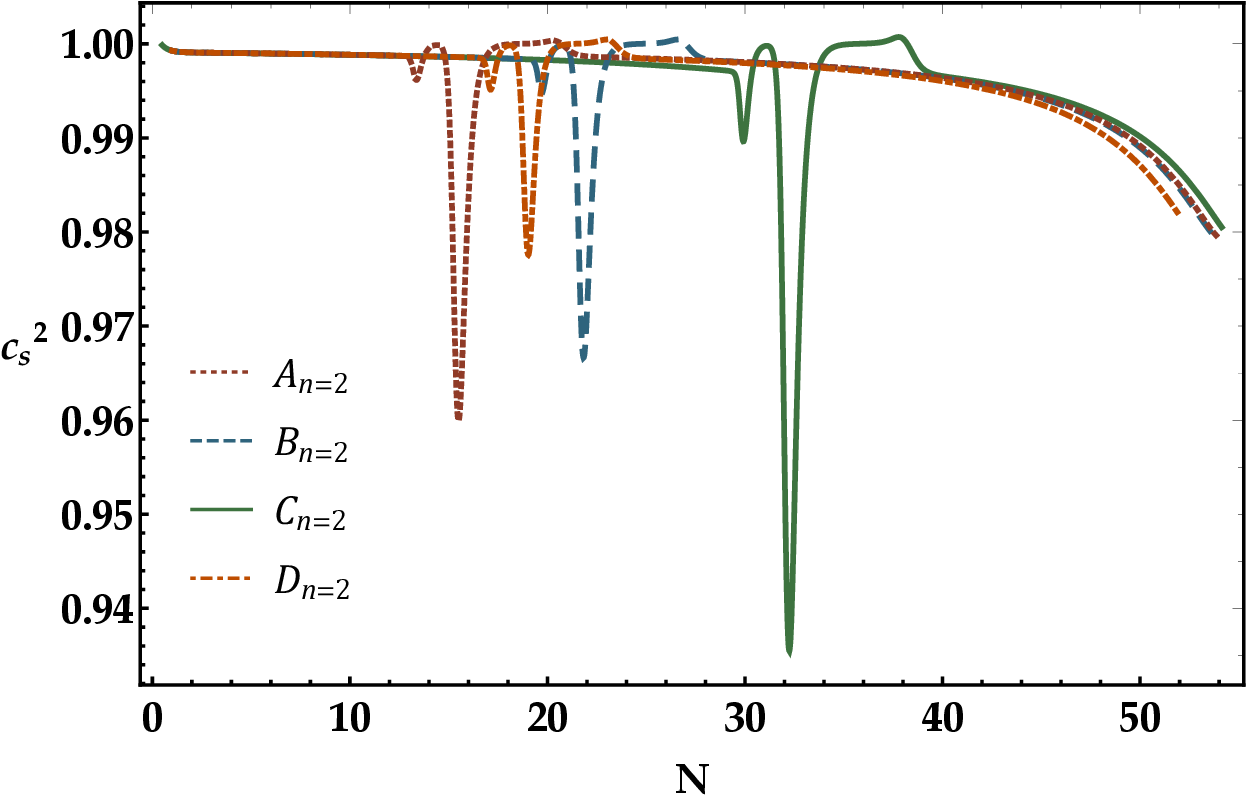}}
\end{minipage}
\caption{Same as Fig. \ref{linear}, but for the case $n = 2$. These evolutions are obtained using the parameter set of cases $A_{\rm n=2}$, $B_{\rm n=2}$, $C_{\rm n=2}$, and $D_{\rm n=2}$ in Table \ref{tab1-phi}, with $N_{*}=0$ at the horizon exit.}
\label{quadratic}
\end{figure}

\begin{figure}[H]
\centering
\begin{minipage}[b]{1\textwidth}
\subfigure[\label{fig-phi} ]{ \includegraphics[width=0.44\textwidth]%
{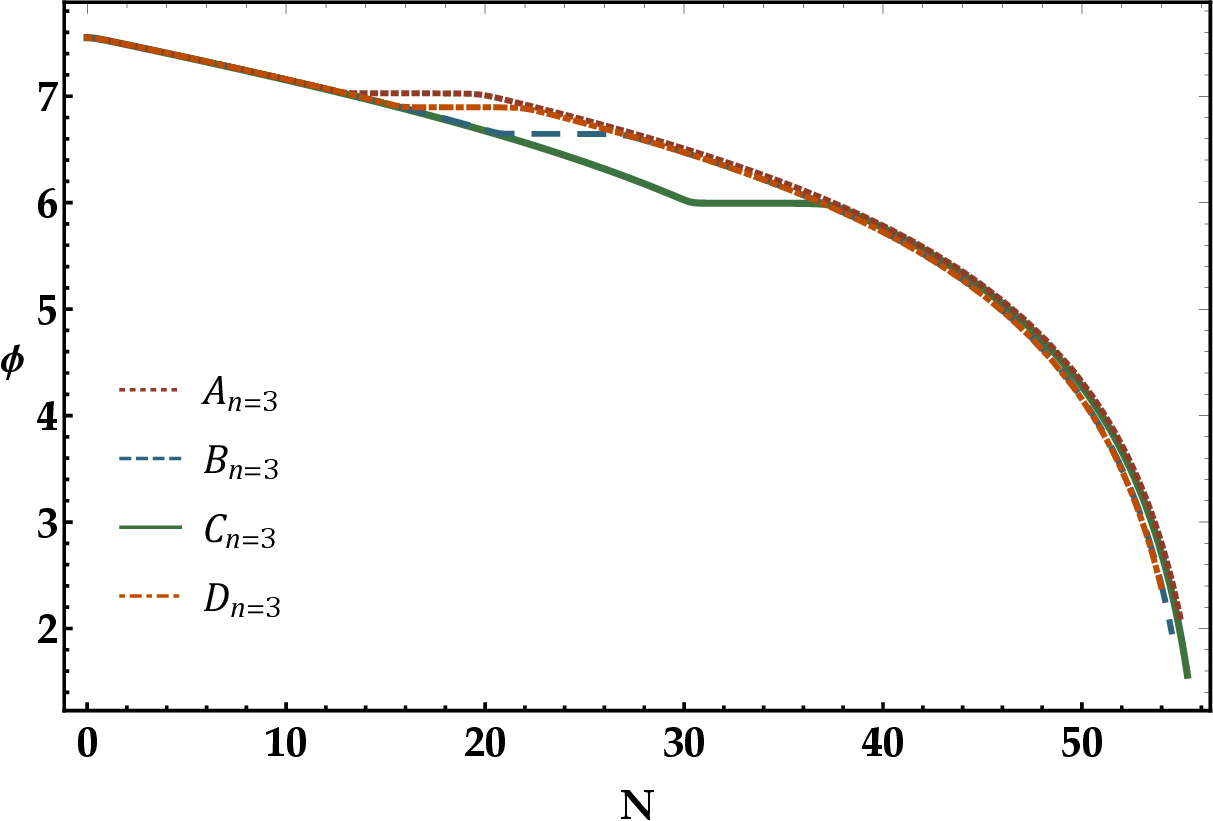}}\hspace*{0.3cm}
\subfigure[\label{phi-e1n} ]{ \includegraphics[width=0.48\textwidth]%
{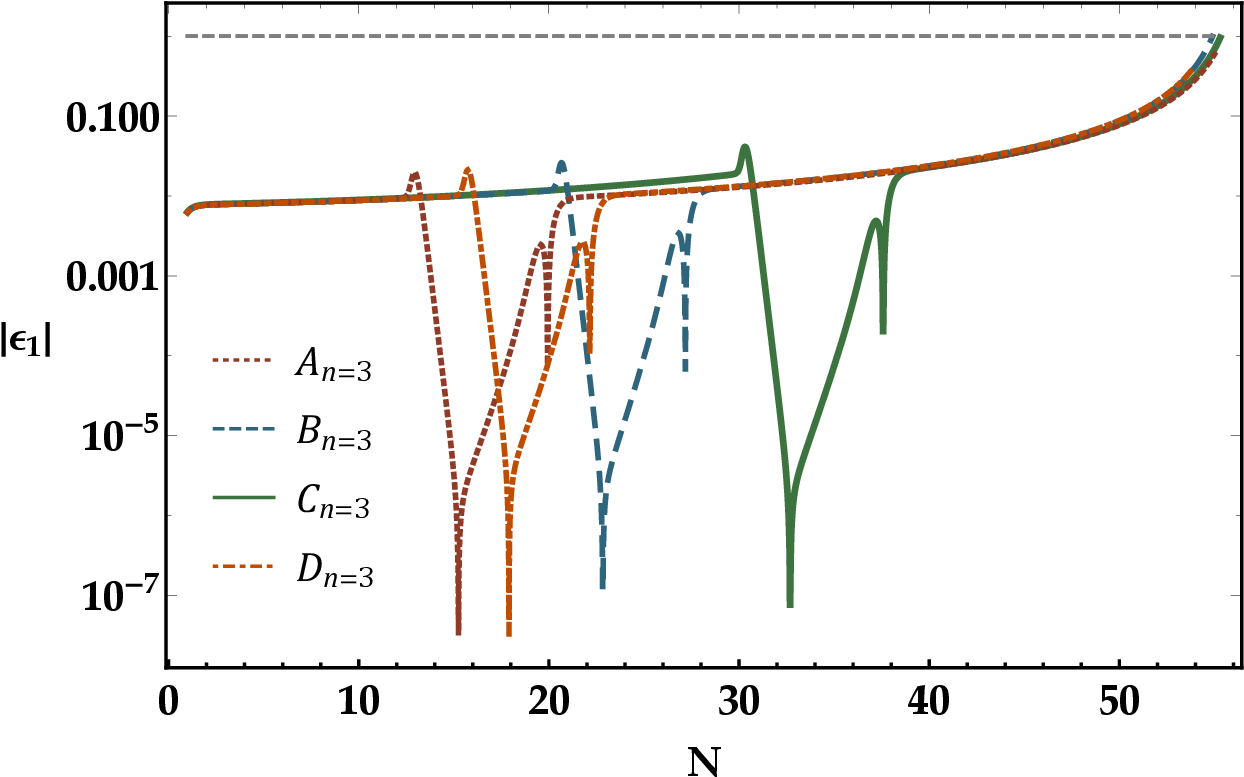}}\\
\hspace*{-0.5cm}
\subfigure[\label{phi-e2}]{
 \includegraphics[width=.46\textwidth]%
{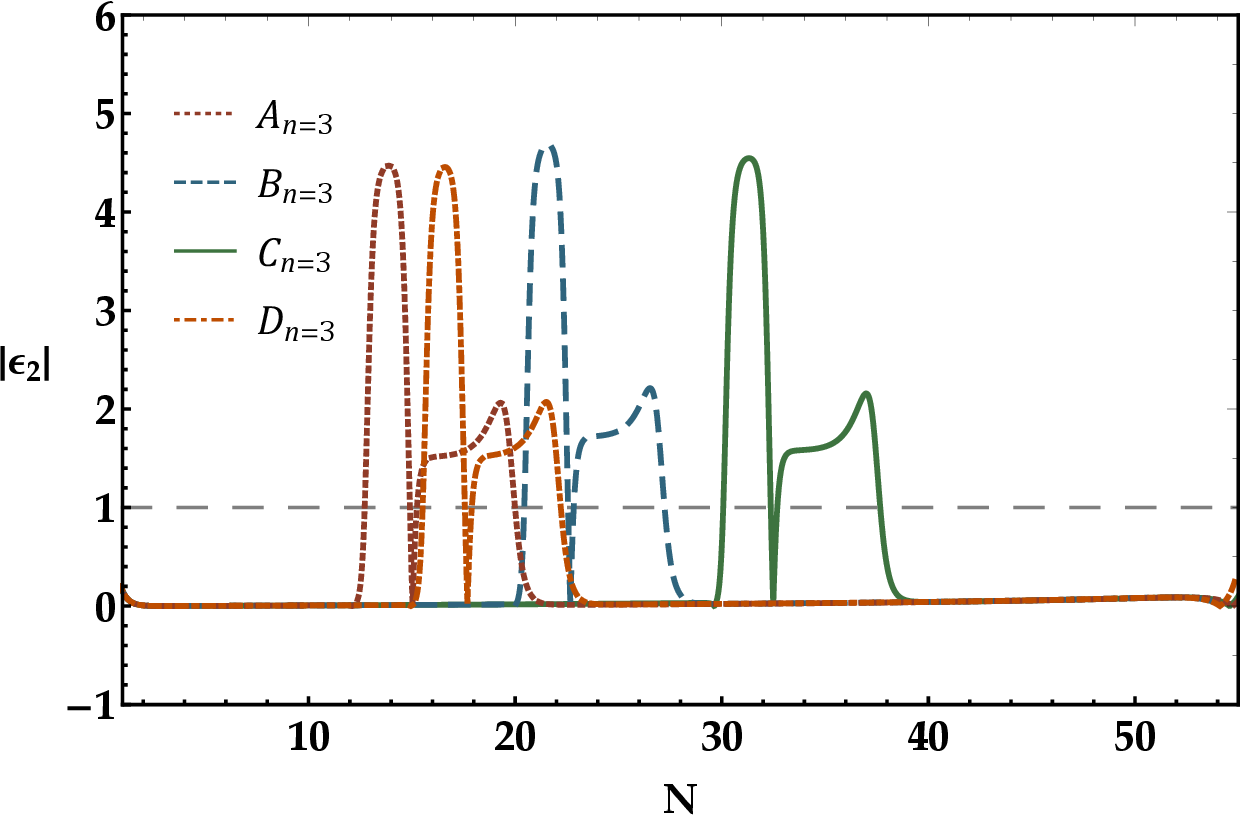}}\hspace*{0.3cm}
\subfigure[\label{phi-cs}]{
 \includegraphics[width=.46\textwidth]%
{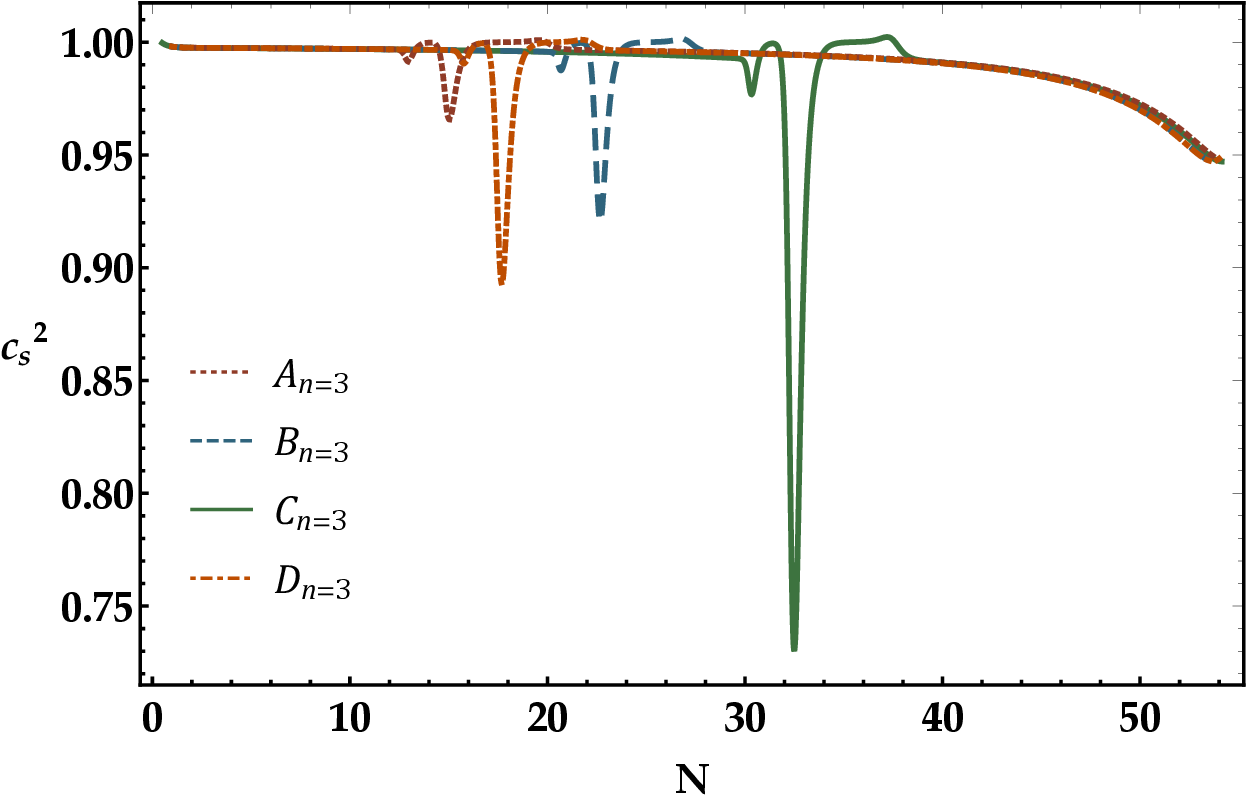}}
\end{minipage}
\caption{Same as Fig. \ref{linear}, but for the case $n = 3$. These evolutions are obtained using the parameter set of cases $A_{\rm n=3}$, $B_{\rm n=3}$, $C_{\rm n=3}$, and $D_{\rm n=3}$ in Table \ref{tab1-phi}, with $N_{*}=0$ at the horizon exit.}
\label{cubic}
\end{figure}

\begin{figure}[H]
\centering
\begin{minipage}[b]{1\textwidth}
\subfigure[\label{fig-phi} ]{ \includegraphics[width=0.44\textwidth]%
{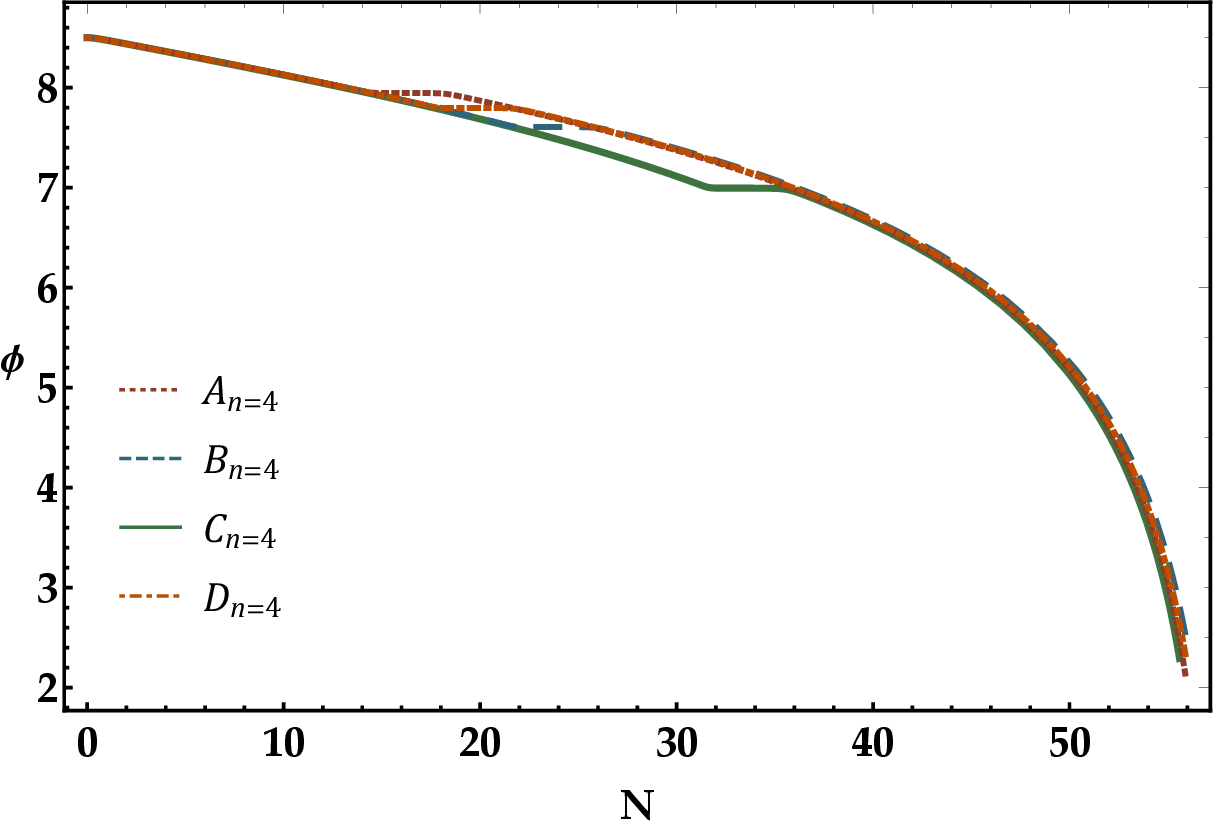}}\hspace*{0.3cm}
\subfigure[\label{phi-e1n} ]{ \includegraphics[width=0.48\textwidth]%
{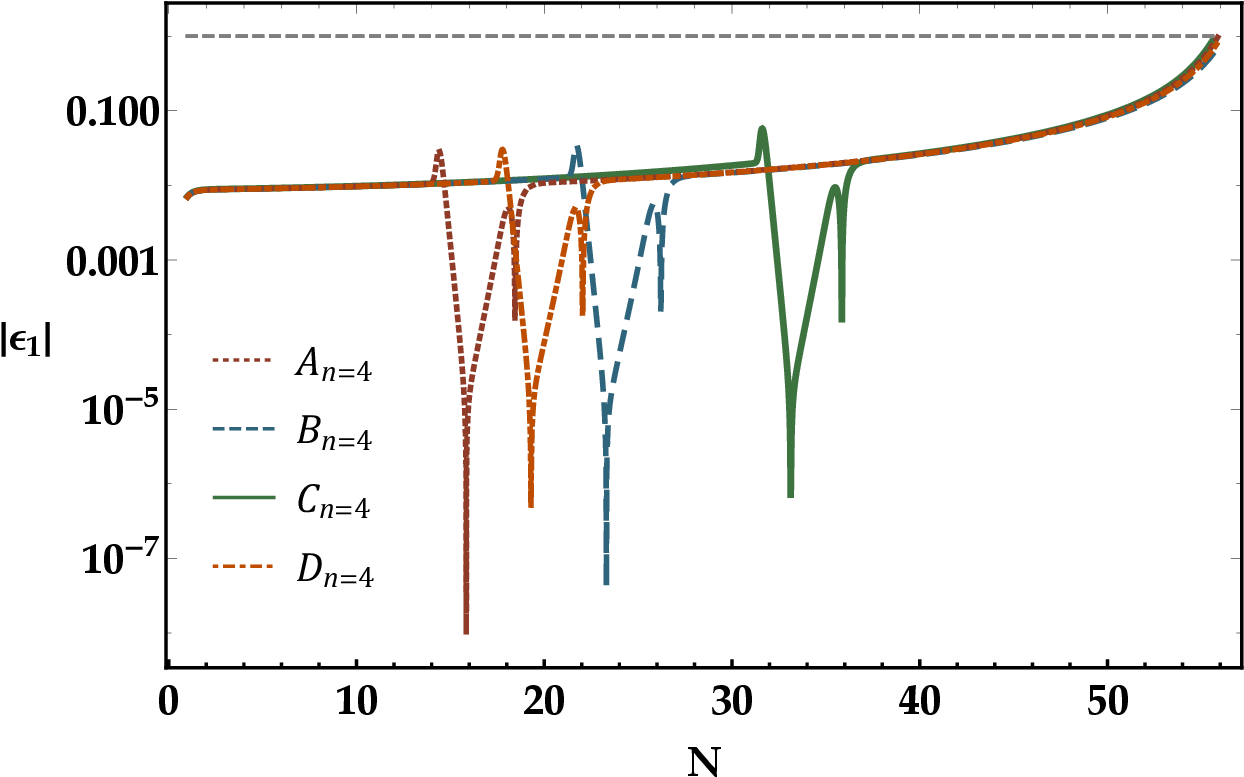}}\\
\hspace*{-0.5cm}
\subfigure[\label{phi-e2}]{
 \includegraphics[width=.46\textwidth]%
{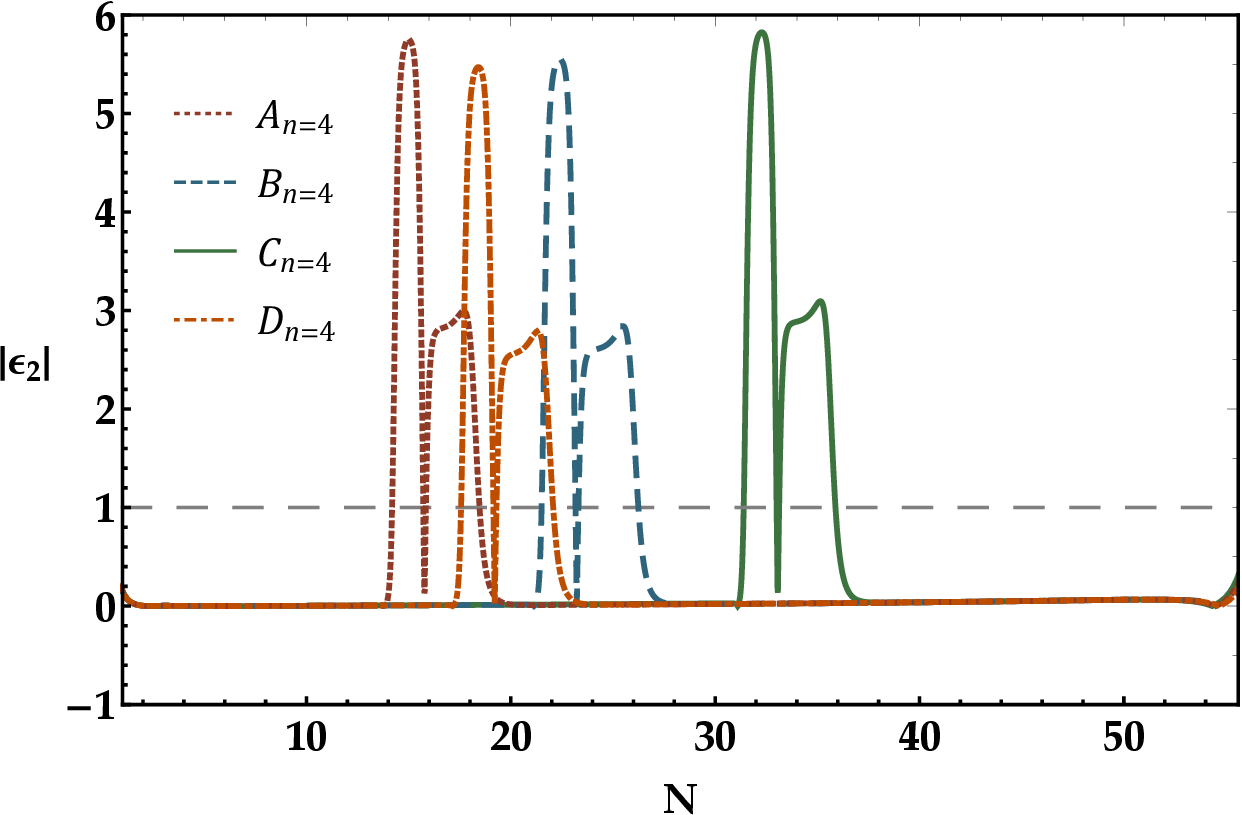}}\hspace*{0.3cm}
\subfigure[\label{phi-cs}]{
 \includegraphics[width=.46\textwidth]%
{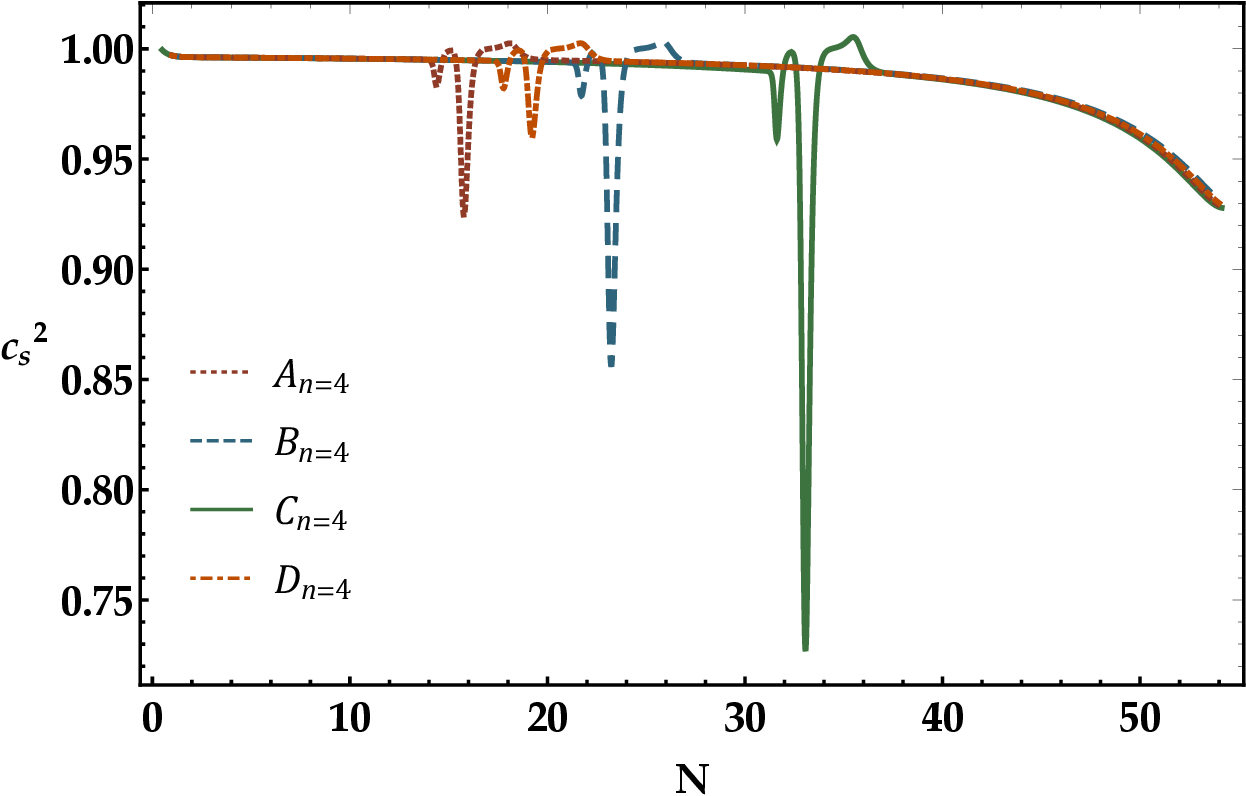}}
\end{minipage}
\caption{Same as Fig. \ref{linear}, but for the case $n = 4$. These evolutions are obtained using the parameter set of cases $A_{\rm n=4}$, $B_{\rm n=4}$, $C_{\rm n=4}$, and $D_{\rm n=4}$ in Table \ref{tab1-phi}, with $N_{*}=0$ at the horizon exit.}
\label{quartic}
\end{figure}

Finding appropriate values for $d$, $c$, and $\phi_c$ can be a challenging process. Fortunately, in the GB gravity, it is possible to utilize an auxiliary condition to approximate these parameters values.
As mentioned before, during the USR phase, the primordial curvature can undergo significant enhancement, subsequently leading to the generation of PBHs.
In the presence of the GB coupling term, the appearance of a non-trivial fixed point is possible.
Hence, the inflaton can enter the USR regime when its trajectory approaches a fixed point.
Around a fixed point, the parameters $\dot{H}$, $\dot{\phi}$, and $\ddot{\phi}$ become zero. Therefore, by utilizing Eq. (\ref{bge3}), one can derive the following condition at $\phi=\phi_c$ \cite{Kawai:2021,Kawaguchia:2023,Zhang:2022} as
\begin{align}
	\Bigg[ V_{,\phi}+\frac{4 \xi_{,\phi}{V(\phi)}^2}{3\Mpl^4}\Bigg]_{\phi=\phi_c}
	=0\,.
	\label{EqC}
\end{align}
We can achieve both objectives of this study by utilizing Eqs. (\ref{xip}) and (\ref{EqC}) and adjusting the parameters.
In the vicinity of a fixed point, when the inflaton undergoes the USR phase $(\e_2 > 1)$, the validity of the slow-roll approximation is compromised, as depicted in panel (c) of Figs. \ref{linear}-\ref{quartic}.
As a result, we cannot use Eq. (\ref{ps-sr}) to calculate the power spectrum of curvature fluctuations during this phase.
Instead, we must numerically solve Eqs. (\ref{bge3}) and (\ref{Hdot}) to estimate the evolution of $H$ and $\phi$ with respect to the $e$-fold number $N$.
Ultimately, we can accurately determine the value of the curvature power spectrum by employing Eqs. (\ref{M.S}) and (\ref{powerspectraZeta}).

In the present work, we investigate the possibility of the PBH formation in our setup in the presence of the power-law potentials as
\begin{align}\label{eqv}
V(\phi)=V_0 \phi^n,
\end{align}
where $V_0$ is a constant parameter with $\Mpl^{4-n}$ dimension. Here, we consider the cases $n=$1, 2, 3 and 4. Note that in the framework of standard inflation, the power-law potentials with $n=$ 1, 2, 3, and 4 are completely rolled out by Planck 2018 TT, TE, EE + lowE + lensing + BK18 + BAO \cite{akrami:2018,bk18}.
This motivates our investigation to assess the compatibility of the power-law potentials with the observational constraints at CMB (large) scale in this scenario.

Therefore, our model is described with six free parameters ($V_0$, $M$, $\alpha$, $d$, $c$, $\phi_c$). The value of $V_0$ is determined by fixing the amplitude of the scalar power spectrum at the pivot scale as ${\cal P}_s(k_{*}) = 2.1 \times 10^{-9}$.
Furthermore, adjusting the parameters $M$ and $\alpha$ can ensure that the observational parameters $n_s$ and $r$ of the model are compatible with Planck 2018 data.
Hence, PBHs formation can be influenced by the remaining parameters in the model, i.e. $d$, $c$, and $\phi_c$.
To investigate PBH formation within specific mass ranges relevant to cosmological phenomena, we strategically select the field value, $\phi_c$, at which the curvature perturbation is enhanced.
This approach allows us to focus on the desired PBH mass window.
Finding the proper value of the parameter $c$ necessitates balancing two important considerations.
First, the total number of $e$-folds must be maintained within the established range (50-60) to ensure a viable inflationary scenario.
Second, we require that the derivative of $\xi_{2}(\phi)$ far away from the peak be negligible to uphold the validity of the slow-roll approximation.
Finally, to determine the value of $d$, we utilize the auxiliary condition (\ref{EqC}) to obtain an initial estimate.
The approximated values are then fine-tuned to ensure they align with observational constraints on the abundance of PBHs
\cite{Clark,Laha:2019,EGG,subaro,Icarus,Kepler,EORS,Boehm:2021,CMB,Kavanagh:2018,Chen:2022,Alcock:2001,OGLE}.

In what follows, for the power-law potentials with $n=$ 1, 2, 3, and 4, we investigate the possibility of PBH formation at small scales and also the model compatibility with Planck observations at the CMB scale. Here, we set $N_{*}=0$ at the horizon exit.

Table \ref{tab1-phi} shows the set of model parameters in our study. Here, we set parameters $M$ and $\alpha$ to keep our model predictions consistent with Planck measurements for $n_s$ and $r$ \cite{akrami:2018,bk18}.
In contrast to the standard model, the estimated values of $n_s$ and $r$ in our model, as shown in Table \ref{tab2-phi}, are consistent with the 68\% CL region of Planck 2018 TT, TE, EE + lowE + lensing + BK18 + BAO data \cite{akrami:2018,bk18}.
%
Note that since the slow-roll inflation occurs before the USR attractor phase, hence the initial conditions are estimated under the slow-roll approximation. Therefore, we estimate the initial value of the scalar field at the pivot scale $\phi_{*}$ under the slow-roll approximation. The corresponding values of $\phi_{*}$ for all cases have been provided in Table \ref{tab2-phi}.

The panel (a) in Figs. \ref{linear}-\ref{quartic} illustrates the evolution of the scalar field $\phi$ as a function of the $e$-fold number. The flat era, as shown in these figures, describes the USR phase.
It is important to note that, during the USR phase, although the value of $\epsilon_1$ remains below one (as depicted in panel (b) of Figs. \ref{linear}-\ref{quartic}), the value of $\epsilon_2$ increases and exceeds one.
This signifies the violation of the slow-roll condition, as shown in panel (c) of Figs. \ref{linear}-\ref{quartic}.
Furthermore, as depicted in Figs. \ref{linear}-\ref{quartic}, during the USR phase, the value of $c_s^2$ undergoes a significant reduction which results in an enhancement in the scalar power spectrum, see Eq. (\ref{ps-sr}).

\begin{table}[H]
\centering
\caption{The parameter set for all cases investigated in this study. The subscript $n$ corresponds to the power index in the power-law potential.}
\resizebox{\textwidth}{!}{%
\begin{tabular}{cccccccc}
\thickhline
Sets \quad &
\quad $M$ \quad &
\quad $\alpha$ \quad &
\quad $V_{0}^{1/4} \Mpl^{n/4} (\Mpl)$ \quad &
\quad $\phi_{c}/ \Mpl$ \quad &
\quad $c/ \Mpl$\quad &
$d$\quad &
$d$ ({\footnotesize From Eq. (\ref{EqC})})\quad  \\ [0.5ex]
\thickhline
$A_{\rm n=1}$ \quad &
\quad $1.36\times 10^{6}$ \quad &
\quad $0.3$ \quad &
\quad $0.0216$ \quad &
\quad $5.34$ \quad &
\quad $70$\quad &
\quad $-0.00205$&
\quad $-0.00193$ \quad\\[0.5ex]
\hline
$B_{\rm n=1}$ \quad &
\quad $1.36\times 10^{6}$ \quad &
\quad $0.3$ \quad &
\quad $0.0216$ \quad &
\quad $4.95$ \quad &
\quad $60$\quad &
\quad $-0.00266$&
\quad $-0.00250$ \quad\\[0.5ex]
\hline
$C_{\rm n=1}$ \quad &
\quad $1.36\times 10^{6}$ \quad &
\quad $0.3$ \quad &
\quad $0.0216$ \quad &
\quad $4.20$ \quad &
\quad $45 $ \quad &
\quad $-0.00474$&
\quad $-0.00443$ \quad\\[0.5ex]
\hline
$D_{\rm n=1}$ \quad &
\quad $1.36\times 10^{6}$ \quad &
\quad $0.3$ \quad &
\quad $0.0216$ \quad &
\quad $5.15$ \quad &
\quad $70$ \quad &
\quad $-0.00219$&
\quad $-0.00203$ \quad\\[0.5ex]
\thickhline
$A_{\rm n=2}$ \quad &
\quad $7.5\times 10^{6}$ \quad &
\quad $0.4$ \quad &
\quad $0.0119$ \quad &
\quad $6.5$ \quad &
\quad $80 $ \quad &
\quad $-0.00121$&
\quad $-0.00113$ \quad\\[0.5ex]
\hline
$B_{\rm n=2}$ \quad &
\quad $7.5\times 10^{6}$ \quad &
\quad $0.4$ \quad &
\quad $0.0119$ \quad &
\quad $6.15$ \quad &
\quad $70$\quad &
\quad $-0.00158$&
\quad $-0.00147$ \quad\\[0.5ex]
\hline
$C_{\rm n=2}$ \quad &
\quad $7.5\times 10^{6}$ \quad &
\quad $0.4$ \quad &
\quad $0.0119$ \quad &
\quad $5.4$ \quad &
\quad $50$\quad &
\quad $-0.00355$&
\quad $-0.00335$ \quad\\[0.5ex]
\hline
$D_{\rm n=2}$ \quad &
\quad $7.5\times 10^{6}$ \quad &
\quad $0.4$ \quad &
\quad $0.0119$ \quad &
\quad $6.3$ \quad &
\quad $80$\quad &
\quad $-0.00133$&
\quad $-0.00121$ \quad\\[0.5ex]
\thickhline
$A_{\rm n=3}$ \quad &
\quad $3\times 10^{7}$ \quad &
\quad $0.521$ \quad &
\quad $0.0067$ \quad &
\quad $7.03$ \quad &
\quad $95$ \quad &
\quad $-0.00088$&
\quad $-0.00082$ \quad\\[0.5ex]
\hline
$B_{\rm n=3}$ \quad &
\quad $3\times 10^{7}$ \quad &
\quad $0.521$ \quad &
\quad $0.0067$ \quad &
\quad $6.65$ \quad &
\quad $85$\quad &
\quad $-0.00119$&
\quad $-0.00108$ \quad\\[0.5ex]
\hline
$C_{\rm n=3}$ \quad &
\quad $3\times 10^{7}$ \quad &
\quad $0.521$ \quad &
\quad $0.0067$ \quad &
\quad $6$ \quad &
\quad $55$\quad &
\quad $-0.00273$&
\quad $-0.00251$ \quad\\[0.5ex]
\hline
$D_{\rm n=3}$ \quad &
\quad $3\times 10^{7}$ \quad &
\quad $0.521$ \quad &
\quad $0.0067$ \quad &
\quad $6.9$ \quad &
\quad $90$\quad &
\quad $-0.00098$&
\quad $-0.00090$ \quad\\[0.5ex]
\thickhline
$A_{\rm n=4}$ \qquad &
\quad $7.5\times 10^{7}$ \quad &
\quad $0.585$ \quad &
\quad $0.0038$ \quad &
\quad$7.95$ \quad &
\quad$150$ \quad &
\quad $-0.00061$ \quad &
\quad $-0.00049$ \\[0.5ex]
\hline
$B_{\rm n=4}$ \quad &
\quad $7.5\times 10^{7}$ \quad &
\quad $0.585$ \quad &
\quad $0.0038$ \quad &
 \quad$7.61$ \quad &
\quad$120$ \quad &
\quad $-0.00088$ \quad &
\quad $-0.00073$ \\[0.5ex]
\hline
$C_{\rm n=4}$ \quad &
\quad $7.5\times 10^{7}$ \quad &
\quad $0.585$ \quad &
\quad $0.0038$ \quad &
\quad$7$ \quad &
\quad$90$ \quad &
\quad $-0.00180$ \quad &
\quad $-0.00143$ \quad\\[0.5ex]
\hline
$D_{\rm n=4}$ \quad &
\quad $7.5\times 10^{7}$ \quad &
\quad $0.585$ \quad &
\quad $0.0038$ \quad &
\quad$7.8$ \quad &
\quad$130$ \quad &
\quad $-0.00073$ \quad &
\quad $-0.00061$ \quad\\[0.5ex]
\thickhline
\end{tabular}}
\label{tab1-phi}
\end{table}

\begin{table}[H]
\centering
\caption{The values of $\phi_{*}$, $n_s$, $r$, $k_{\text{peak}}$, ${\cal P}_{s}^\text{peak}$, $M_{\text{PBH}}^{\text{peak}}$ and $f_{\text{PBH}}^{\text{peak}}$ for the cases listed in Table \ref{tab1-phi}.}
\begin{tabular}{cccccccc}
\thickhline
Sets \quad &
\quad$\phi_{*}/\Mpl$ \quad &
\quad$n_s$ \quad &
\quad$r$ \quad &
\quad$k_{\text{peak}}/\text{\rm Mpc}^{-1}$ \quad &
\quad ${\cal P}_{s}^\text{peak}$ \quad &
\quad  $M_{\text{PBH}}^{\text{peak}}/M_{\odot}$ \quad &
\quad $f_{\text{PBH}}^{\text{peak}}$\\ [0.5ex]
\thickhline
$A_{\rm n=1}$ \qquad &
\quad$6.0$ \quad &
\quad$0.9681$ \quad &
\quad$0.018$ \quad &
\quad $6.9\times10^{5}$ \quad &
\quad $0.052$ \quad &
\quad $7.98$ \quad &
\quad$0.003$ \\[0.5ex]
\hline
$B_{\rm n=1}$ \quad &
\quad$6.0$ \quad &
\quad$0.9681$ \quad &
\quad$0.018$ \quad &
\quad $4.4\times10^{8}$ \quad &
\quad $0.042$ \quad &
\quad $1.9\times10^{-5}$ \quad &
\quad$0.022$ \\[0.5ex]
\hline
$C_{\rm n=1}$ \quad &
\quad$6.0$ \quad &
\quad$0.9680$ \quad &
\quad$0.019$ \quad &
\quad $7.0\times10^{12}$ \quad &
\quad $0.035$ \quad &
\quad $7.7 \times 10^{-14}$ \quad &
\quad$0.936$\\[0.5ex]
\hline
$D_{\rm n=1}$ \quad &
\quad$6.0$ \quad &
\quad$0.9681$ \quad &
\quad$0.018$ \quad &
\quad $1.7\times10^{7}$ \quad &
\quad $0.019$ \quad &
\quad $7.6\times10^{-3}$ \quad &
\quad$\sim 0$\\[0.5ex]
\thickhline
$A_{\rm n=2}$ \qquad &
\quad$7.1$ \quad &
\quad$0.9632$ \quad &
\quad$0.017$ \quad &
\quad $6.7\times10^{5}$ \quad &
\quad $0.052$ \quad &
\quad $13.8$ \quad &
\quad $0.0026$ \\[0.5ex]
\hline
$B_{\rm n=2}$ \quad &
\quad$7.1$ \quad &
\quad$0.9632$ \quad &
\quad$0.017$ \quad &
\quad $4.2\times10^{8}$ \quad &
\quad $0.045$ \quad &
\quad $2.1\times10^{-5}$ \quad &
\quad $0.03$ \\[0.5ex]
\hline
$C_{\rm n=2}$ \quad &
\quad$7.1$ \quad &
\quad$0.9630$ \quad &
\quad$0.018$ \quad &
\quad $1.1\times10^{13}$ \quad &
\quad $0.033$ \quad &
\quad $3.4 \times 10^{-14}$ \quad &
\quad $0.99$\\[0.5ex]
\hline
$D_{\rm n=2}$ \quad &
\quad$7.1$ \quad &
\quad$0.9633$ \quad &
\quad$0.017$ \quad &
\quad $2.8\times10^{7}$ \quad &
\quad $0.025$ \quad &
\quad $3.0 \times 10^{-3}$ \quad &
\quad $\sim 0$\\[0.5ex]
\thickhline
$A_{\rm n=3}$ \qquad &
\quad$7.5$ \quad &
\quad$0.9630$ \quad &
\quad$0.017$ \quad &
\quad $5.1\times10^{5}$ \quad &
\quad $0.049$ \quad &
\quad $23.9$ \quad &
\quad$0.0007$ \\[0.5ex]
\hline
$B_{\rm n=3}$ \quad &
\quad$7.5$ \quad &
\quad$0.9629$ \quad &
\quad$0.017$ \quad &
\quad $8.5\times10^{8}$ \quad &
\quad $0.044$ \quad &
\quad $5.2\times10^{-6}$ \quad &
\quad $0.074$ \\[0.5ex]
\hline
$C_{\rm n=3}$ \quad &
\quad$7.5$ \quad &
\quad$0.9626$ \quad &
\quad$0.018$ \quad &
\quad $1.6\times10^{13}$ \quad &
\quad $0.036$ \quad &
\quad $2.3 \times 10^{-14}$ \quad &
\quad$0.98$\\[0.5ex]
\hline
$D_{\rm n=3}$ \quad &
\quad$7.5$ \quad &
\quad$0.9629$ \quad &
\quad$0.017$ \quad &
\quad $6.1\times10^{6}$ \quad &
\quad $0.012$ \quad &
\quad $6.2 \times 10^{-2}$ \quad &
\quad$\sim 0$\\[0.5ex]
\thickhline
$A_{\rm n=4}$ \qquad &
\quad$8.5$ \quad &
\quad$0.9635$ \quad &
\quad$0.018$ \quad &
\quad $3.0\times10^{5}$ \quad &
\quad $0.058$ \quad &
\quad $25.2$ \quad &
\quad$0.001$ \\[0.5ex]
\hline
$B_{\rm n=4}$ \quad &
\quad$8.5$ \quad &
\quad$0.9634$ \quad &
\quad$0.018$ \quad &
\quad $5.0\times10^{8}$ \quad &
\quad $0.046$ \quad &
\quad $1.5\times10^{-5}$ \quad &
\quad $0.019$ \\[0.5ex]
\hline
$C_{\rm n=4}$ \quad &
\quad$8.5$ \quad &
\quad$0.9631$ \quad &
\quad$0.019$ \quad &
\quad $7.9\times10^{12}$ \quad &
\quad $0.038$ \quad &
\quad $7.0 \times 10^{-14}$ \quad &
\quad$0.98$\\[0.5ex]
\hline
$D_{\rm n=4}$ \quad &
\quad$8.5$ \quad &
\quad$0.9634$ \quad &
\quad$0.018$ \quad &
\quad $9.8\times10^{6}$ \quad &
\quad $0.024$ \quad &
\quad $1.7 \times 10^{-2}$ \quad &
\quad$\sim 0$\\[0.5ex]
\thickhline
\end{tabular}
\label{tab2-phi}
\end{table}

Under the slow-roll conditions, the scalar power spectrum cannot increase sufficiently to be suitable for PBH formation.
However, during the USR phase, decreasing the inflaton velocity provides suitable time for the primordial curvatures to amplify and subsequently generate PBHs after horizon re-entry.

\begin{figure}[H]
\centering
\begin{minipage}[b]{1\textwidth}
\subfigure[\label{phi-ps} ]{ \includegraphics[width=0.47\textwidth]%
{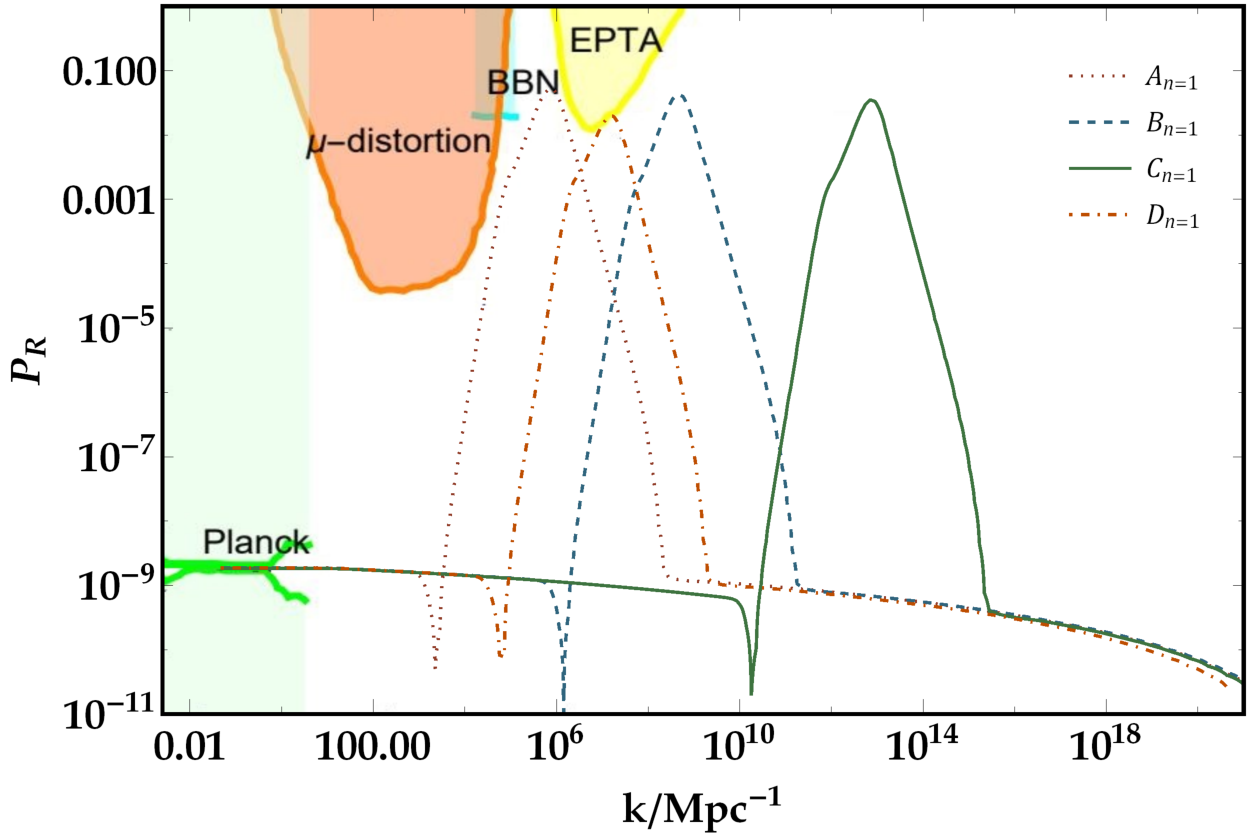}}\hspace*{0.3cm}
\subfigure[\label{phi2-ps} ]{ \includegraphics[width=0.47\textwidth]%
{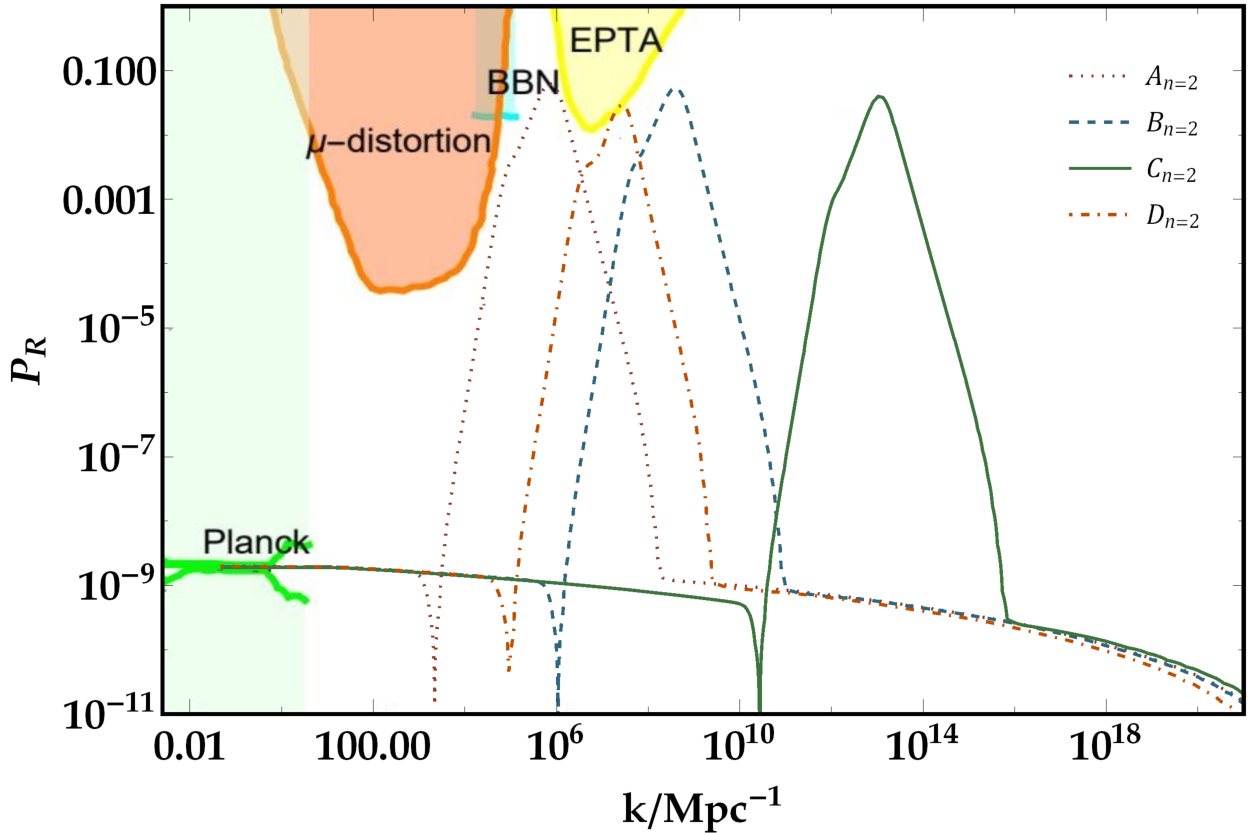}}\\
\subfigure[\label{phi3-ps}]{
 \includegraphics[width=.47\textwidth]%
{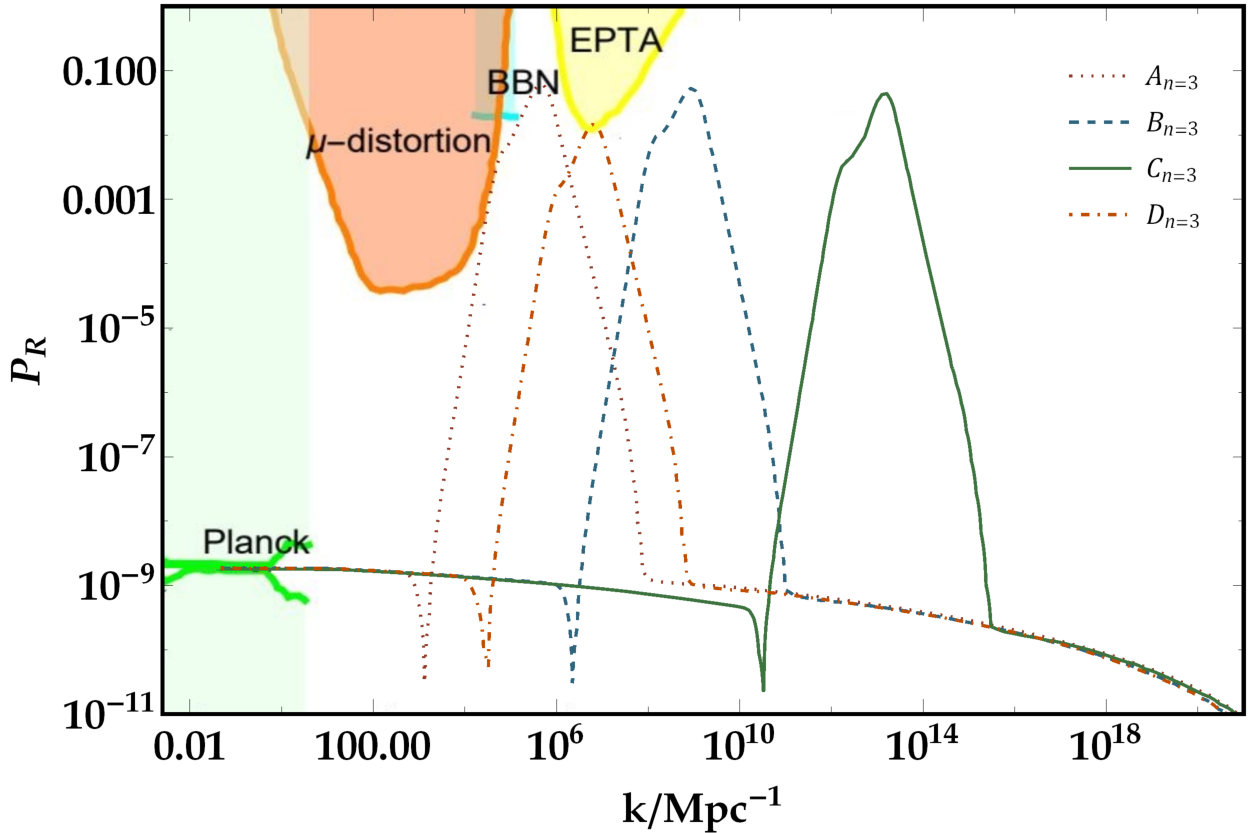}}\hspace*{0.3cm}
\subfigure[\label{phi4-ps}]{
 \includegraphics[width=.47\textwidth]%
{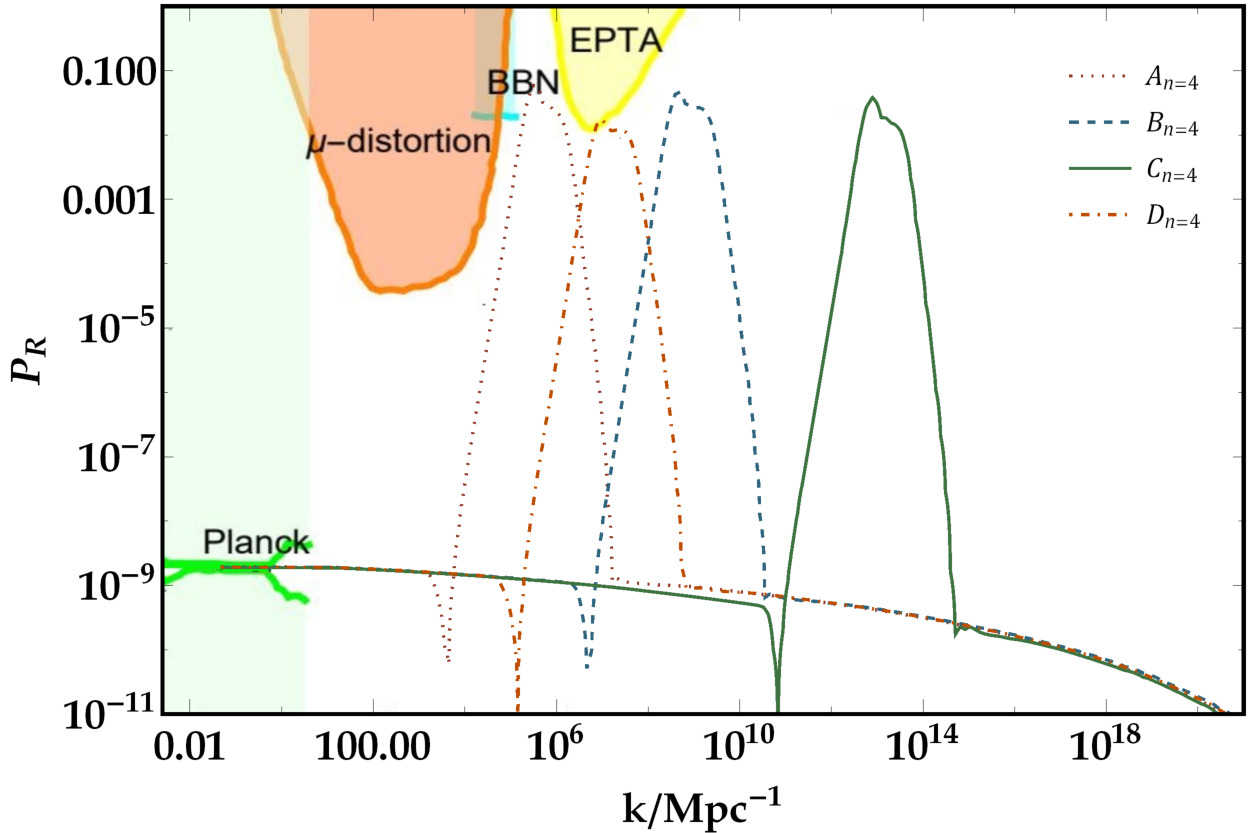}}
\end{minipage}
\vspace{-0.5em}
\caption{The scalar power spectrum ${\cal P}{s}$ as a function of comoving wavenumber $k$ for our model. The figure illustrates the evolution of ${\cal P}_{s}$ for (a) $n=1$, (b) $n=2$, (c) $n=3$, and (d) $n=4$. The dotted, dashed, solid, and dash-dotted lines correspond to the cases $A_{\rm n}$, $B_{\rm n}$, $C_{\rm n}$, and $D_{\rm n}$, respectively. The CMB observations exclude the light-green shaded area \cite{akrami:2018}. The orange zone shows the $\mu$-distortion of CMB \cite{Fixsen:1996}. The cyan area represents the effect on the ratio between neutron and proton during the big bang nucleosynthesis (BBN) \cite{Inomata:2016}. The yellow region demonstrates the constraint from the PTA observations \cite{Inomata:2019-a}. }
\label{fig-ps}
\end{figure}

Upon solving the MS equation (\ref{M.S}), we estimate the value of the scalar power spectrum at the peak position ${\cal P}_{s}^\text{peak}$.
The value of ${\cal P}_{s}^\text{peak}$ for all cases of Table \ref{tab1-phi} are presented in Table \ref{tab2-phi}.
The evolution of ${\cal P}_{s}$ for all parameter sets listed in Table \ref{tab1-phi} is depicted in Fig. \ref{fig-ps}.
The results demonstrate that the scalar power spectrum is in agreement with CMB measurements at the pivot scale.
Additionally, at small scales, ${\cal P}_{s}$ experiences a substantial increase of approximately seven orders of magnitude, which is completely suitable for the formation of PBHs.

\begin{figure}[H]
\begin{minipage}[b]{1\textwidth}
\centering
\subfigure[\label{phi_sc} ]{ \includegraphics[width=0.45\textwidth]%
{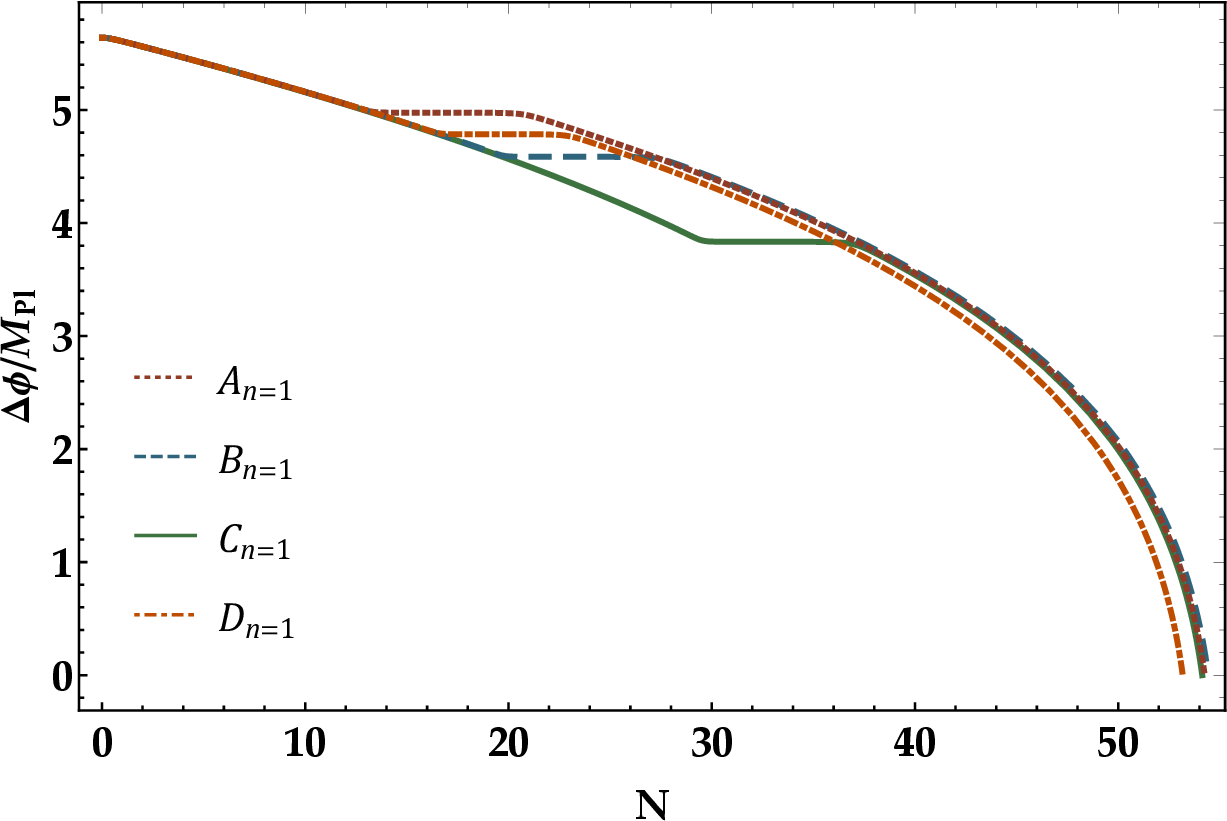}}\hspace{.1cm}
\subfigure[\label{phi2_sc}]{ \includegraphics[width=.45\textwidth]%
{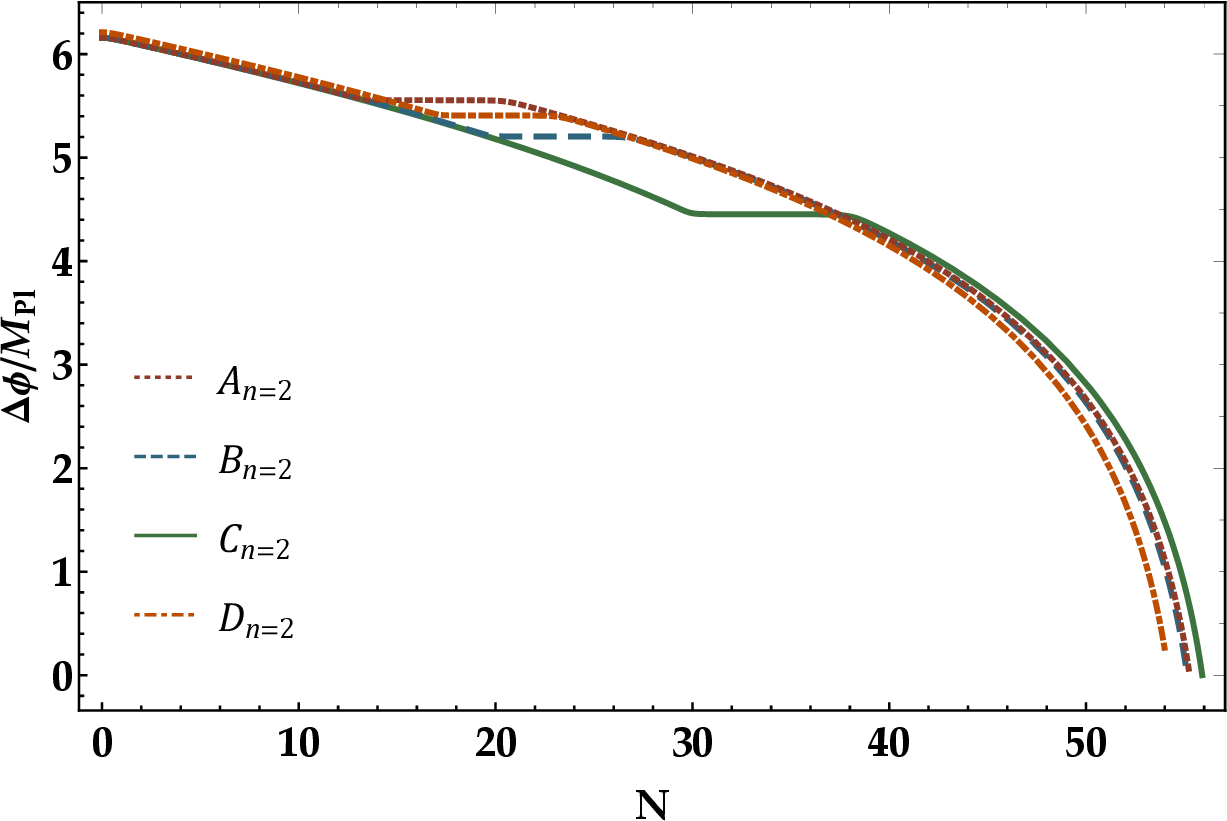}}
\centering
\subfigure[\label{phi3_sc} ]{ \includegraphics[width=0.45\textwidth]%
{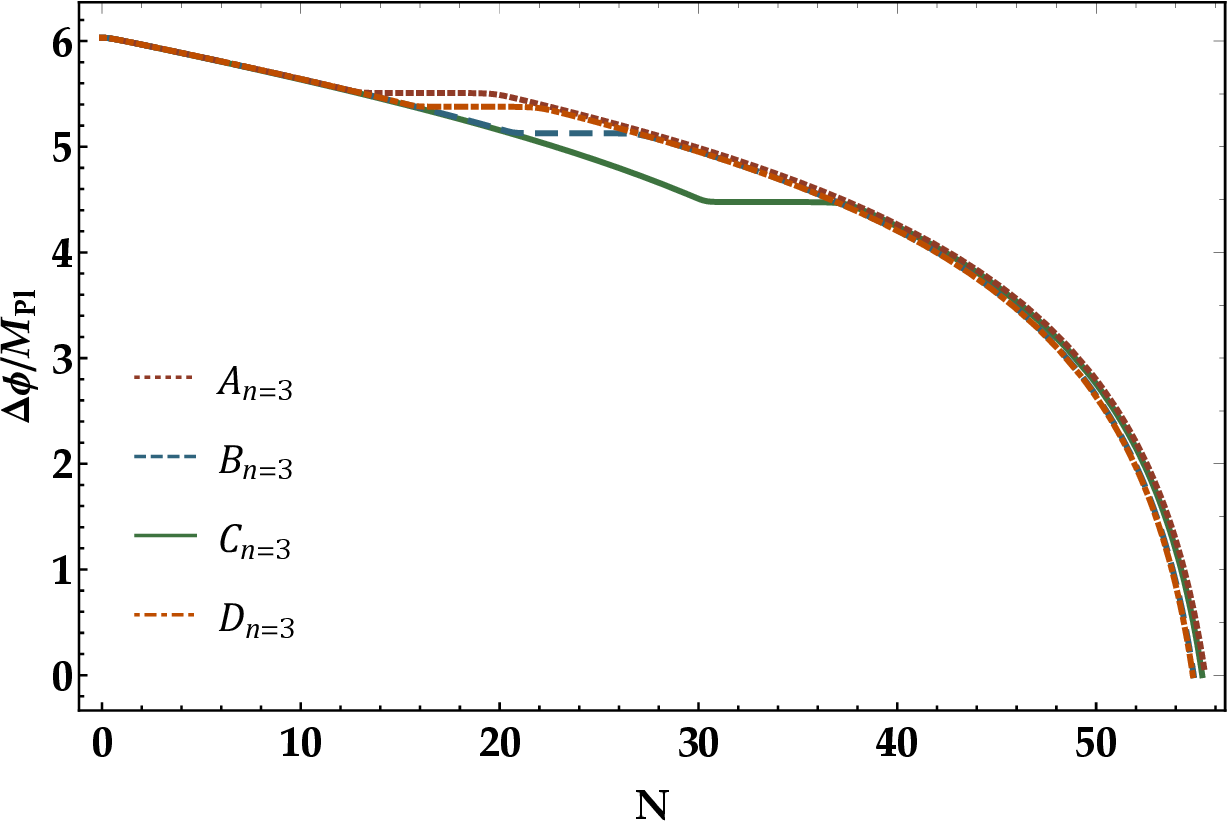}}\hspace{.1cm}
\subfigure[\label{phi4_sc}]{ \includegraphics[width=.45\textwidth]%
{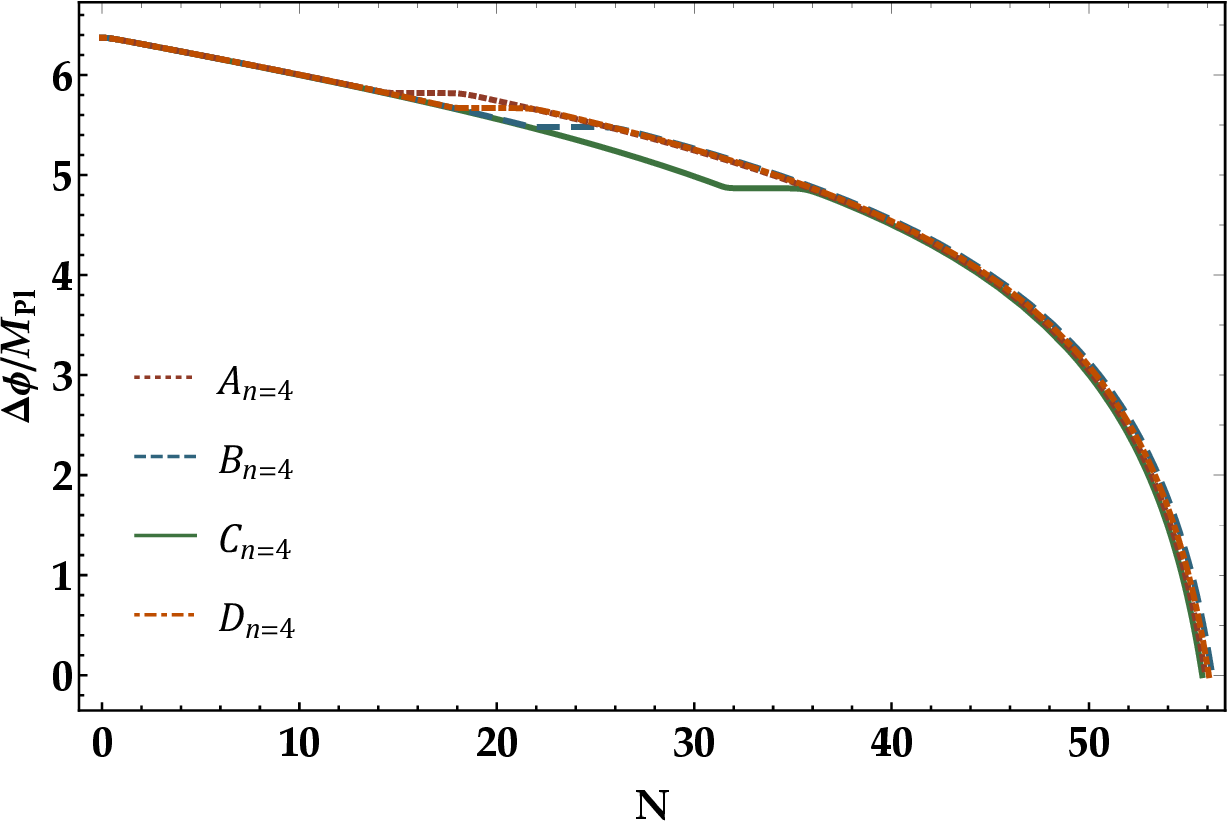}}
\end{minipage}
\caption{Confirming the validity of the first swampland criteria for our model. The figure demonstrates the fulfilment of the distance conjecture, where $\Delta\phi / \Mpl \sim \mathcal{O}(1)$ for (a) $n=1$, (b) $n=2$, (c) $n=3$, and (d) $n=4$. Here $\Delta\phi = \phi(N) - \phi(N_{\rm end}) $.}\label{fig_sc_phi}
\end{figure}

\begin{figure}[H]
\begin{minipage}[b]{1\textwidth}
\centering
\subfigure[\label{vphi_sc} ]{ \includegraphics[width=0.48\textwidth]%
{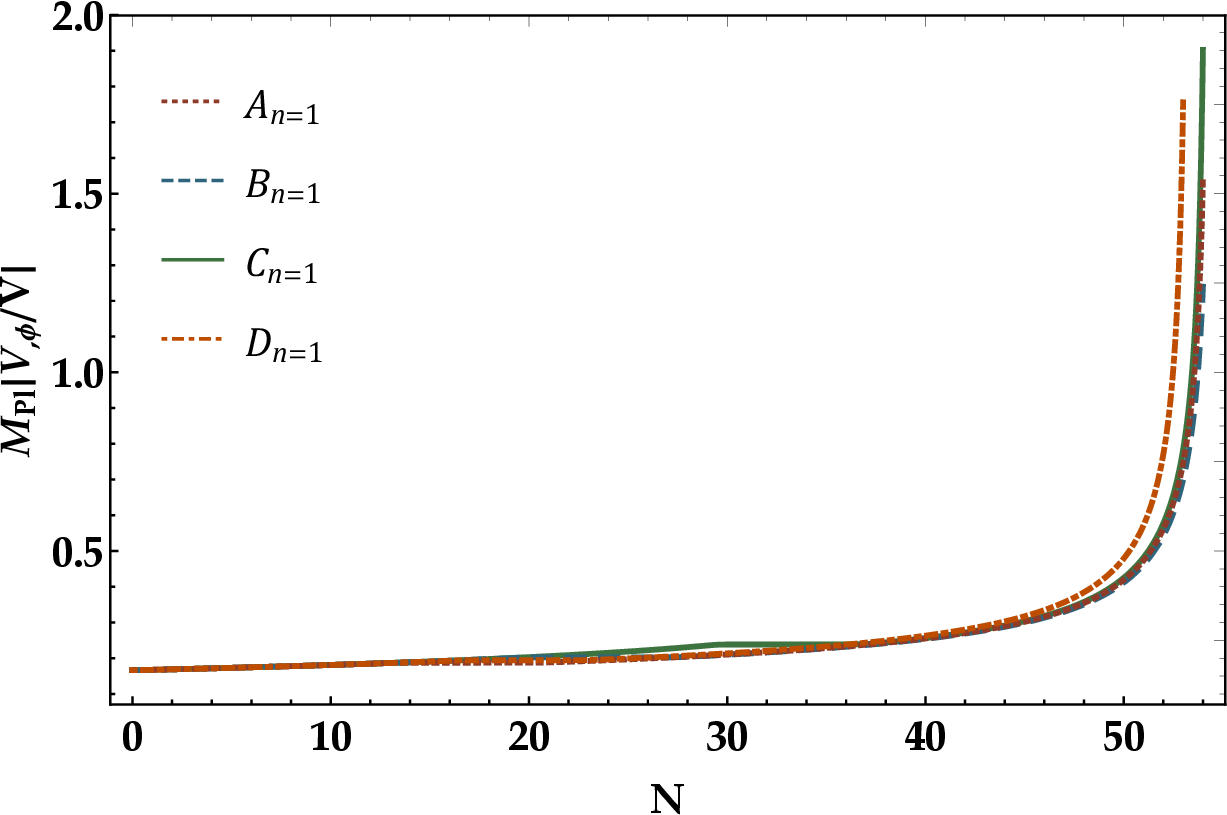}}\hspace{.1cm}
\subfigure[\label{vphi2_sc}]{ \includegraphics[width=.48\textwidth]%
{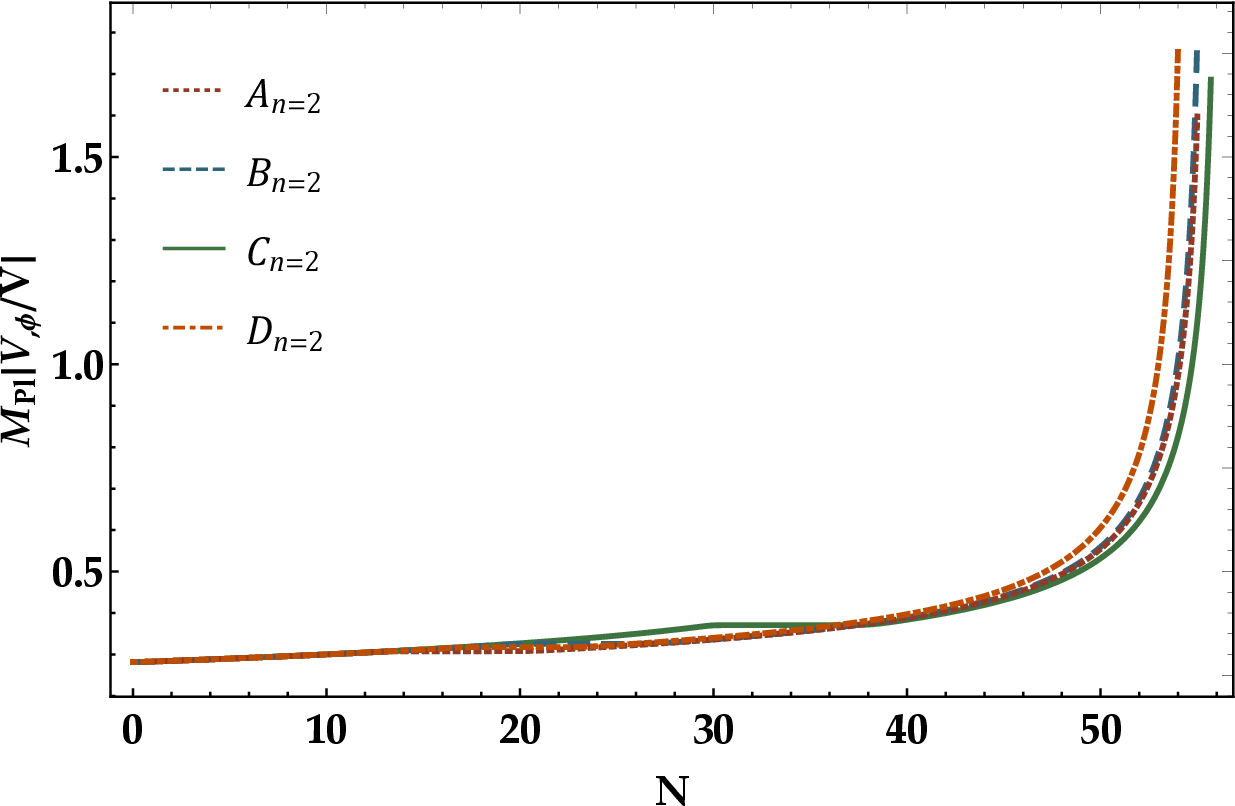}}
\centering
\subfigure[\label{vphi3_sc} ]{ \includegraphics[width=0.48\textwidth]%
{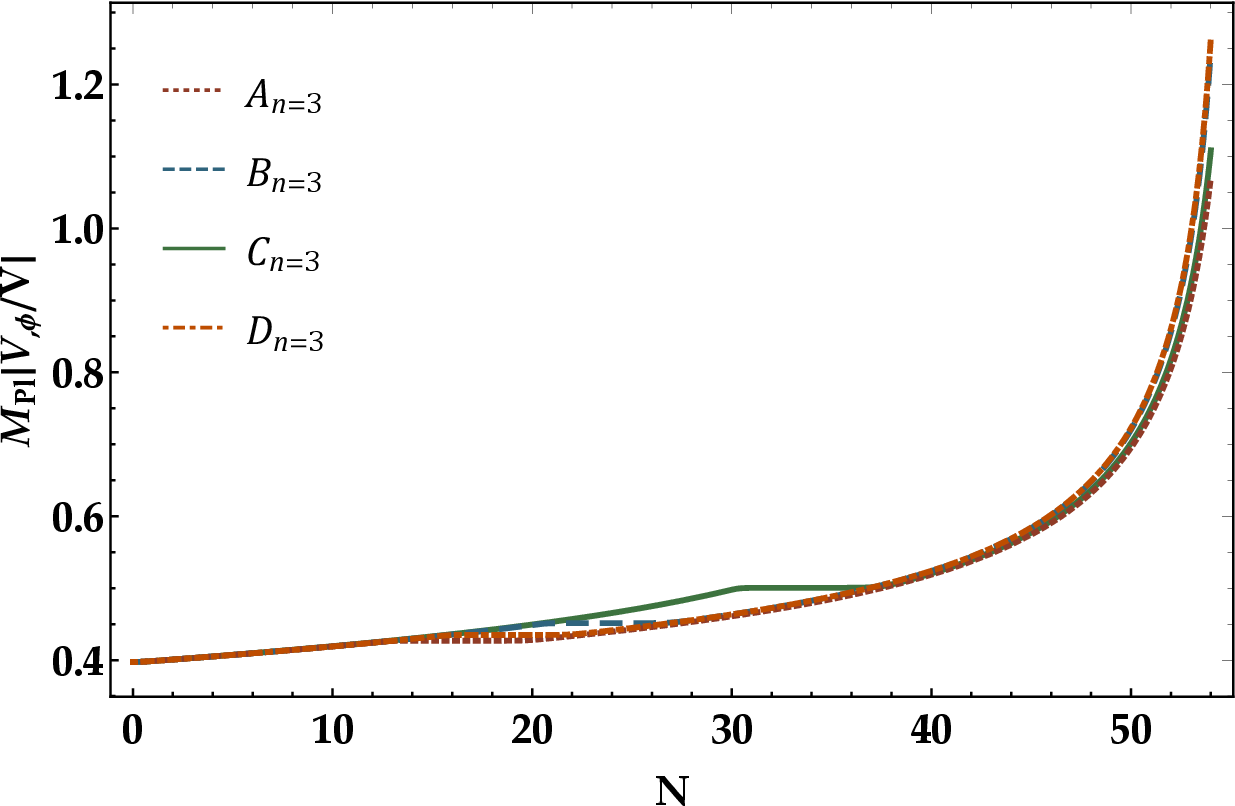}}\hspace{.1cm}
\subfigure[\label{vphi4_sc}]{ \includegraphics[width=.48\textwidth]%
{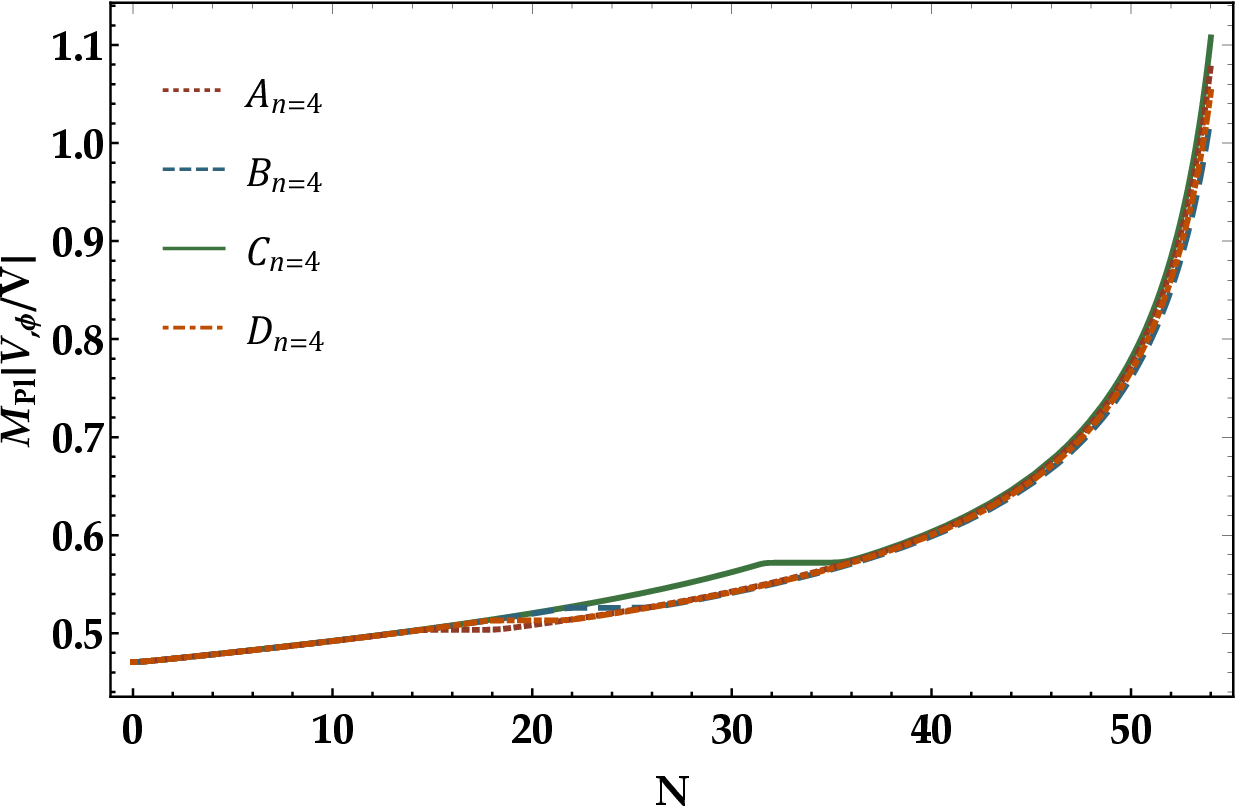}}
\end{minipage}
\caption{Confirming the validity of the second swampland criteria for our model. The figure demonstrates the fulfilment of the de Sitter conjecture, where $\Mpl |V_{,\phi}/V|$ is of the order of $\mathcal{O}(10^{-1} - 1)$ for (a) $n=1$, (b) $n=2$, (c) $n=3$, and (d) $n=4$.}\label{fig_sc_vphi}
\end{figure}

Before concluding this section, let's briefly discuss the swampland criteria and why it is important to consider this criteria when studying inflationary models.
The swampland criteria originate from string theory and contain two main theoretical conjectures, which are called the distance and de Sitter conjectures. They are defined by
\begin{equation}\label{sw_eq}
\frac{|\Delta\phi|}{{\Mpl}} \lesssim \Delta,
\quad
\Mpl \Big| \frac{V_{,\phi}}{V} \Big|\gtrsim c,
\end{equation}
respectively, where $(\Delta,~c) \sim \mathcal{O}(1)$ \cite{Garg:2019,Ooguri:2019,Kehagias:2018,Das:2019}.
The conflict between the swampland criteria and standard inflationary model is obvious.
The slow-roll condition in the standard inflationary model is given by $\e_V \equiv \left( {1}/{2}\right) \left( {V_{,\phi}}/{V}\right)^2<1$, which is in complete contradiction with the de Sitter conjecture.
On the other hand, in \cite{Das:2019}, it is indicated that the value of parameter $c$ can be smaller than unity as long as it is positive.
More precisely, $c \sim \mathcal{O}(10^{-1})$ is as good as $c \sim \mathcal{O}(1)$ \cite{Das:2019}.
For the standard inflationary model, the ratio of tensor to scalar perturbations is estimated by $r=16\epsilon_V$. Hence, considering the definition of $\epsilon_V$ and the de Sitter conjecture, one can deduce that $r \gtrsim 8c^2$.
Given that the parameter $c\sim \mathcal{O}(10^{-1})$ leads to $r \sim 0.08$, which is not consistent with Planck measurements at the CMB scale \cite{akrami:2018,bk18}.
Therefore, the conflict between the swampland criteria and the standard inflationary model motivates us to investigate the compatibility of the present model with the swampland criteria.
Here, we examine the swampland criteria for all cases in this study.
Our estimations show that in the present model, $\Delta \sim \mathcal{O}(1)$ and $c \sim \mathcal{O}(10^{-1})$, which are in complete agreement with the swampland criteria (See Figs. \ref{fig_sc_phi} and \ref{fig_sc_vphi}).

\section{Implications of reheating}\label{sec4}
At the end of inflation, the scalar field begins to oscillate nearby the minimum potential value.
Accordingly, inflaton will decay to the standard model particles. The reheating is the name of this process.
At the end of inflation, the universe became supercooled, and at the radiation-dominated (RD) epoch, the universe is thermalized. Reheating joins these two eras, and it means that after reheating stage, the hot Big Bang will begin \cite{mahbub:2020,Dalianis:2019}.

The inflaton decay rate can alter the length of the reheating stage. In addition, the decay rate of the inflaton is dependent on the reheating temperature $T_{\rm reh}$. Hence, The highest reheating temperature announces the shortest reheating period. A prolonged reheating stage can lead to the structures re-enter the horizon before the RD era. Thus, recognizing the horizon re-entry era is an essential issue. It can affect the mathematical formalisms utilized to estimate the physical quantities like the mass fraction of PBHs and the energy density of GWs  \cite{mahbub:2020}.

In the following, we investigate the reheating stage using the described method in \cite{Dalianis:2019,mahbub:2020} and specify the horizon re-entry epoch. If a scale such as $k^{-1}$ exits the horizon with $\Delta N_{k}$ $e$-fold before the end of inflation, we can formulate
\begin{equation}\label{deltank}
\left( \frac{a_{k,\text{re}}}{a_{\text{end}}} \right)^{\frac{1}{2}(1+3w_{\rm reh})}=e^{\Delta N_{k}},
\end{equation}
where $a_{k,\text{re}}$ indicates the scale factor at the horizon re-entry and $w_{\rm reh}$ is the equation of state parameter during the reheating epoch.
The number of $e$-folds between the end of inflation and horizon re-entry is given by
\begin{equation}\label{nk}
\tilde{N}_{k}\equiv \ln\left( \frac{a_{k,\text{re}}}{a_{\text{end}}} \right).
\end{equation}
Accordingly, Eqs. (\ref{deltank}) and (\ref{nk}) are associated as follows
\begin{equation}\label{deltaNtoN}
\tilde{N}_{k}=\frac{2}{1+3w_{\rm reh}}\Delta N_{k},
\end{equation}
where $w_{\rm reh}>-1/3$. The number of $e$-folds during the reheating phase is
 $\tilde{N}_{\text{reh}}\equiv \ln\left( a_{\text{reh}}/a_{\text{end}} \right)$,
where $a_{\rm reh}$ implies the scale factor at the end of reheating.

At the end of reheating era, the energy density takes the form $\rho_{\text{reh}}=\rho_{\text{end}}e^{-3\tilde{N}_{\text{reh}}(1+w_{\text{reh}})}$, in which $\rho_{\text{end}}=3H_{\text{end}}^{2}M_{\rm pl}^2$ indicates the energy density at the end of inflation \cite{mahbub:2020,Dalianis:2019}.
Then, $\tilde{N}_{\rm reh}$ can be asserted in terms of the energy density as follows
\begin{equation}\label{nktilde}
\tilde{N}_{\rm reh}=\frac{1}{3(1+w_{\text{reh}})}\ln\left( \frac{\rho_{\text{end}}}{\rho_{\text{reh}}} \right).
\end{equation}
Using the values of $\tilde{N}_{k}$ and $\tilde{N}_{\text{reh}}$, one can determine the time of horizon re-entry.
The scale $k^{-1}$ re-enters during the RD epoch when  $\tilde{N}_{k}>\tilde{N}_{\text{reh}}$, and horizon re-entry occurs in the reheating phase if $\tilde{N}_{k}<\tilde{N}_{\text{reh}}$.
In general, the reheating phase can be considered as an early matter-dominated era, i.e. $w_{\rm reh}\simeq 0$, then
the relation between $\Delta N$ and $\tilde{N}_{\rm reh}$ is given by \cite{mahbub:2020}
\begin{equation}\label{deltaNfinal}
\Delta N\simeq 57.3+\frac{1}{4}\ln\left( \frac{\epsilon_{1_{*}}V_{*}}{\rho_{\text{end}}} \right)-\frac{1}{4}\tilde{N}_{\text{reh}},
\end{equation}
where $\Delta N$ is the $e$-fold number of the period in which both the CMB anisotropies and
the PBH are produced. Also, $\epsilon_{1_{*}}$ and $V_{*}$ describe the value of the first slow-roll parameter and potential at the pivot scale, respectively.
The amount of observable inflation can be expressed as $\Delta N=N_{\text{end}}-N_{*}$, where $N_{*}$ is the $e$-fold number at the pivot scale $k_{*}=0.05~{\rm Mpc}^{-1}$, and  $N_{\text{end}}$ shows the $e$-fold number at the end of inflation.
On the other hand, we can write $\Delta N=N_{\text{k}}^{\text{peak}}+\Delta N_{\text{k}}^{\text{peak}}$, where $N_{\text{k}}^{\text{peak}}$ indicates the $e$-fold number that the peak scale $k_{\text{peak}}^{-1}$ exit the horizon.

Placing boundaries on $\Delta N$ can further clarify the current criterion that designates the horizon re-entry epoch. Using Eq. (\ref{deltaNtoN}) and taking $w_{\rm reh}\simeq 0$ one can obtain $\Delta N_k^{\text{peak}}=\tilde{N}_k^{\text{peak}}/2$, where $\tilde{N}_k^{\text{peak}}$ indicates the number of $e$-folds after the end of inflation until $k^{-1}_{\text{peak}}$ re-enters the horizon.
If horizon re-entry of the peak occurs at the moment the reheating phase ends, we can write  $\Delta N_{\text{peak}}^{(\rm cr)}=\tilde{N}_{\text{reh}}/2$, and as a consequence, we have $\Delta N=N_k^{\text{peak}}+\tilde{N}_{\text{reh}}/2$. Then, $\Delta N_{\text{peak}}^{(\rm cr)}$ can be evaluated using (\ref{deltaNfinal}), as follows \cite{mahbub:2020}

\begin{equation}\label{deltaNpeak}
\Delta N_{\text{peak}}^{(\rm cr)}=\frac{2}{3}\left[ 57.3 +\frac{1}{4}\ln\left( \frac{\epsilon_{1_{*}}V_{*}}{\rho_{\text{end}}} \right)-N_k^{\text{peak}} \right].
\end{equation}
Now, it is possible to compute the values of $\Delta N_{\text{peak}}^{(\text{cr})}$ and $\tilde{N}_{\text{reh}}$ for all parameter sets listed in Table \ref{tab1-phi}.
Based on the findings, it can be concluded that the scale $k_{\text{peak}}^{-1}$ re-enters the horizon during the RD era if $\Delta N_{\text{peak}}^{(\text{cr})} > \tilde{N}_{\text{reh}}/2$ or $\Delta N > N_k^{\text{peak}}+\Delta N_{\text{peak}}^{(\text{cr})}$.
Our results, listed in Table \ref{tab9}, indicate that the horizon re-entry for the scale $k^{-1}_{\text{peak}}$ occurs after the reheating phase for all cases in the present study.

\begin{figure}[H]
\begin{minipage}[b]{1\textwidth}
\centering
\subfigure[\label{phi1_ps_reh}]{ \includegraphics[width=0.4\textwidth]%
{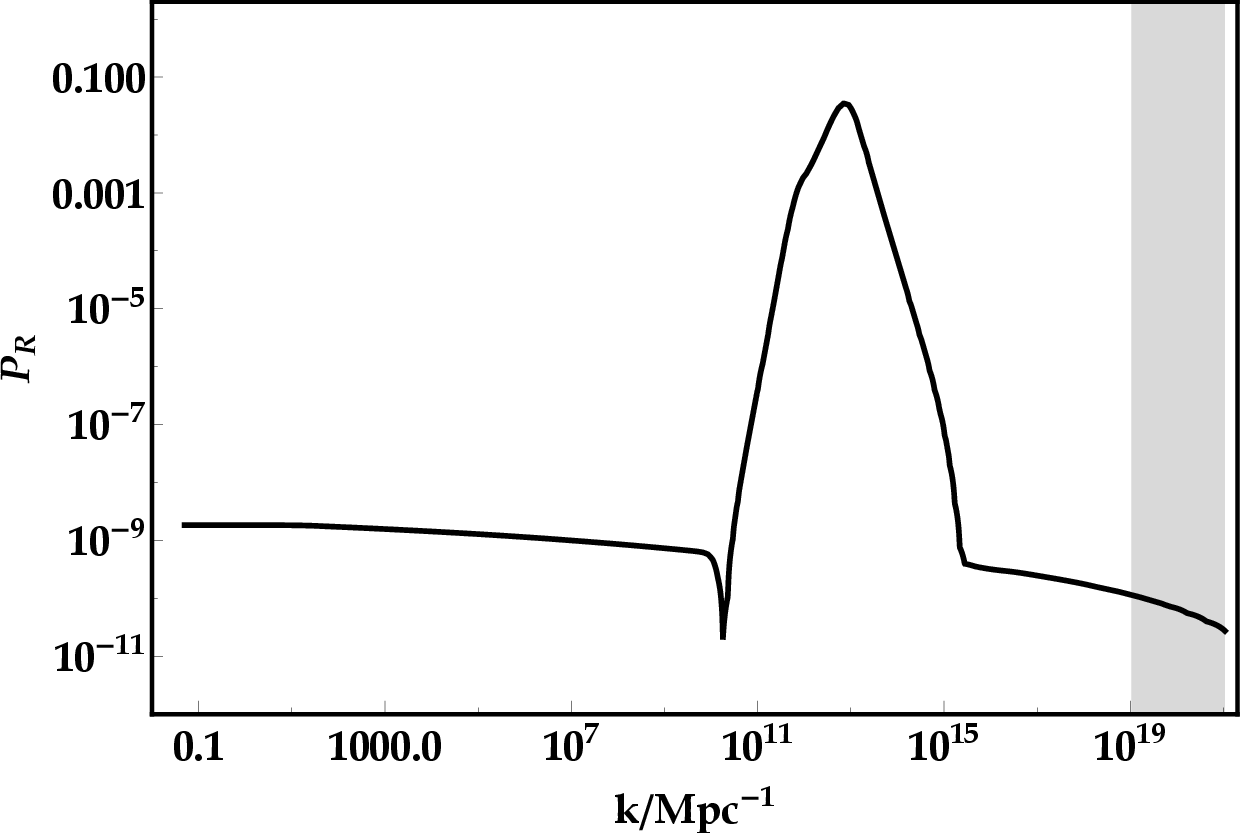}}\hspace{.1cm}
\subfigure[\label{phi1_com_reh}]{ \includegraphics[width=.4\textwidth]%
{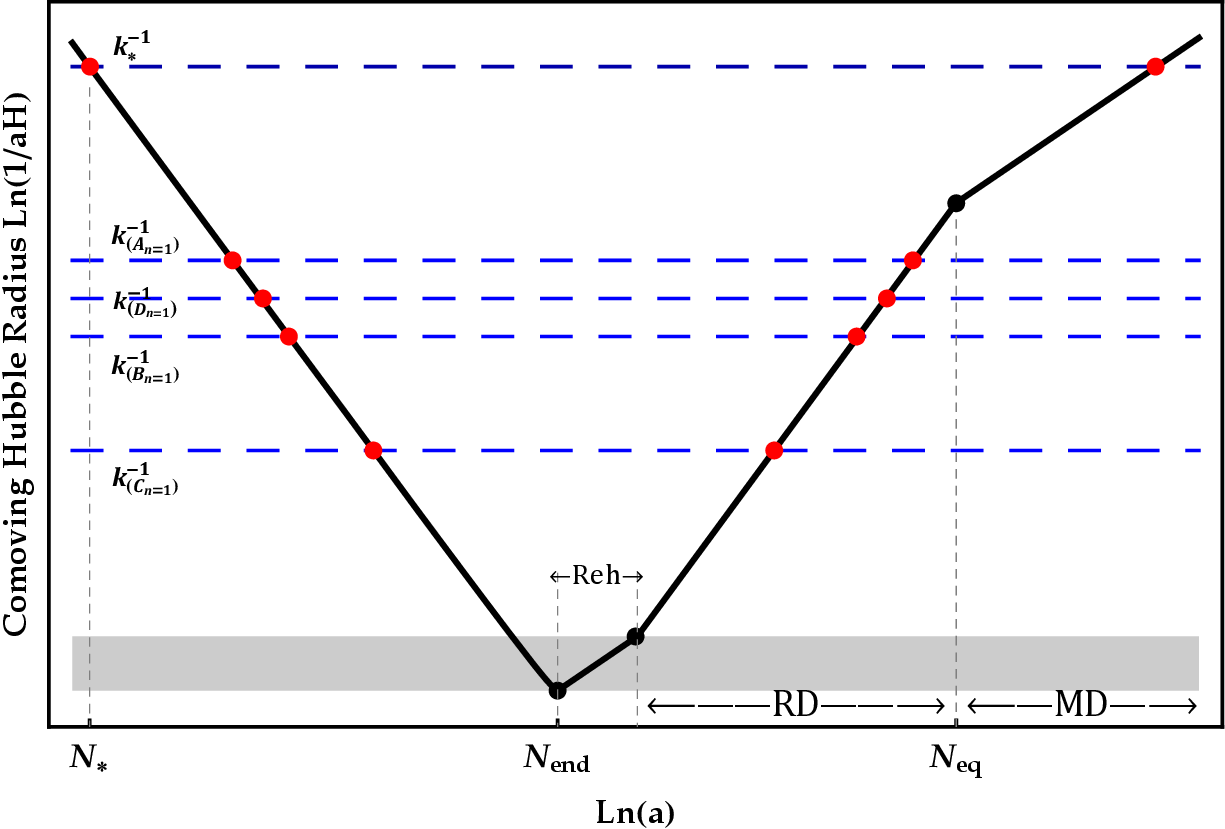}}
\centering
\subfigure[\label{phi2_ps_reh} ]{ \includegraphics[width=0.4\textwidth]%
{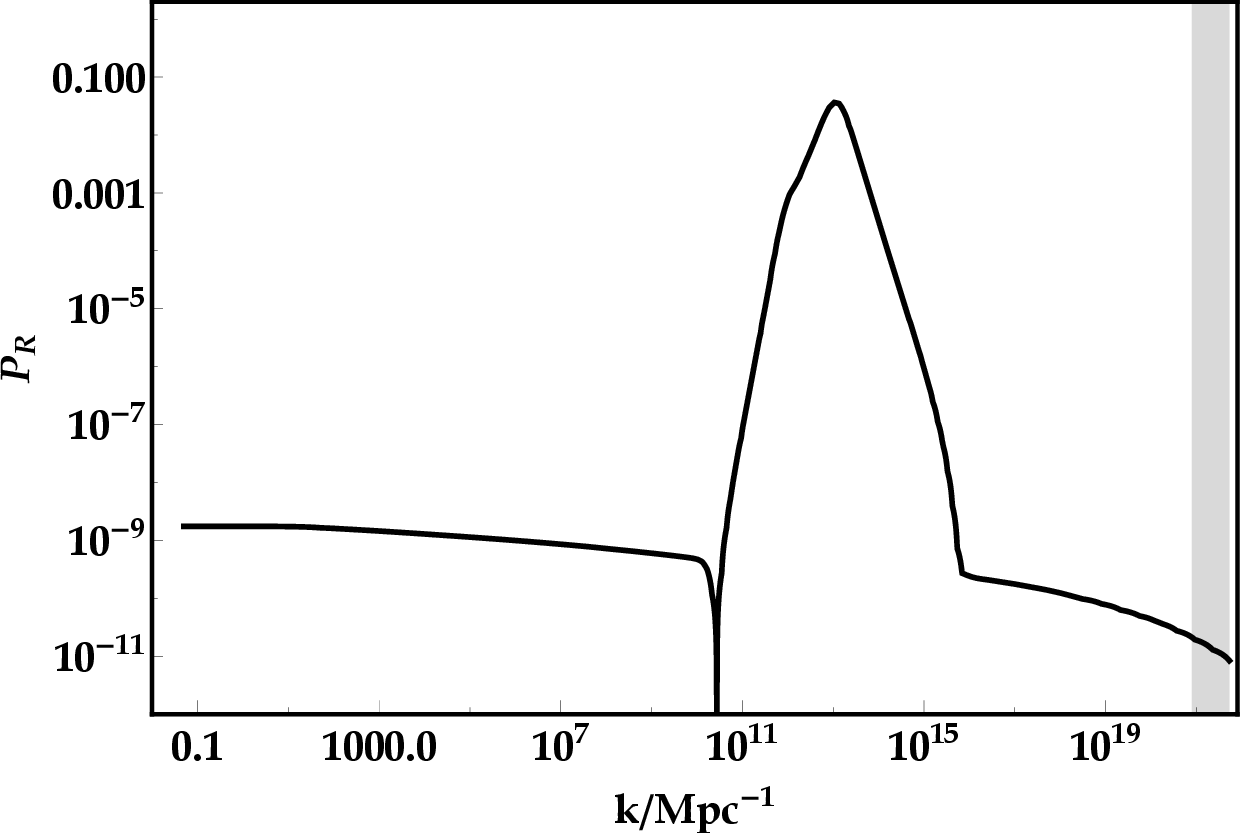}}\hspace{.1cm}
\subfigure[\label{phi2_com_reh}]{ \includegraphics[width=.4\textwidth]%
{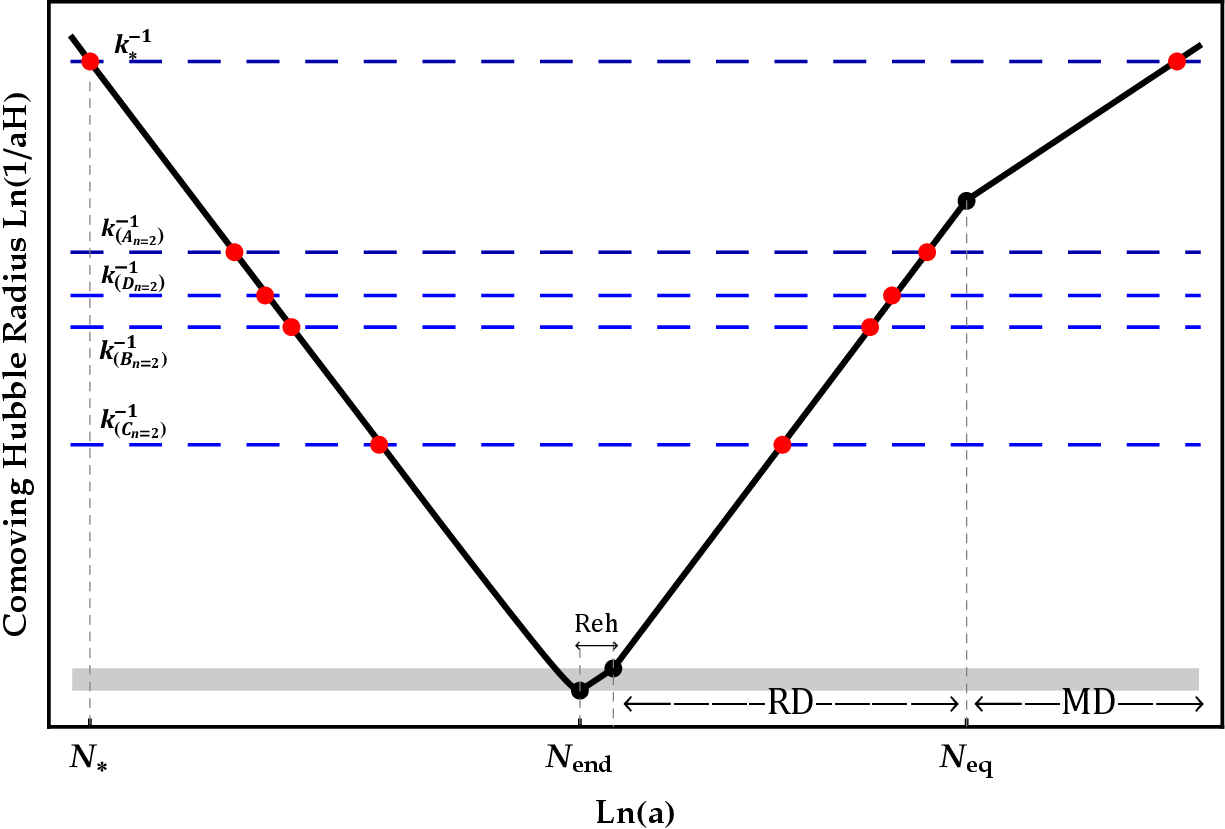}}
\centering
\subfigure[\label{phi3_ps_reh}]{ \includegraphics[width=0.4\textwidth]%
{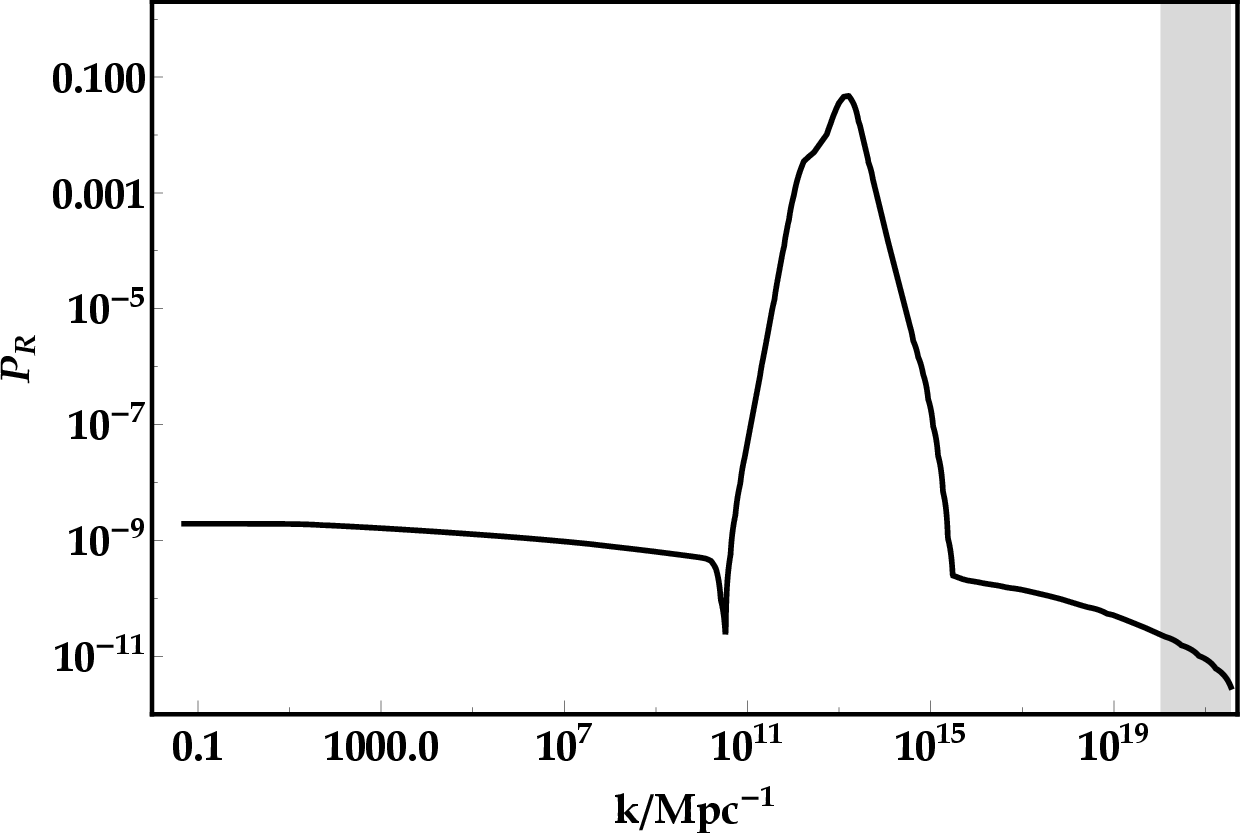}}\hspace{.1cm}
\subfigure[\label{phi3_com_reh}]{ \includegraphics[width=.4\textwidth]%
{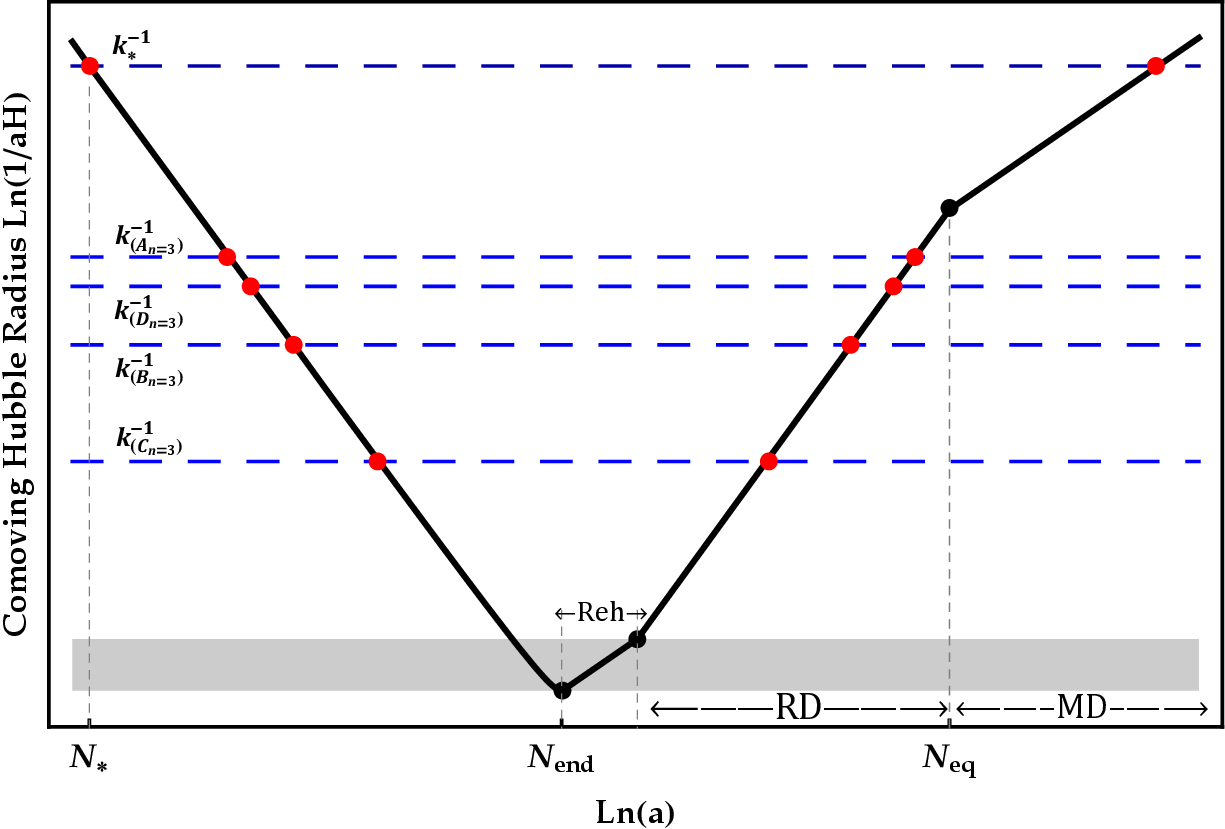}}
\centering
\subfigure[\label{phi4_ps_reh} ]{ \includegraphics[width=0.4\textwidth]%
{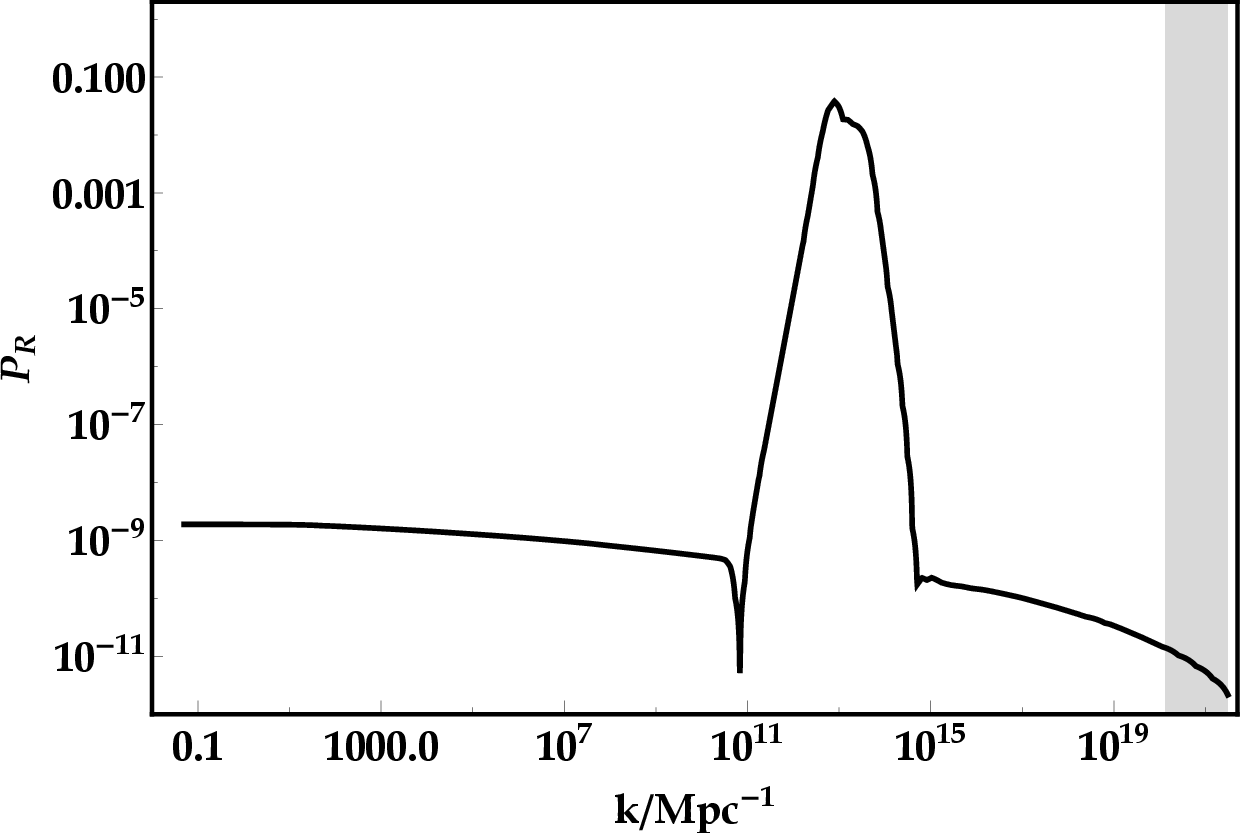}}\hspace{.1cm}
\subfigure[\label{phi4_com_reh}]{ \includegraphics[width=.4\textwidth]%
{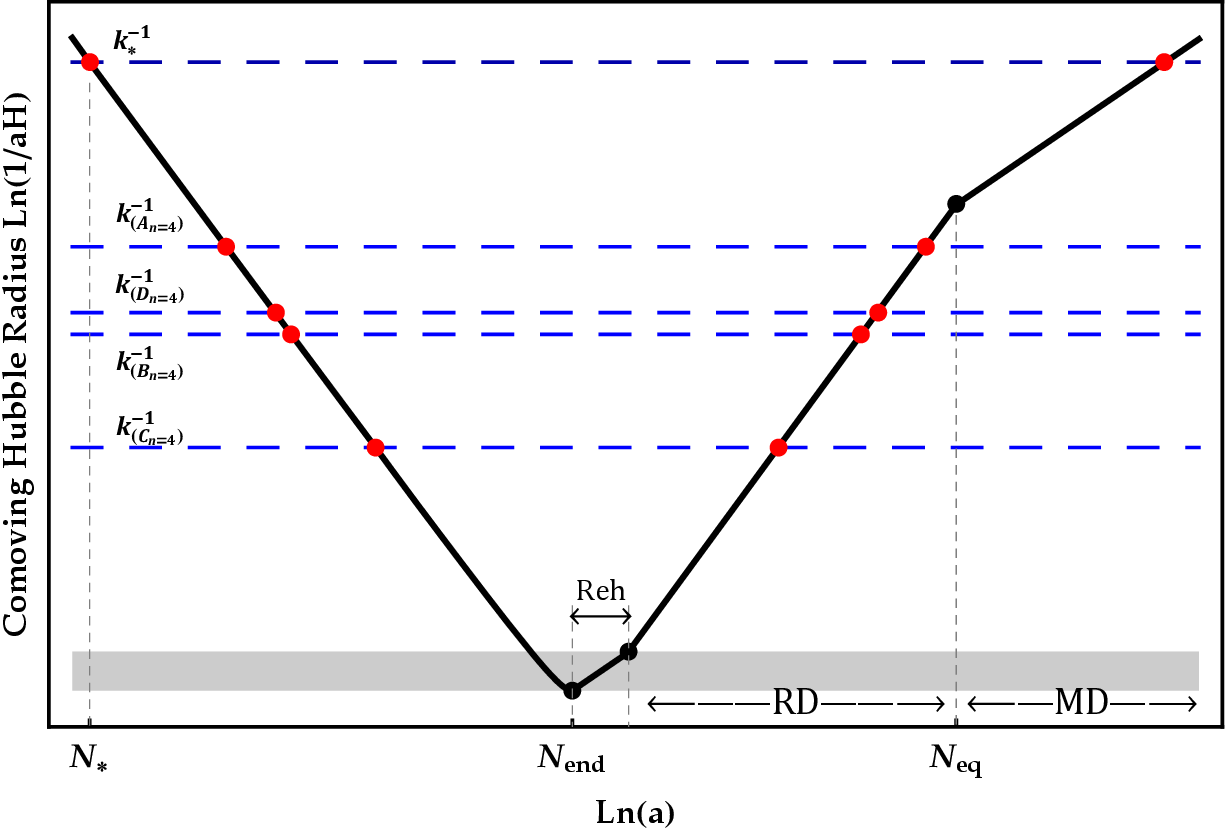}}
\end{minipage}
\caption{
(Left) The scalar power spectrum for the cases $C_{\rm n=1}$,$C_{\rm n=2}$, $C_{\rm n=3}$ and $C_{\rm n=4}$, respectively.
(Right) The comoving Hubble radius in versus $e$-fold numbers. Here, the red points represent the horizon exit and re-entry for different scales, which are indicated by blue dashed lines.
The shaded areas comprise the scales that re-enter the horizon during the reheating phase.
}\label{fig_reh}
\end{figure}

The evolution of the scalar power spectrum for the cases $C_n$, for instance, is depicted in the left panels of Fig. \ref{fig_reh}.
The shaded regions illustrate the scales that re-enter the horizon throughout the reheating era, revealing that the peak scales re-enter the horizon during the RD era.
Furthermore, in the rights panel of Fig. \ref{fig_reh}, we have depicted the comoving Hubble radius versus $e$-fold number.
In these figures, the red points indicate the time of horizon exit and subsequent re-entry of peak scales.
The shaded regions represent the reheating phase.
Therefore, it is evident that the peak scale re-enters the horizon after the reheating era.
Hence, we can use the mathematical formalisms governing the RD epoch, to calculate PBHs abundance and energy density of the scalar-induced GWs in the next sections.

\begin{table}[H]
\centering
\caption{The values of $\Delta N$,  $k_{\text{peak}}$, $N_k^{\text{peak}}$, $\Delta N_{\text{peak}}^{(\rm cr)}$
and $\tilde{N}_{\text{reh}}$ for all parameter sets in Table \ref{tab1-phi}.}
\begin{tabular}{ccccccc}
\thickhline
 Sets \quad &
\quad $\Delta N$\quad &
\quad $k_{\text{peak}}/{\rm Mpc}^{-1}$ \quad &
\quad $N_k^{\text{peak}}$ \quad &
\quad $\Delta N_{\text{peak}}^{(\rm cr)}$ \quad &
\quad$\tilde{N}_{\text{reh}}$ \quad\\ [0.5ex]
\thickhline
$A_{\rm n=1}$ \quad &
\quad 54.25 \quad &
\quad $6.9\times 10^{5}$ \quad &
\quad 16.5 \quad &
\quad 26.62 \quad &
\quad 8.76 \\ \hline
$B_{\rm n=1}$ \quad &
\quad 54.43 \quad &
\quad $4.4\times 10^{8}$ \quad &
\quad 23.0 \quad &
\quad 22.29  \quad &
\quad 8.05  \\ \hline
$C_{\rm n=1}$\quad &
\quad 54.13 \quad &
\quad $7.0\times 10^{12}$ \quad &
\quad 31.8 \quad &
\quad 15.79 \quad &
\quad 9.24  \\ \hline
$D_{\rm n=1}$\quad &
\quad 53.20 \quad &
\quad $1.7\times 10^{7}$ \quad &
\quad 19.7 \quad &
\quad 24.47 \quad &
\quad 12.96  \\
\thickhline
$A_{\rm n=2}$ \quad &
\quad 55.26 \quad &
\quad $6.7\times 10^{5}$ \quad &
\quad 16.5 \quad &
\quad 26.90 \quad &
\quad 6.36 \\ \hline
$B_{\rm n=2}$ \quad &
\quad 55.14 \quad &
\quad $4.2\times 10^{8}$ \quad &
\quad 23.0 \quad &
\quad 22.57  \quad &
\quad 6.84 \\ \hline
$C_{\rm n=2}$\quad &
\quad 55.90 \quad &
\quad $1.0\times 10^{13}$ \quad &
\quad 33.2 \quad &
\quad 15.73 \quad &
\quad 3.81 \\ \hline
$D_{\rm n=2}$\quad &
\quad 54.11 \quad &
\quad $2.8\times 10^{7}$ \quad &
\quad 20.2 \quad &
\quad 24.38 \quad &
\quad 10.87 \\
\thickhline
$A_{\rm n=3}$ \quad &
\quad 55.48 \quad &
\quad $5.1\times 10^{5}$ \quad &
\quad 16.0 \quad &
\quad 27.40 \quad &
\quad 6.45  \\ \hline
$B_{\rm n=3}$ \quad &
\quad 54.91 \quad &
\quad $8.5\times 10^{8}$ \quad &
\quad 23.7 \quad &
\quad 22.23  \quad &
\quad 8.77  \\ \hline
$C_{\rm n=3}$\quad &
\quad 55.33 \quad &
\quad $1.6\times 10^{13}$ \quad &
\quad 33.7 \quad &
\quad 15.56 \quad &
\quad 7.07 \\ \hline
$D_{\rm n=3}$\quad &
\quad 54.90 \quad &
\quad $6.1\times 10^{6}$ \quad &
\quad 18.7 \quad &
\quad 25.58 \quad &
\quad 8.86 \\
\thickhline
$A_{\rm n=4}$ \quad &
\quad 55.90 \quad &
\quad $3.0\times 10^{5}$ \quad &
\quad 15.7 \quad &
\quad 27.71 \quad &
\quad 5.64  \\ \hline
$B_{\rm n=4}$ \quad &
\quad 56.24 \quad &
\quad $5.0\times 10^{8}$ \quad &
\quad 23.2 \quad &
\quad 22.70  \quad &
\quad 4.26 \\ \hline
$C_{\rm n=4}$\quad &
\quad 55.73 \quad &
\quad $7.3\times 10^{12}$ \quad &
\quad 33.0 \quad &
\quad 16.21 \quad &
\quad 6.31 \\ \hline
$D_{\rm n=4}$\quad &
\quad 56.01 \quad &
\quad $9.8\times 10^{6}$ \quad &
\quad 19.2 \quad &
\quad 25.35 \quad &
\quad 5.10 \\
\thickhline
\end{tabular}
\label{tab9}
\end{table}

\vspace*{-2cm}

\section{The abundance of PBHs}
\label{sec5}
As mentioned before, the gravitational collapse of the over-dense regions at horizon re-entry can lead to the formation of PBHs.
On the other hand, for our model, the horizon re-entry occurs during the RD era based on the obtained results in the section \ref{sec4}.
Hence, if the gravity of this over-dense region overcomes the RD pressure at the horizon re-entry, gravitational collapse, and consequently, PBH production would be possible.

The PBH mass at the production time is related to the horizon mass $M_{\rm H}$ and can be estimated by $M=\gamma M_{\rm H}$.
Here, the parameter $\gamma=0.2$ represents the collapse efficiency  \cite{Inomata:2017,Sasaki:2018}.
The ratio of the density parameter of the PBHs to the total dark matter at the present time can be calculated as follows \cite{Sasaki:2018}
\begin{align}
\label{fpbheq}
f_{\rm{PBH}}(M)\simeq \frac{\Omega_{\rm {PBH}}}{\Omega_{\rm{DM}}}= \frac{\beta(M)}{1.84\times10^{-8}}\left(\frac{\gamma}{0.2}\right)^{3/2}\left(\frac{g_*}{10.75}\right)^{-1/4}
\left(\frac{0.12}{\Omega_{\rm{DM}}h^2}\right)
\left(\frac{M}{M_{\odot}}\right)^{-1/2},
\end{align}
where $g_{*}\simeq106.75$ stands for the effective degrees of freedom during the PBH formation process \cite{Motohashi:2017}. The parameter $\beta$ in Eq. (\ref{fpbheq}) represents the mass fraction of PBH and is given by \cite{Sasaki:2018,Motohashi:2017,young:2014,harada:2013,Musco:2013,Shibata:1999,Polnarev:2007,Musco:2009}
\begin{align}
\label{betta}
\beta(M)=\int_{\delta_{c}}\frac{{\rm d}\delta}{\sqrt{2\pi\sigma^{2}(M)}}e^{-\frac{\delta^{2}}{2\sigma^{2}(M)}}=\frac{1}{2}~ {\rm erfc}\left(\frac{\delta_{c}}{\sqrt{2\sigma^{2}(M)}}\right),
\end{align}
where $\delta_{\rm th}$ is the threshold density contrast for PBH formation.
Numerical investigations have confirmed that the value of $\delta_{\rm th}$ for PBH formation in the RD era can range from 0.3 to 0.66 \cite{Shibata:1999,Polnarev:2007,Musco:2009,harada:2013,Musco:2013}.
Here, we set $\delta_{\rm th} = 0.4$, which aligns with the calculations in \cite{Musco:2013,harada:2013}.
Furthermore, $\sigma_{M_{\rm PBH}}$ describes the variance of density contrast at the comoving horizon scale and it is estimated as follows \cite{young:2014}
\begin{align}	
\label{Sigma}
\sigma_{k}^{2}=\left(\frac{4}{9} \right)^{2} \int \frac{{\rm d}q}{q} W^{2}(q/k)(q/k)^{4} {\cal P}_{s}(q),
\end{align}
where
$W(x)=\exp{\left(-x^{2}/2 \right)} $
is the Gaussian window function.
Also, the mass of the PBHs in terms of wavenumber can be estimated as follows \cite{Motohashi:2017,mishra:2020,Sasaki:2018}
\begin{align}
\label{masseq}
M_{\rm PBH}(k)= M_{\odot} \left(\frac{\gamma}{0.2} \right) \left(\frac{10.75}{g_{*}} \right)^{1/6} \left(\frac{k}{1.9\times 10^{6}\rm Mpc^{-1}} \right)^{-2}.
\end{align}
Utilizing the Eqs. (\ref{fpbheq}) and (\ref{masseq}), one can estimate the abundance of PBHs in terms of their mass for all cases in this study as listed in Table \ref{tab2-phi}.

\begin{figure}[H]
\begin{minipage}[b]{1\textwidth}
\subfigure[\label{fig-phi-f} ]{ \includegraphics[width=0.48\textwidth]{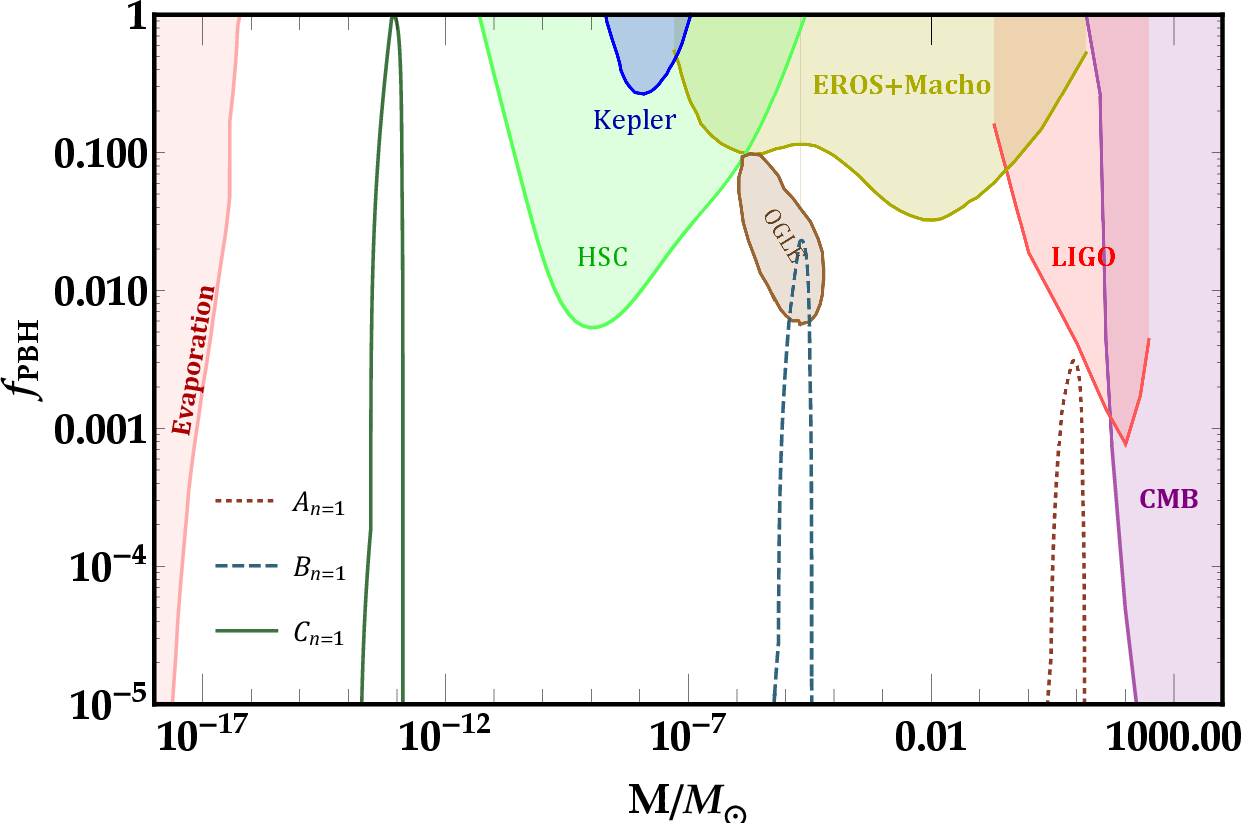}}
\subfigure[\label{fig-phi2-f}]{ \includegraphics[width=.48\textwidth]%
{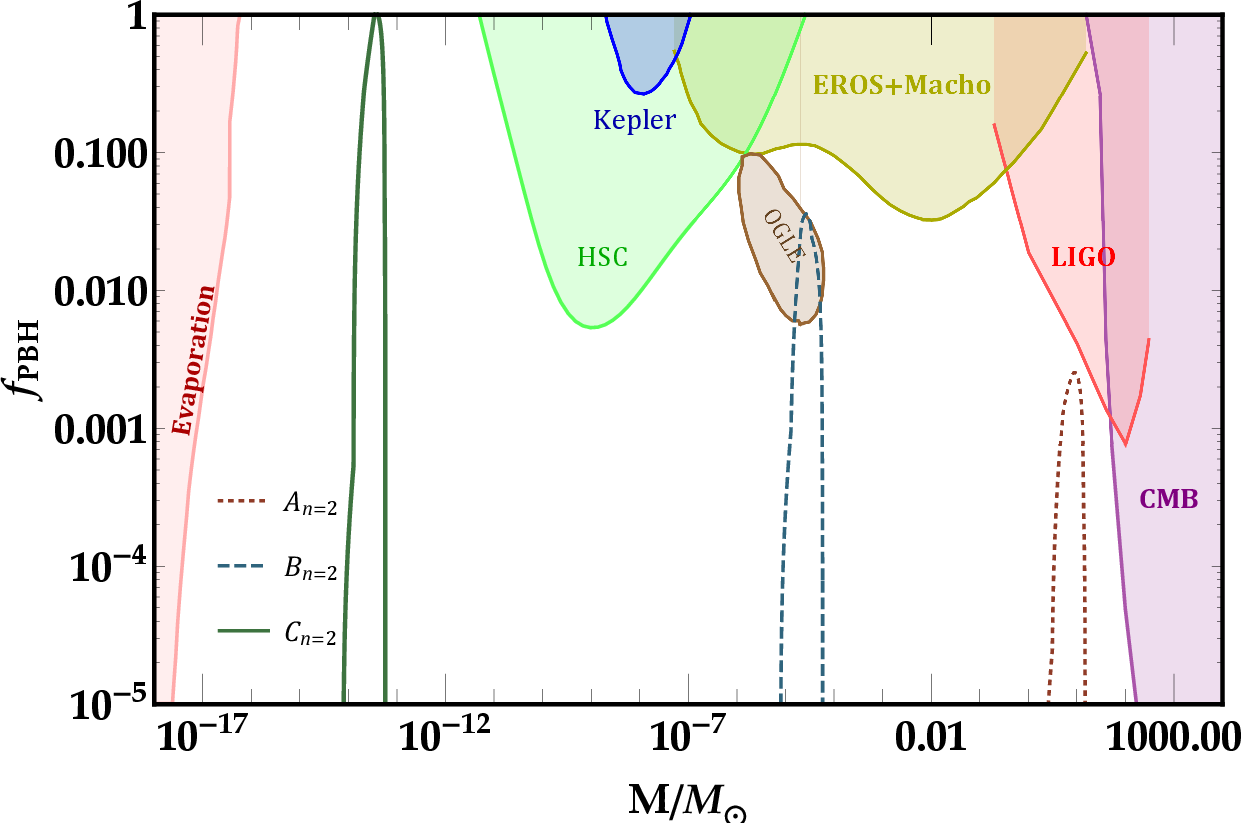}}
\subfigure[\label{fig-phi3-f}]{
 \includegraphics[width=.48\textwidth]%
{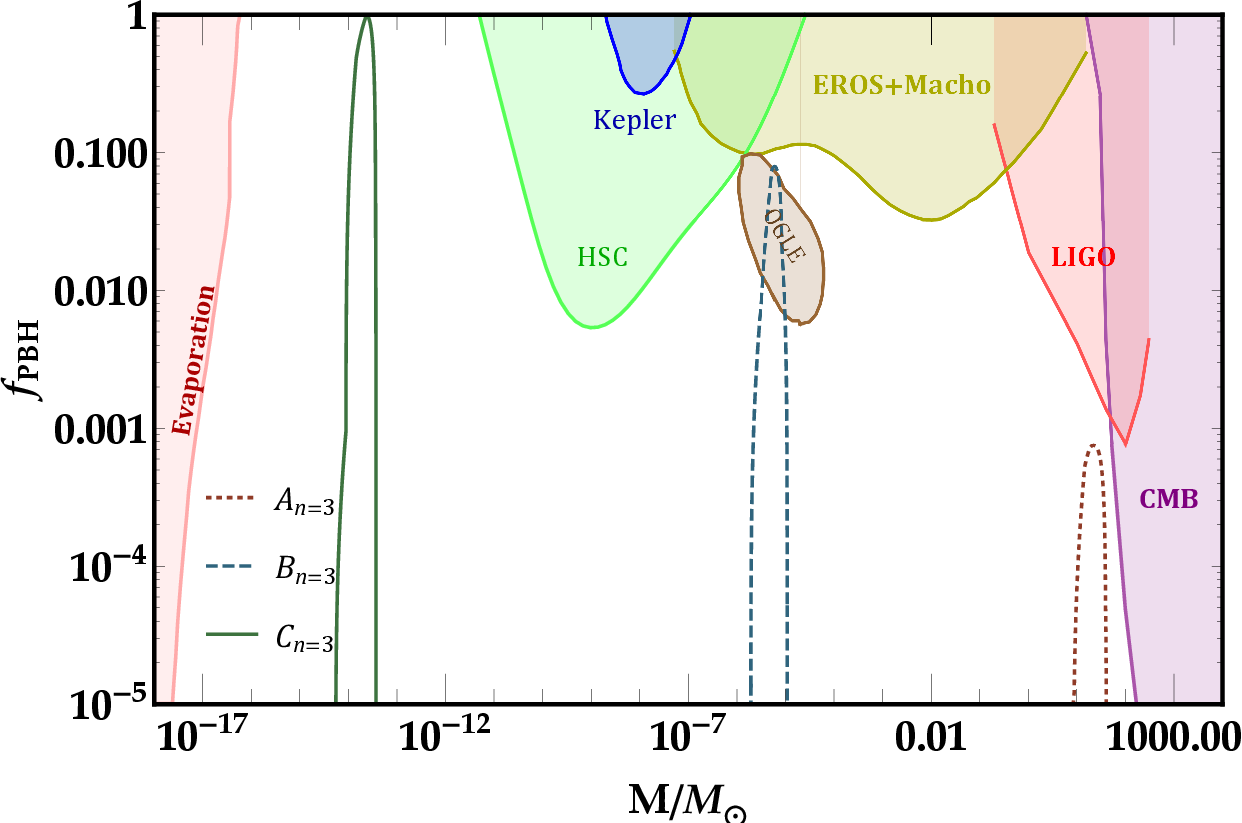}}
\subfigure[\label{fig-phi4-f}]{
 \includegraphics[width=.49\textwidth]%
{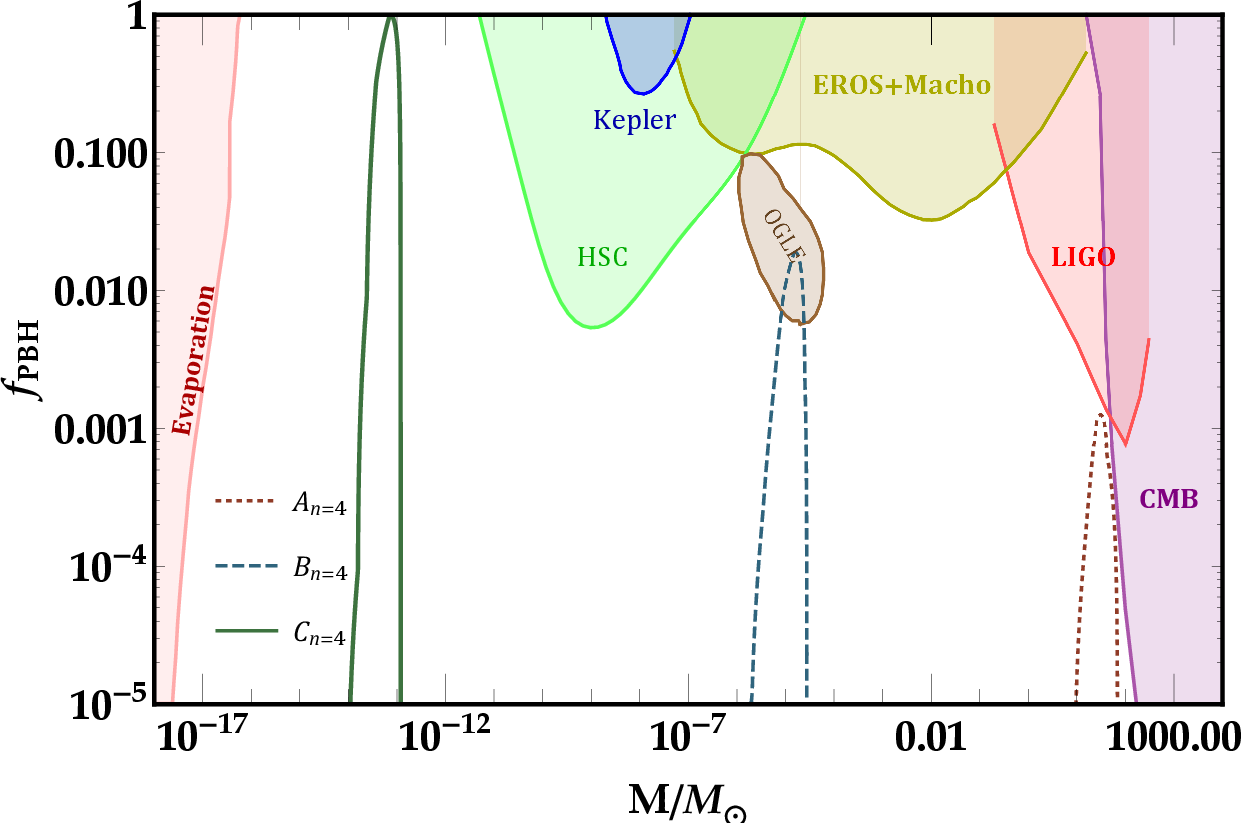}}
\end{minipage}
\caption{The abundances of PBHs with respect to their masses $M$ have been shown for three parameter sets corresponding to (a) $n=1$, (b) $n=2$, (c) $n=3$, and (d) $n=4$. The dotted, dashed, and solid lines in each figure describe cases $A_{\rm n}$, $B_{\rm n}$, and $C_{\rm n}$, respectively.
The shaded regions represent recent observational constraints on the fractional abundance of PBHs. The purple area depicts the accretion constraints by CMB \cite{CMB}. The border of the red shaded region describes the upper bound on the abundance of PBHs derived from the LIGO-Virgo event consolidation rate \cite{Abbott:2019,Boehm:2021,Chen:2022,Kavanagh:2018}. The brown shaded region represents the allowed region for abundance of PBHs in the OGLE data \cite{OGLE}. The yellow, blue, and green areas respectively represent the observational constraints resulting from EROS/MACHO \cite{Alcock:2001,EORS}, Kepler \cite{Kepler}, and Subaru-HSC \cite{subaro}. The pink shaded region describes constraints related to PBHs evaporation \cite{EGG,Laha:2019,Clark}.} \label{fpbh-figs}
\end{figure}

As mentioned before, there are a lot of observational constraints on PBH abundances. These constraints are depicted as shaded regions in Fig. \ref{fpbh-figs}.
For the case $A_{\rm n=1}$, the mass of generated PBH is around the solar mass, while the masses of PBHs corresponding to the cases $A_{\rm n=2}$, $A_{\rm n=3}$, and $A_{\rm n=4}$ are of order of $10M_{\odot}$.
The PBHs produced at these mass scales can be bounded by the upper limits on the LIGO merger rate, as shown in Fig. \ref{fpbh-figs}.
Hence, these PBHs can be the source of the GWs observed by the LIGO and Virgo collaboration.
For the cases  $B_{\rm n=1}$, $B_{\rm n=2}$, $B_{\rm n=3}$ and $B_{\rm n=4}$, which have masses ranging from $\mathcal{O}(10^{-6})M_\odot$ to $\mathcal{O}(10^{-5})M_\odot$, the peak of the PBH abundance $f_{\text{PBH}}^{\text{peak}}$ increases to the $\mathcal{O}(10^{-2})$.
As shown in Fig. \ref{fpbh-figs}, these PBHs are completely compatible with the allowed region of microlensing events in the OGLE data.
The most interesting cases in the present study are PBHs with masses in the rang
$10^{-17} \leq M_{\text{PBH}}^{\text{peak}}\leq 10^{-12}$.
There is no upper limit within this mass interval. Consequently, PBHs with masses in this range could explain most of the dark matter in the universe.
Our results indicate that the cases $C_{\rm n=1}$, $C_{\rm n=2}$, $C_{\rm n=3}$ and $C_{\rm n=4}$ fall within this mass range. As shown in Fig. \ref{fpbh-figs}, these cases can account for approximately all of the dark matter in the universe.
Our estimations indicate that the abundance of PBHs for the cases $D_{\rm n}$ is negligible. This is due to the upper limit imposed on the amplitude of the scalar power spectrum by PTA observations \cite{Inomata:2019-a}.
It is necessary to mention that the values of the masses and abundances of the PBHs are listed in Table \ref{tab2-phi}.

\section{Scalar-induced gravitational waves}
\label{sec6}
At the horizon re-entry, simultaneous with PBHs formation, scalar-induced GWs can be generated \cite{Matarrese:1998,Mollerach:2004,Saito:2009,Garcia:2017,Cai:2019-a,Cai:2019-b,Cai:2019-c,Bartolo:2019-a,Bartolo:2019-b,Wang:2019,Fumagalli:2020b,Domenech:2020a,Domenech:2020b,Hajkarim:2019,Kohri:2018,Xu:2020,Fu:2020}.
The detection of these generated GWs is possible through the various GWs observatory.
In this section, we will study the generation of scalar-induced GWs within the framework GB gravity in the presence of the power-law potentials.
The transition from the inflationary era to the RD epoch necessitates the thermalization of the universe. This thermalization occurs through the decay of the inflaton field during the reheating phase.
As a result, the influence of the inflaton field on the cosmic evolution
becomes insignificant during RD epoch.
Therefore, the standard Einstein formulation can be utilised to study the generation of scalar-induced GWs during this era.
Thus during RD era, the energy density parameter of scalar-induced GWs can be calculated as follows  \cite{Kohri:2018,Lu:2019}
%
\begin{equation}
 \label{OGW}
 \Omega_{\rm GW}(k,\eta)=\frac{1}{6}\left(\frac{k}{aH}\right)^{2}\int_{0}^{\infty}dv\int_{|1-v|}^{|1+v|}du\left(\frac{4v^{2}-\left(1-u^{2}+v^{2}\right)^{2}}{4uv}\right)^{2}\overline{I_{RD}^{2}(u,v,x)}\mathcal{P}_{s}(ku)\mathcal{P}_{s}(kv),
\end{equation}
where $\eta$ denotes the conformal time, and the time average of the source terms is given by
\begin{align}
 \overline{I_{\rm RD}^{2}(u,v,x\to\infty)}= & \frac{1}{2x^{2}}\Bigg[\left(\frac{3\pi\left(u^{2}+v^{2}-3\right)^{2}\Theta\left(u+v-\sqrt{3}\right)}{4u^{3}v^{3}}+\frac{T_{c}(u,v,1)}{9}\right)^{2}
 \nonumber\\
 & +\left(\frac{\tilde{T}_{s}(u,v,1)}{9}\right)^{2}\Bigg],
 \label{IRD2b}
\end{align}
where $\Theta$ stands for the Heaviside theta function.
In addition, the functions employed within Eq. (\ref{Tc}) are defined as follows

\begin{align}
T_{c}= & -\frac{27}{8u^{3}v^{3}x^{4}}\Bigg\{-48uvx^{2}\cos\left(\frac{ux}{\sqrt{3}}\right)\cos\left(\frac{vx}{\sqrt{3}}\right)\left(3\sin(x)+x\cos(x)\right)+
\nonumber\\
& 48\sqrt{3}x^{2}\cos(x)\left(v\sin\left(\frac{ux}{\sqrt{3}}\right)\cos\left(\frac{vx}{\sqrt{3}}\right)+u\cos\left(\frac{ux}{\sqrt{3}}\right)\sin\left(\frac{vx}{\sqrt{3}}\right)\right)+
\nonumber\\
& 8\sqrt{3}x\sin(x)\Bigg[v\left(18-x^{2}\left(u^{2}-v^{2}+3\right)\right)\sin\left(\frac{ux}{\sqrt{3}}\right)\cos\left(\frac{vx}{\sqrt{3}}\right)+
\nonumber\\
& u\left(18-x^{2}\left(-u^{2}+v^{2}+3\right)\right)\cos\left(\frac{ux}{\sqrt{3}}\right)\sin\left(\frac{vx}{\sqrt{3}}\right)\Bigg]+
\nonumber\\
& 24x\cos(x)\left(x^{2}\left(-u^{2}-v^{2}+3\right)-6\right)\sin\left(\frac{ux}{\sqrt{3}}\right)\sin\left(\frac{vx}{\sqrt{3}}\right)+
\nonumber\\
& 24\sin(x)\left(x^{2}\left(u^{2}+v^{2}+3\right)-18\right)\sin\left(\frac{ux}{\sqrt{3}}\right)\sin\left(\frac{vx}{\sqrt{3}}\right)\Bigg\}
\nonumber\\
& -\frac{\left(27\left(u^{2}+v^{2}-3\right)^{2}\right)}{4u^{3}v^{3}}\Bigg\{\text{Si}\left[\left(\frac{u-v}{\sqrt{3}}+1\right)x\right]-\text{Si}\left[\left(\frac{u+v}{\sqrt{3}}+1\right)x\right]
\nonumber\\
& +\text{Si}\left[\left(1-\frac{u-v}{\sqrt{3}}\right)x\right]-\text{Si}\left[\left(1-\frac{u+v}{\sqrt{3}}\right)x\right]\Bigg\},
\label{Tc}
\end{align}

\begin{align}
T_{s}= & \frac{27}{8u^{3}v^{3}x^{4}}\Bigg\{48uvx^{2}\cos\left(\frac{ux}{\sqrt{3}}\right)\cos\left(\frac{vx}{\sqrt{3}}\right)\left(x\sin(x)-3\cos(x)\right)-
\nonumber\\
& 48\sqrt{3}x^{2}\sin(x)\left(v\sin\left(\frac{ux}{\sqrt{3}}\right)\cos\left(\frac{vx}{\sqrt{3}}\right)+u\cos\left(\frac{ux}{\sqrt{3}}\right)\sin\left(\frac{vx}{\sqrt{3}}\right)\right)+
\nonumber\\
& 8\sqrt{3}x\cos(x)\Bigg[v\left(18-x^{2}\left(u^{2}-v^{2}+3\right)\right)\sin\left(\frac{ux}{\sqrt{3}}\right)\cos\left(\frac{vx}{\sqrt{3}}\right)+
\nonumber\\
& u\left(18-x^{2}\left(-u^{2}+v^{2}+3\right)\right)\cos\left(\frac{ux}{\sqrt{3}}\right)\sin\left(\frac{vx}{\sqrt{3}}\right)\Bigg]+
\nonumber\\
& 24x\sin(x)\left(6-x^{2}\left(-u^{2}-v^{2}+3\right)\right)\sin\left(\frac{ux}{\sqrt{3}}\right)\sin\left(\frac{vx}{\sqrt{3}}\right)+
\nonumber\\
& 24\cos(x)\left(x^{2}\left(u^{2}+v^{2}+3\right)-18\right)\sin\left(\frac{ux}{\sqrt{3}}\right)\sin\left(\frac{vx}{\sqrt{3}}\right)\Bigg\}-\frac{27\left(u^{2}+v^{2}-3\right)}{u^{2}v^{2}}+
\nonumber\\
& \frac{\left(27\left(u^{2}+v^{2}-3\right)^{2}\right)}{4u^{3}v^{3}}\Bigg\{-\text{Ci}\left[\left|1-\frac{u+v}{\sqrt{3}}\right|x\right]+\ln\left|\frac{3-(u+v)^{2}}{3-(u-v)^{2}}\right|+
\nonumber\\
& \text{Ci}\left[\left(\frac{u-v}{\sqrt{3}}+1\right)x\right]-\text{Ci}\left[\left(\frac{u+v}{\sqrt{3}}+1\right)x\right]+\text{Ci}\left[\left(1-\frac{u-v}{\sqrt{3}}\right)x\right]\Bigg\}.
\label{Ts}
\end{align}
Furthermore, the sine-integral $\text{Si}(x)$ and cosine-integral $\text{Ci}(x)$ functions are defined as
\begin{equation}
 \label{SiCi}
 \text{Si}(x) \equiv \int_{0}^{x}\frac{\sin(y)}{y}dy,\qquad\text{Ci}(x) \equiv -\int_{x}^{\infty}\frac{\cos(y)}{y}dy.
\end{equation}
The function $\tilde{T}_{s}(u,v,1)$ which also appears in Eq. \eqref{IRD2b} is given by
\begin{equation}
 \label{Tst}
 \tilde{T}_{s}(u,v,1)=T_{s}(u,v,1)+\frac{27\left(u^{2}+v^{2}-3\right)}{u^{2}v^{2}}-\frac{27\left(u^{2}+v^{2}-3\right)^{2}}{4u^{3}v^{3}}\ln\left|\frac{3-(u+v)^{2}}{3-(u-v)^{2}}\right|.
\end{equation}
The current energy spectra of the scalar-induced GWs is given by \cite{Inomata:2019-a}
\begin{align}
\label{OGW0}
\Omega_{\rm GW_0}h^2 = 0.83\left( \frac{g_{*}}{10.75} \right)^{-1/3}\Omega_{\rm r_0}h^2\Omega_{\rm{GW}}(\eta_c,k)\,,
\end{align}
where $\Omega_{\rm{r_0}}h^2\simeq 4.2\times 10^{-5}$ is the radiation density parameter at the present time, while $g_*\simeq106.75$ is the effective degrees of freedom in the energy density at $\eta_c$. The frequency in terms of the wavenumber can be written as
\begin{align}
\label{k_to_f}
f=1.546 \times 10^{-15}\left( \frac{k}{{\rm Mpc}^{-1}}\right){\rm Hz}.
\end{align}

\begin{figure}[H]
\begin{minipage}[b]{1\textwidth}
\subfigure[\label{fig-phi-o} ]{ \includegraphics[width=0.48\textwidth]%
{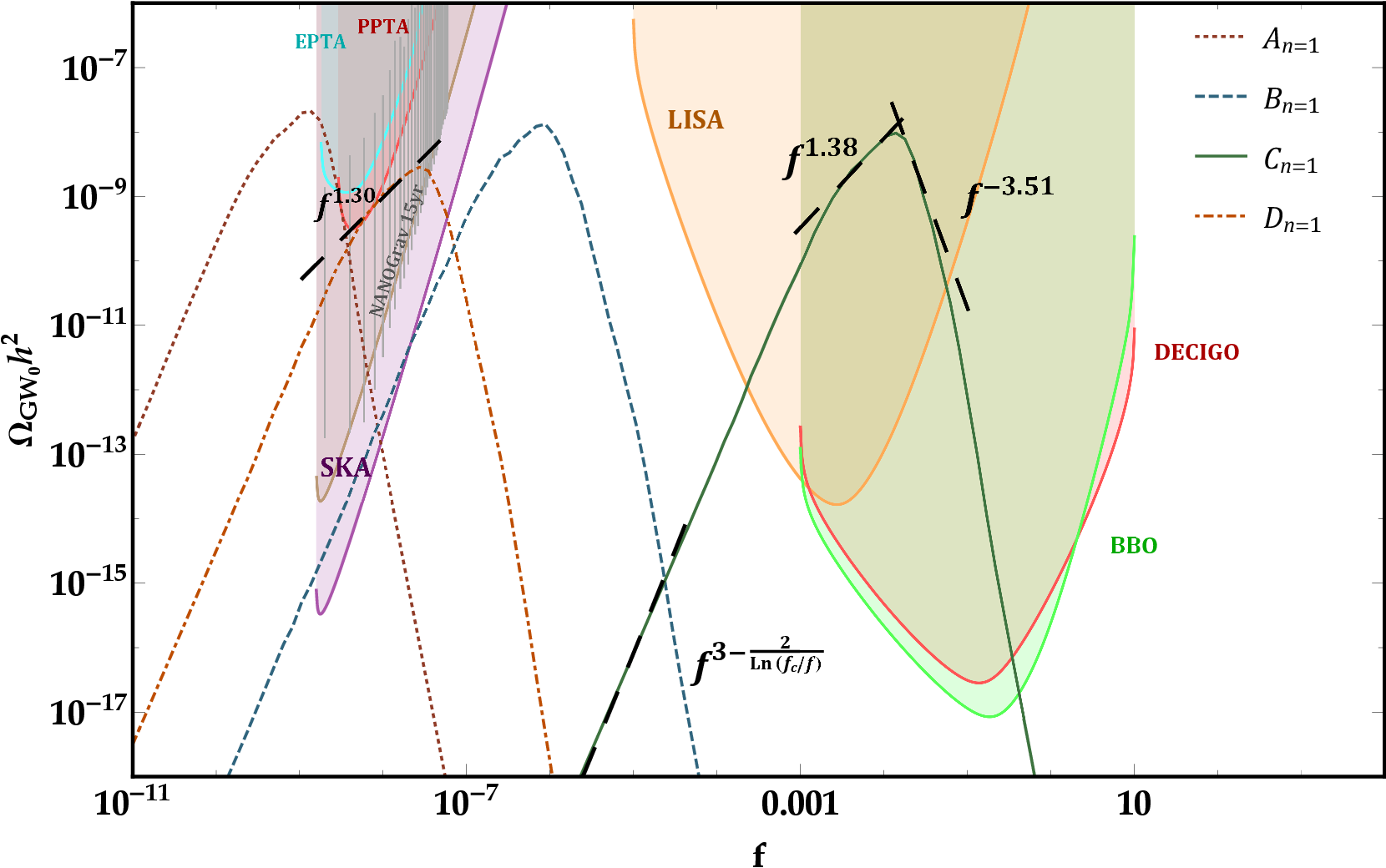}}
\subfigure[\label{fig-phi2-o}]{ \includegraphics[width=.48\textwidth]%
{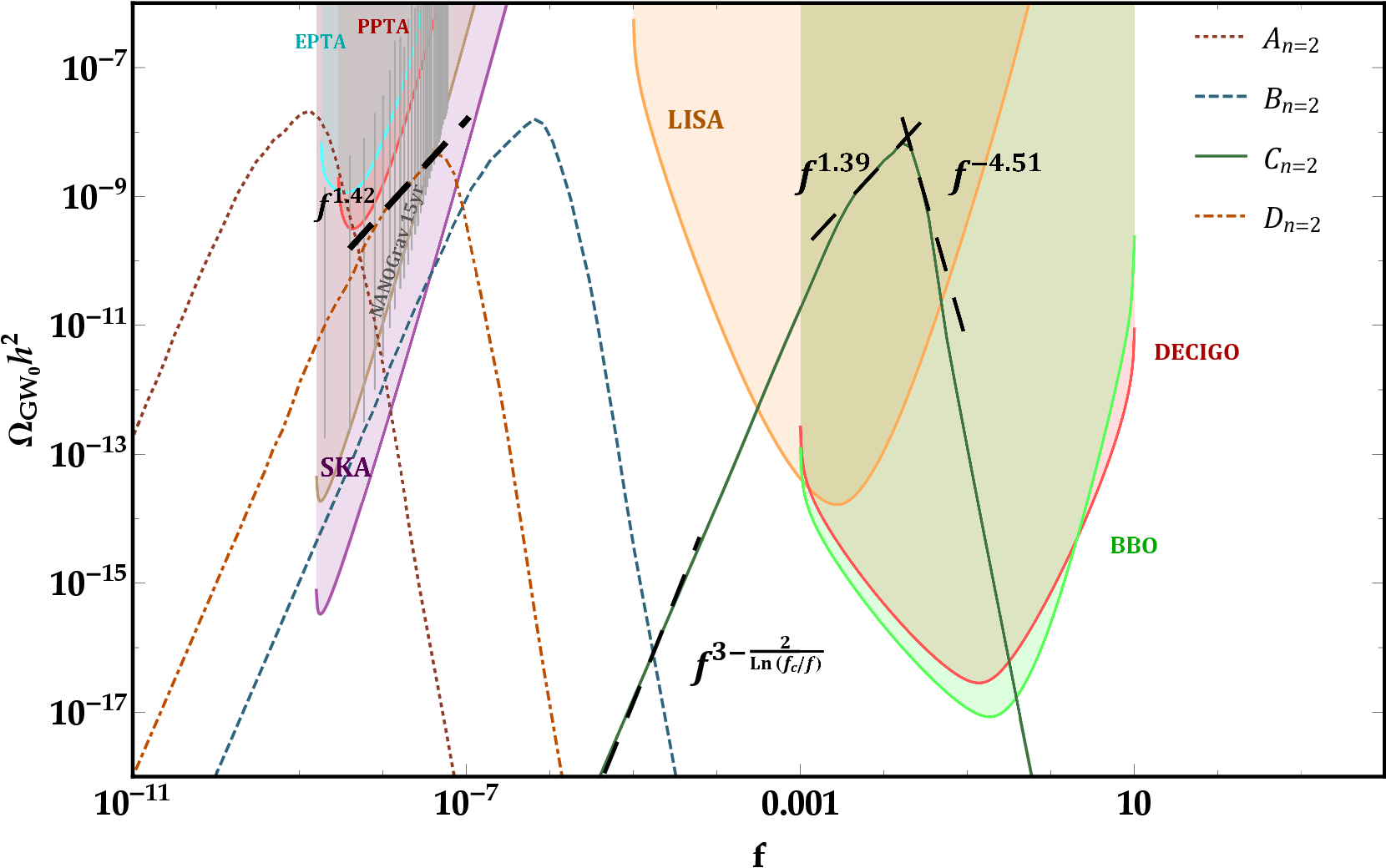}}
\subfigure[\label{fig-phi3-o}]{
 \includegraphics[width=.48\textwidth]%
{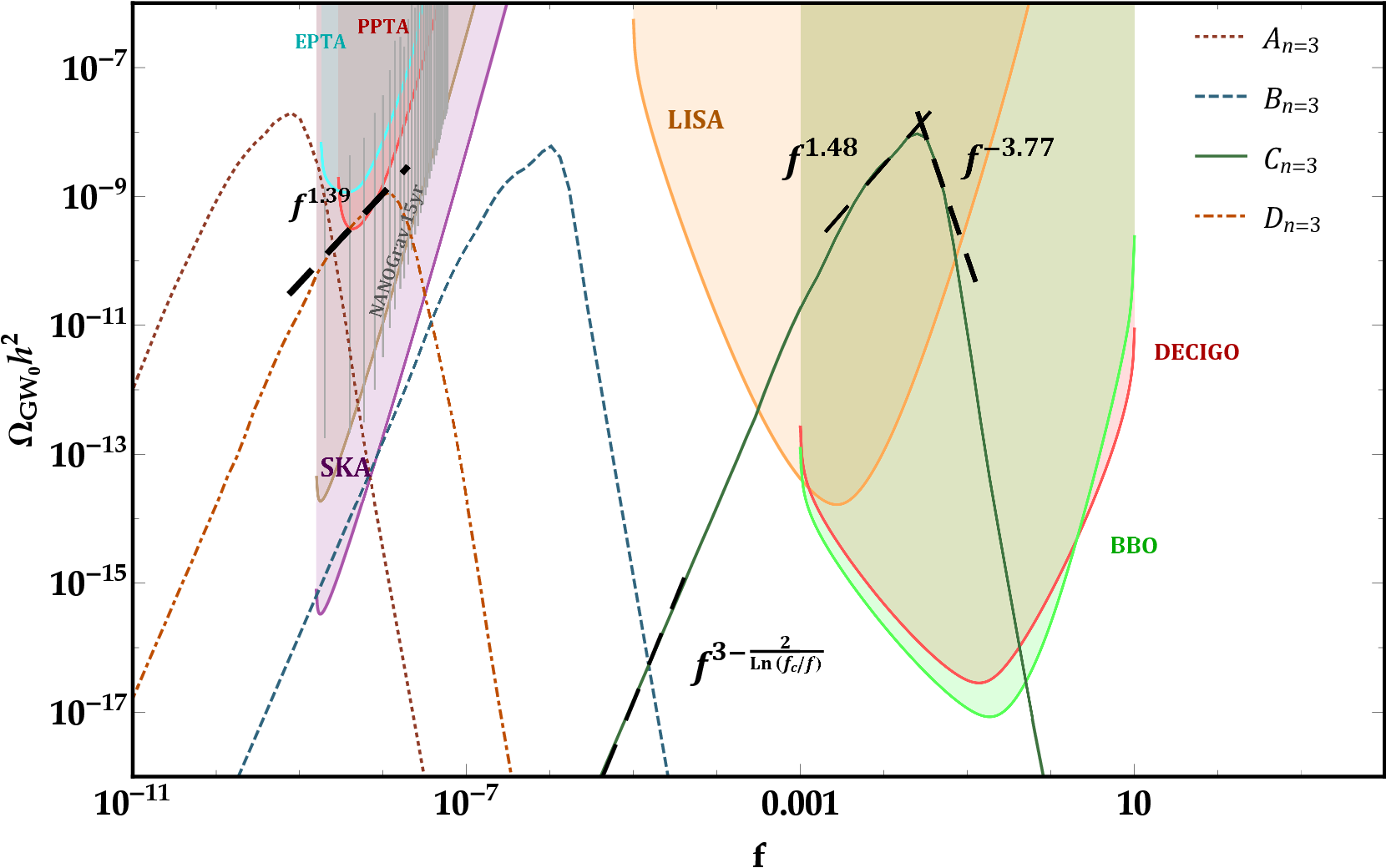}}
\subfigure[\label{fig-phi4-o}]{
 \includegraphics[width=.49\textwidth]%
{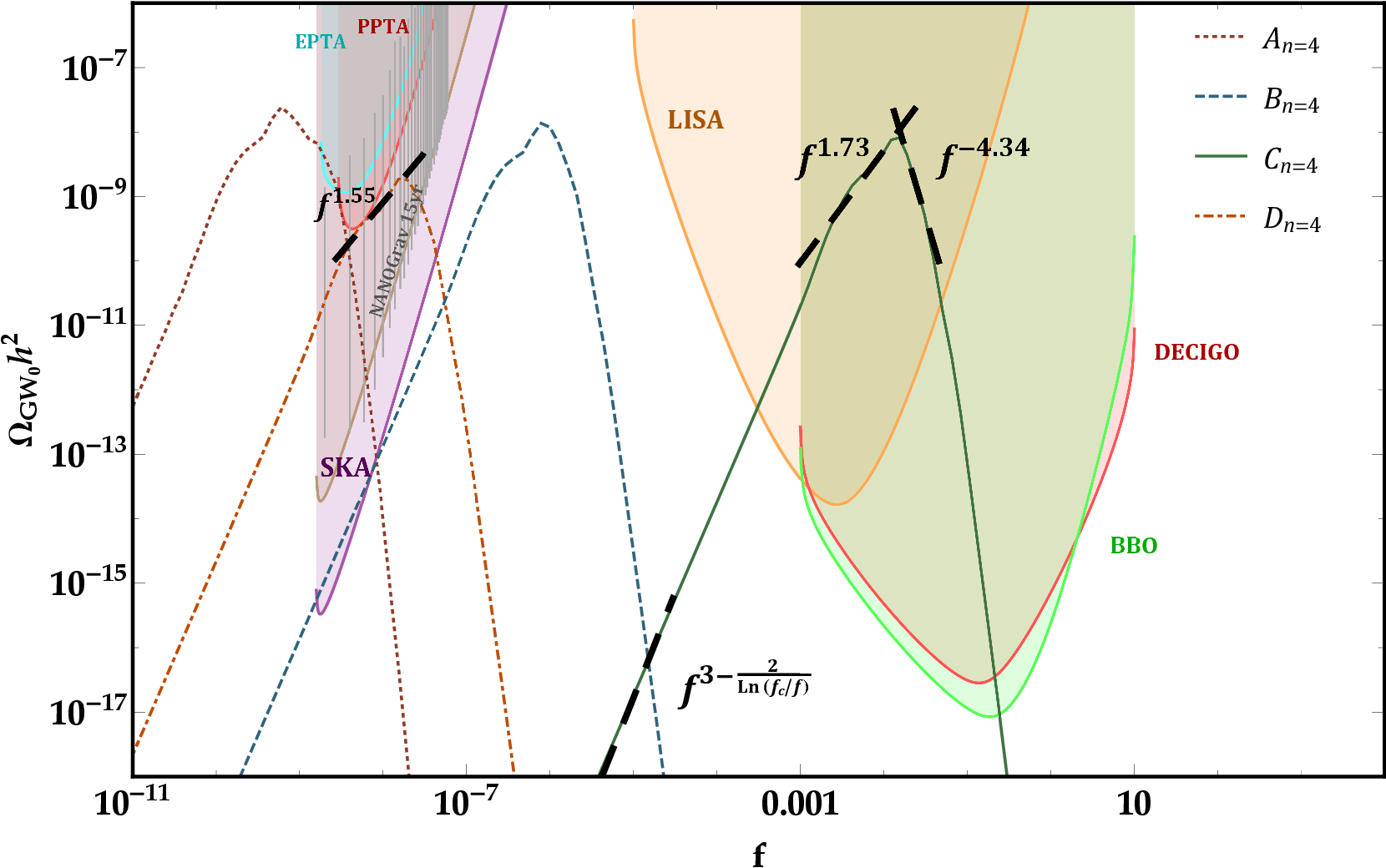}}
\end{minipage}
\caption{The induced GWs energy density parameter $\Omega_{\rm GW{_0}}$ versus the frequency pertinent to the fractional power-law potentials with (a) $n=1$ (b) $n=2$ (c) $n=3$ and (d) $n=4$ for parameter sets of Table \ref{tab1-phi}. The dotted, dashed, solid, and dash-dotted lines in each figure are corresponding to cases $A_n$, $B_n$, $C_n$, and $D_n$, respectively. The accuracy of our predictions can be tested by Pulsar Timing Array (PTA) collaboration \cite{NG15a,NG15b,NG15c,NG15d,epta1:add,epta2:add,epta3:add,epta4:add,epta5:add}, the Square Kilometer Array (SKA) \cite{ska,skaCarilli:2004,skaWeltman:2020}, Laser Interferometer Space Antenna (LISA) \cite{ligo-a,ligo-b,lisa,lisa-a}, BBO \cite{Yagi:2011BBODECIGO,Yagi:2017BBODECIGO,Harry:2006BBO,Crowder:2005BBO,Corbin:2006BBO}, and DECIGO observatories \cite{Yagi:2011BBODECIGO,Yagi:2017BBODECIGO,Seto:2001DECIGO,Kawamura:2006DECIGO,Kawamura:2011DECIGO}.
  }\label{fig-omega}
\end{figure}

Now, we can estimate the energy density of scalar-induced GWs at the present time using Eqs. (\ref{OGW}) to (\ref{k_to_f}) and by employing the power spectrum obtained from the MS equation (\ref{M.S}).
The results of $\Omega_{\rm{GW}_0}$ for all cases in this study are depicted in Fig. \ref{fig-omega}.

Figure \ref{fig-omega} shows that the energy density of scalar-induced gravitational waves for the cases $A_{\rm{n}}$ and $B_{\rm{n}}$ have peaks around the $10^{-9}$ and $10^{-6}\rm{Hz}$, respectively. These gravitational waves can be detected through the SKA observations.
In addition, for the cases $C_{\rm{n}}$, the peaks of $\Omega_{\rm GW_0}$ fall in the $\rm{dHz}$ band and can be examined by observatories such as LISA, DECIGO, and BBO.
More interestingly enough, is that the peaks of $\Omega_{\rm GW_0}$ for the cases $D_{\rm{n}}$ locate around the range of $f \sim 10^{-8} \rm{Hz}$, which are aligned with the frequency band probed by the PTA observations, as shown in Fig. \ref{fig-omega}.
Therefore, these scalar-induced GWs can be regarded as the origin of the detected signals by the PTA observations.

Recent studies have confirmed that the energy density of scalar-induced GWs can be parameterized as  $\Omega_{\rm GW_0} (f) \sim f^{\beta} $ \cite{Xu:2020,Fu:2020}.
It is conventional to parameterize the power index of the PTA GWs signal as $\beta = 5 - \gamma$  \cite{NG15a}. Furthermore, the best-fit value of $\gamma$ based on the PTA data reads $\gamma = 3.2 \pm 0.6$ \cite{NG15a}. Our numerical estimations for the cases $D_{\rm n}$, which have a peak in this specific area, are completely compatible with the observational data. The obtained values for $\gamma$ have been listed in Table \ref{tabNG}.

\begin{table}[H]
\centering
\caption{The values of $\beta$ and the spectral index $\gamma$ in $\Omega_{\rm{GW_0}}(f)\sim f^{5-\gamma}$ for the cases ${\rm D}_{\rm n}$.}
\scalebox{1}[1]
{\begin{tabular}{ccc}
\thickhline
Sets  & $\qquad\qquad$ $\beta$ $\qquad\qquad$ & $\quad$ $\gamma=5-\beta$ $\quad$\tabularnewline
\thickhline
\hline
$\rm{Case~}{\rm D}_{\rm n=1}$ & $1.30$ & $3.70$\\
\hline
$\rm{Case~}{\rm D}_{\rm n=2}$ & $1.42$ & $3.58$\\
\hline
$\rm{Case~}{\rm D}_{\rm n=3}$ & $1.39$ & $3.61$\\
\hline
$\rm{Case~}{\rm D}_{\rm n=4}$ & $1.55$ & $3.45$\\
\thickhline
\end{tabular}}
\label{tabNG}
\end{table}

In addition, the behaviour of $\Omega_{\rm GW_0}$ in other cases can also be characterized as a power-law function $\Omega_{\rm GW_0} (f) \sim f^{\beta} $.
Here, we consider the cases $C_{\rm n}$ for $n=1,\,2,\,3$, and $4$ to investigate the behaviour of $\Omega_{\rm GW_0}$.
Our results, listed in Table \ref{tableGWs}, verify that $\Omega_{\rm GW_0}$ can be parameterized as a power-law function in terms of frequency.
Furthermore, the results confirm that $\Omega_{\rm GW_0} \sim f^{3-2/\ln(f_c/f)}$ in the infrared limit $f\ll f_{c}$, which is consistent with the results of \cite{Yuan:2020,shipi:2020}.

%
%
%
%

%

\begin{table}[H]
\centering
\caption{The values of $f_{c}$, $\Omega_{\rm GW_0}(f_{c})$, and the power indices $\beta$ for the cases $C_{\rm n}$ in frequency ranges $f\ll f_{c}$, $f<f_{c}$, and $f>f_{c}$.
}
\scalebox{1}[1] {
\begin{tabular}{cccccc}
\thickhline
Sets  & $\qquad\qquad$ $f_{c}$ $\qquad\qquad$ & $\quad$ $\Omega_{\rm GW_0}\left(f_{c}\right)$ $\quad$ & $\quad$ $\beta_{f\ll f_{c}}$ $\quad$ & $\quad$ $\beta_{f<f_{c}}$ $\quad$ & $\quad$ $\beta_{f>f_{c}}$\\
\thickhline
$C_{\rm n=1}$ & $1.4\times10^{-2}$ & $9.59\times10^{-9}$ & $3.04$ & $1.38$ & $-3.51$\\
\hline
$C_{\rm n=2}$ & $1.6\times10^{-2}$ & $6.57\times10^{-9}$ & $3.03$ & $1.39$ & $-3.03$ \\
\hline
$C_{\rm n=3}$ & $2.5\times10^{-2}$ & $9.37\times10^{-9}$ & $3.02$ & $1.48$ & $-3.77$ \\
\hline
$C_{\rm n=4}$ & $1.5\times10^{-2}$ & $8.09\times10^{-9}$ & $3.09$ & $1.73$ & $-4.34$ \\
\thickhline
\end{tabular}
}
\label{tableGWs}
\end{table}

\section{Conclusions}\label{sec7}
Here, we investigated the PBHs formation in non-minimally coupling Gauss-Bonnet inflationary model driven by power-law potentials.
During inflation, if the scalar field enters the USR regime, the primordial curvature can significantly increase, consequently leading to the formation of PBHs.
At the USR phase ($\epsilon_2 > 1$), the slow-roll approximation is violated, as shown in Figs.~\ref{phi-e2}.
Hence, to estimate the exact value of the scalar power spectrum, the MS equation must be numerically solved.
To examine the versatility of the provided approach, we explored different values of the power index $n$ in the potential function, in particular $n = 1, 2, 3, 4$.
The motivation for considering the power-law potentials in GB gravity arises from the fact that these potentials are completely ruled out by Planck measurements in the standard inflationary model \cite{akrami:2018}.

In GB inflation, selecting an adequate coupling function can lead to a proper enhancement in the scalar power spectrum at small scales.
However, the selected function must ensure compatibility between the model and Planck observations at the CMB scale simultaneously.
To achieve both of these aims, our approach is to employ a two-part coupling function, given by $\xi (\phi) = \xi_{1}(\phi) \big( 1 + \xi_{2}(\phi)\big)$.
Here, $\xi_1(\phi)$ is responsible for ensuring that our model be compatible with the CMB constraints on $n_s$ and $r$. Additionally, $\xi_{2}(\phi)$ is responsible for enhancing the scalar curvature at small scales.

For each potential, we found four sets of parameters, which are reported in Table \ref{tab1-phi}.
The results of the numerical calculations listed in Table \ref{tab2-phi} specify that the present model is in good agreement with observations at the CMB scales.
Simultaneously, the present model shows a significant enhancement at small scales and can lead to PBH generation.
Moreover, our analysis demonstrated that our model satisfies the swampland criteria.
In addition, we studied the implications of the reheating stage.
The obtained results listed in Table \ref{tab9} proved that the peak scales re-enter the horizon after the reheating stage.
Hence, we could employ the mathematical formalism, which is valid during the RD era.

For the cases $A_{\rm n}$, the generated PBHs have masses in the range of $(1-10)M_{\odot}$.
The abundance of PBHs on this mass scale can be bounded by the upper limits on the LIGO merger rate, as shown in Fig. \ref{fpbh-figs}.
The cases $B_{\rm n}$ have masses ranging from $\mathcal{O}(10^{-6})M_{\odot}$ to $\mathcal{O}(10^{-5})M_\odot$, with an abundance on the order of $\mathcal{O}(10^{-2})$. These values are completely consistent with the allowed area of microlensing events in the OGLE data.
Furthermore, our results show that the cases $C_{\rm n}$ with masses around $\mathcal{O}(10^{-14})M_\odot$ can take into account for almost all of the dark matter in the universe (see Fig. \ref{fpbh-figs}).
In addition, the abundance of PBHs for the cases $D_{n}$ with masses around the $\mathcal{O}(10^{-3} - 10^{-2})M_\odot$ are negligible.

In the next, we investigated the generation of scalar-induced GWs in our model.
Our calculations indicated that the current energy density parameter $\Omega_{\rm GW_0}$ of the scalar-induced GWs associated with the cases $A_{\rm n}$ and $B_{\rm n}$ fall within the sensitivity region of SKA observations.
Furthermore, the peaks of $\Omega_{\rm GW_0}$ for the cases $C_{\rm n}$ are generated in the dHz frequency band.
Hence, as depicted in Fig. \ref{fig-omega}, observatories such as LISA, DECIGO, and BBO can detect these GWs.
However, the most intriguing case lies with the $D_{\rm{n}}$ scenarios.
The peaks of $\Omega_{\rm GW_0}$ for cases $D_{\rm n}$ fall within the sensitivity region of PTA observations. Consequently, these scalar-induced GWs can be considered as the source of observed signal in PTA data set.

Also, the spectrum of $\Omega_{\rm GW_0}$ can be parameterized as a power-law function $\Omega_{\rm GW_0} (f) \sim f^{\beta}$.
It is worth noting that our results in cases $D_{\rm n}$ can satisfy the best-fit value of spectral index $\gamma=5-\beta$ reported by the PTA collaboration, as shown in Table \ref{tabNG}.
In addition, we investigated the power-law behavior of $\Omega_{\rm GW_0}$ for the cases $C_{\rm n}$ as shown in Table \ref{tableGWs}.
Moreover, in the infrared region $f\ll f_{c}$, we obtained $\Omega_{\rm GW_0} \sim f^{3-2/\ln(f_c/f)}$, which is consistent with the results of \cite{Yuan:2020,shipi:2020}.

\subsection*{Acknowledgements}
The authors thank the referee for his/her valuable comments.


\end{document}